\documentclass[onecolumn]{aastex631}
\usepackage{longtable}
\usepackage{gensymb}
\usepackage{amsmath}
\usepackage{natbib}
\usepackage[utf8]{inputenc}
\DeclareUnicodeCharacter{03C3}{$\sigma$}
\DeclareUnicodeCharacter{200B}{\hspace{0pt}}
\DeclareUnicodeCharacter{00B1}{\pm}
\DeclareUnicodeCharacter{2212}{-} 
\shorttitle{Searching TTVs in Exoplanet WASP-19b}
\shortauthors{Biswas et al.}

\begin{document}

\title{Investigating Transit Timing Variations in the Ultra-short Period Exoplanet WASP-19b}

\author[0009-0003-8446-4557]{Shraddha Biswas}
\affiliation{Indian Centre For Space Physics \\
466 Barakhola, Singabari road, Netai Nagar, Kolkata, West Bengal 700099}
\email{hiyabiswas12@gmail.com}

\author[0000-0001-7359-3300]{Ing-Guey Jiang}
\affiliation{Department of Physics and Institute of Astronomy \\
National Tsing Hua University, Hsinchu 30013, Taiwan}
\email{jiang@phys.nthu.edu.tw}

\author[0000-0001-8677-0521]{Li-Chin Yeh}
\affiliation{Institute of Computational and Modeling Science \\
National Tsing Hua University, Hsinchu 30013, Taiwan}
\email{lichinyeh@mx.nthu.edu.tw}

\author{Hsin-Min Liu}
\affiliation{Department of Physics and Institute of Astronomy \\
National Tsing Hua University, Hsinchu 30013, Taiwan}

\author{Kaviya Parthasarathy}
\affiliation{Department of Physics and Institute of Astronomy \\
National Tsing Hua University, Hsinchu 30013, Taiwan}

\author[0000-0001-8452-7667]{Devesh P. Sariya}
\affiliation{Department of Physics and Institute of Astronomy \\
National Tsing Hua University, Hsinchu 30013, Taiwan}

\author[0000-0002-8988-8434]{D. Bisht}
\affiliation{Indian Centre For Space Physics \\
466 Barakhola, Singabari road, Netai Nagar, Kolkata, West Bengal 700099}
\email{devendrabisht297@gmail.com}

\author{Mohit Singh Bisht}
\affiliation{Indian Centre For Space Physics \\
466 Barakhola, Singabari road, Netai Nagar, Kolkata, West Bengal 700099}

\author{A. Raj}
\affiliation{Indian Centre For Space Physics \\
466 Barakhola, Singabari road, Netai Nagar, Kolkata, West Bengal 700099}
\affiliation{Uttar Pradesh State Institute of Forensic Science (UPSIFS) \\
Aurawan, P.O. Banthra, Lucknow 226401, U.P, India}

%% Note that the \and command from previous versions of AASTeX is now
%% depreciated in this version as it is no longer necessary. AASTeX 
%% automatically takes care of all commas and "and"s between authors names.

%% AASTeX 6.31 has the new \collaboration and \nocollaboration commands to
%% provide the collaboration status of a group of authors. These commands 
%% can be used either before or after the list of corresponding authors. The
%% argument for \collaboration is the collaboration identifier. Authors are
%% encouraged to surround collaboration identifiers with ()s. The 
%% \nocollaboration command takes no argument and exists to indicate that
%% the nearby authors are not part of surrounding collaborations.

%% Mark off the abstract in the ``abstract'' environment. 
\begin{abstract}
In this study, we present a comprehensive analysis of transit timing variations (TTVs) in the ultra-short-period gas giant WASP-19b, which orbits a G-type main-sequence star. Our analysis is based on a dataset comprising 204 transit light curves obtained from the Transiting Exoplanet Survey Satellite (TESS), the Exoplanet Transit Database (ETD), and the ExoClock project, supplemented by 18 publicly available light curves. Mid-transit times were extracted from these data, and an additional 98 mid-transit times compiled from the literature were incorporated, resulting in a combined dataset spanning approximately 14 years. After excluding light curves significantly impacted by stellar activity, such as starspot anomalies, the final dataset consisted of 252 high-quality mid-transit times. Initial inspection of the transit timing residuals using an apsidal precession model suggested the possible presence of an additional planetary companion. However, subsequent frequency analysis and sinusoidal model fitting indicate that the observed TTVs are more consistently explained by apsidal precession of WASP-19b's orbit. We also considered alternative mechanisms, including the Applegate mechanism and the Shklovskii effect. Our findings suggest that stellar magnetic activity, potentially linked to the Applegate mechanism, may also contribute to the observed timing variations. To further constrain the origin of the TTVs and assess the contributions of these mechanisms, continued high-precision photometric monitoring of the WASP-19 system is strongly recommended.
\end{abstract}

%% Keywords should appear after the \end{abstract} command. 
%% The AAS Journals now uses Unified Astronomy Thesaurus concepts:
%% https://astrothesaurus.org
%% You will be asked to selected these concepts during the submission process
%% but this old "keyword" functionality is maintained in case authors want
%% to include these concepts in their preprints.

%% From the front matter, we move on to the body of the paper.
%% Sections are demarcated by \section and \subsection, respectively.
%% Observe the use of the LaTeX \label
%% command after the \subsection to give a symbolic KEY to the
%% subsection for cross-referencing in a \ref command.
%% You can use LaTeX's \ref and \label commands to keep track of
%% cross-references to sections, equations, tables, and figures.
%% That way, if you change the order of any elements, LaTeX will
%% automatically renumber them.
%%
%% We recommend that authors also use the natbib \citep
%% and \citet commands to identify citations.  The citations are
%% tied to the reference list via symbolic KEYs. The KEY corresponds
%% to the KEY in the \bibitem in the reference list below. 

\section{Introduction} \label{sec:intro}

In 1995, the Geneva-based astronomers Michel Mayor and Didier Queloz first discovered an exoplanet (\citealt{1995Natur.378..355M}) orbiting the solar-like main-sequence star 51 Pegasi, utilizing the radial velocity method. This exoplanet, designated 51 Pegasi b, is a massive hot Jupiter that completes an orbit around its host star in approximately four days. Since this initial discovery, over two decades have passed, during which a continuous stream of exoplanet discoveries has taken place. This remarkable progress has not only propelled exoplanetary science into one of the fastest-growing fields of astrophysics but has also significantly enhanced our understanding of planetary physics.

According to the NASA Exoplanet Archive, as of March 30, 2025, a total of 5,867 exoplanets have been confirmed through various detection methods. Of these, 4,360 exoplanets (approximately 75\%) were identified using the transit photometry technique (\citealt{1984Icar...58..121B}). An exoplanet transit occurs when a planet passes in front of its host star as observed from Earth, resulting in a temporary decrease in the star's brightness. The first confirmed detection of a complete exoplanet transit was made in the year 2000 in the HD 209458 system (\citealt {2000ApJ...529L..41H, 2000ApJ...529L..45C}). 

Among various types of exoplanets, the transiting hot Jupiters are a ``goldmine" for astronomers and they are the most intensely researched category of exoplanets. Theoretically, it is expected that most hot Jupiters—typically characterized by circular orbits and masses similar to Jupiter—undergo strong tidal interactions with their host stars due to their short orbital periods ($P < 10\,~$days) and close proximities ($a < 0.1\,~$au). These interactions lead to energy dissipation, and the tidal bulges raised by the hot Jupiters on their host stars generate torques that transfer orbital angular momentum from the planet's orbit to the stellar spin (\citealt {1973ApJ...180..307C, 1996ApJ...470.1187R, 2003ApJ...596.1327S, 2009ApJ...692L...9L, 2014ARA&A..52..171O}. Due to the star's rotational angular momentum being less than one-third of its orbital angular momentum \citep{1980A&A....92..167H}, the transfer of orbital angular momentum from the planet to the star can increase the star's rotational velocity, causing the planetary orbit to shrink and potentially leading to tidal disruption, known as orbital decay (\citealt{2017AJ....154....4P, 2020AJ....159..150P, 2020ApJ...888L...5Y}). Recent studies have detected hints of orbital decay in several exoplanets, including Kepler-1658 b a decay rate of $ \dot{P}=-\,131~ \pm ~21$ ms/yr (\citealt{2022ApJ...941L..31V}), WASP-12b at $ \dot{P}=-\,3.7~ \pm ~0.5$  ms/yr (\citealt{2023ApJS..265....4K}), TrES-2b at $ \dot{P}=-\,5.58~ \pm ~1.81$ (\citealt{2024AJ....168..176B} ). There are several other exoplanets that also show potential for orbit decay. These exoplanets include HAT-P-51 b and HAT-P-53 b (\citealt{2024NewA..10602130Y}), HAT-P-19 b and HAT-P-32 b (\citealt{2022AJ....164..220H}), KELT-9 b (\citealt{2023A&A...669A.124H}), TrES-1 b (\citealt{2022AJ....164..220H, 2022ApJS..259...62I}), TrES-5 b (\citealt{2022ApJS..259...62I, 2024NewA..10602130Y}), and XO-3 b (\citealt{2022AJ....164..220H, 2022RAA....22e5005Y}).  Observing transits over a long temporal baseline (usually more than a decade) is particularly useful to detect a change in the dynamics of the star-planet system, resulting into shift in the transit times. These temporal shifts at the mid-point of transit events often lead to the delay or advancements of the mid-transit times and this phenomenon is known as transit timing variations (hereafter TTVs; \citealt{2010ApJ...716.1047B, Wu_2018}). Transit Timing Variations (TTVs) can arise due to various physical mechanisms, including planetary mass loss (\citealt {2015ApJ...813..101V, 2016CeMDA.126..227J}), apsidal precession of the planet's orbit (\citealt{2002ApJ...564.1019M, 2009ApJ...698.1778R}), Applegate mechanism (\citealt{1992ApJ...385..621A, 2010MNRAS.405.2037W}), starspots (\citealt{2004MNRAS.351..110W, 2009A&A...508.1011R, 2016AJ....151..150M}), secular changes in the orbit due to short-term transit-timing variations caused by the gravitational influence of a nearby planetary companion or star in close orbit to the hot Jupiter (\citealt{2009A&A...506..369B, Gibson_2009}), as well as the presence of other planetary bodies, such as exo-moons. TTVs can also arise from the acceleration of wide-orbiting companions either towards or away from the Earth. This acceleration induces a wobble in the transiting planet's motion around its barycenter, thereby altering the observed transit times due to variations in the light-travel time (\citealt{2019AJ....157..217B, 2020ApJ...893L..29B, 2021AJ....161...72T, 2022AJ....163..281T}).

For this study, we have selected WASP-19b with M$_P$ = 1.168\, M$_J$ and R$_P$ = 1.386\, R$_J$ (\citealt{2011ApJ...730L..31H}), an ultra-short-period gas giant with a period of 0.78 days, formally designated as Banksia. This exoplanet was discovered using the transit method by the Wide Angle Search for Planets (WASP)-South observatory in December 2009 (\citealt{2010ApJ...708..224H}). It orbits a bright, active G8V star in the southern hemisphere, with apparent magnitude V = 12.3, stellar mass M$_\ast$ = 0.97\ M$_{\sun}$, and stellar radius R$_{\ast}$= 0.99 \ R$_{\sun}$, in a close-in orbit with a semi-major axis of 0.01655 AU (\citealt{2011ApJ...730L..31H}). Giant exoplanets (M$_P > $ 0.5\, M$_J$) with exceptionally short orbital periods of less than 3 days, located on extremely tight orbits, are classified as Ultra-hot Jupiters (UHJs). Due to their extremely short orbital periods, UHJs may exhibit a higher frequency of transits and eclipses, displaying distinct characteristics compared to typical Hot Jupiters. Being an ultra-short hot Jupiter,  WASP-19b is an interesting candidate for tidal decay studies and also plays a crucial role in long-term dynamical stability. In past, many authors have studied this system and considered it as a strong candidate for exhibiting orbital decay (e.g. \citealt{2009Natur.460.1098H, 2013MNRAS.428.3671T, 2019MNRAS.490.4230S, 2020MNRAS.491.1243P, 2022A&A...668A.114R}).
\citet{2013MNRAS.428.3671T} and \citet{2019MNRAS.482.2065E} proposed the presence of starspots in WASP-19b. Recent studies have raised debates regarding whether WASP-19b is undergoing orbital decay or apsidal precession. While \citet{2020AJ....159..150P}, and \citet{2023Univ...10...12K} found statistically significant evidence of orbital period shrinkage, \citet{2020AJ....159..150P} also cautioned that stellar activity could introduce systematic errors, potentially skewing results. In contrast, \citet{2020MNRAS.491.1243P} and \citet{2022A&A...668A.114R} dismissed orbital shortening as a plausible cause of TTVs. More recently, \citet{2024NewA..10602130Y} and \citet{2025arXiv250105704M} supported these conclusions, while \citet{2024A&A...684A..78B} proposed apsidal precession as a possible explanation.

These conflicting results highlight the importance for long-term, continuous observations of short-period, massive planets, which also increase the likelihood of detecting orbital decay. Regular transit monitoring helps reduce uncertainty in predicted transit times by refining ephemerides. Given WASP-19b's extensive 15-year observation baseline and nearly 7,200 transit epochs, along with its ongoing study of tidal decay, we identified it as an ideal candidate for inclusion in our study. This study aims to investigate Transit Timing Variations (TTVs) in the WASP-19 system by combining ground-based and space-based observations. Space telescopes provide high-precision data across a broad wavelength range, free from atmospheric distortions and light pollution. Ground-based telescopes are essential for continuous observation, filling data gaps and covering longer timescales, which is critical for systems like WASP-19b. While space missions have limited operational lifetimes, ground-based observations can supplement data during off-times or when space telescopes are unavailable, enabling a more comprehensive analysis of TTVs in this system. For space based observational data, we utilize the Transiting Exoplanet Survey Satellite (TESS; \citealt{2014SPIE.9143E..20R}), a multi-sector survey telescope that provides high-quality, high-precision photometric data compared to ground-based telescopes. TESS has already gathered data on the exoplanet WASP-19b across three observational sectors. In this study, along with TESS, we have also integrated data from all available data resources, for e.g., we have also incorporated some light curves from the publicly available databases Exoplanet Transit Database (ETD; \citep{2010NewA...15..297P}), Exoclock network (\citealt{2023ApJS..265....4K}). We have also included all the archival mid-time points to increase our observational time line. The details of all these data are represented in Section \ref{sec:literature observation}. 

The outline of this paper is as follows: Section \ref{sec:observation} details the observational data from TESS, ETD, Exoclock, including their data processing and the transit time lists derived from the literature. Section \ref{sec:light_curve_analysis} outlines the methods and procedures used to analyze the 205 light curves from TESS, ETD, and Exoclock. Section \ref{sec:Stellar Spots} discusses the stellar activity on WASP-19b. Section \ref{sec:timing_analysis} provides a brief overview of the transit timing analysis, where three timing models are fitted to the data, alongside a frequency analysis for potential secondary planetary companions, and the fitting of a sinusoidal model. Section \ref{sec:results} presents the results of the TTV analysis. Finally, in Section \ref{sec:remarks}, we provide conclusions about this work. 

\section{Observational Data} \label{sec:observation}
\subsection{TESS Observations}

    TESS is equipped with four identical refractive cameras, each optimized to capture wide-field images of the sky. Together, these cameras provide a combined field of view of $24 \times 96\ \text{degrees}$, enabling the satellite to conduct a comprehensive sky survey. We have collected total 327 mid-transit times for our work, among which 116 full transit light curves were extracted from TESS with a two-minute cadence in a total of four sectors - 9, 36, 62, and 63 with exposure time varying 2 minutes upto 30 minutes during the time interval 2019, February 28 to 2023, March 10, using the HLSP Photometer and TESS photometer attached to the telescope. Table \ref{tab:1} provides information on the observation time intervals, the number of observed full transits, and the number of data points for each sector obtained from the TESS database. TESS was designed as the successor to the Kepler mission, continuing the search for exoplanets as well as facilitating the follow-up program through the detection of transit events. The exoplanet host star, WASP-19 (TOI 655), is designated as TIC 35516889 in the TESS Input Catalog (\citealt{2018AJ....156..102S}). The TESS light-curve files of WASP-19b have been downloaded from the Mikulski Archive for Space Telescopes (MAST)\footnote{All the TESS data used in this paper can be found in MAST \citep{10.17909/t9-nmc8-f686}.}. , which is a NASA funded astronomical data archive repository that houses and distributes astronomical data from a variety of space-based telescopes. MAST was named after former U. S. Senator Barbara Ann Mikulski. For downloading the TESS data, we applied the same approach as adopted by \citet{2024AJ....168..176B}. 

    \begin{table*} [htbp]
\begin{center}
\caption {Log of Sector specific TESS photometric observations of WASP-19 utilized in our analysis} 
\label{tab:1}
\begin{tabular}{ccccc}
%\bottomrule light curves
%\bottomrule
\hline
Object Name & TESS sector & Time interval of observations & Number of Transits & Data points\\
\hline
 & No. 09  & 2019-02-28 -- 2019-03-26 & 29 &  15846\\
{WASP-19} & No. 36  & 2021-03-07 -- 2021-04-02 & 27		&	15347\\
& No. 62 & 2023-02-12 -- 2023-03-10	 & 27		&	16080\\
& No. 63  & 2023-03-10 -- 2023-04-05 & 33		&	18358\\
\hline
\hline
{Total} & & & 116 & 65631\\
\hline
\hline
\end{tabular}\\
\end{center}
\end{table*}

    To access the TESS light curve files into Presearch Data Conditioning Simple Aperture Photometry (PDCSAP; \citealt{2012PASP..124.1000S, 2012PASP..124..985S, 2014PASP..126..100S}) light curves, we utilized the TESS Science Processing Operations Center (SPOC) pipeline (\citealt{2016SPIE.9913E..3EJ}). In this analysis, the PDCSAP (Pre-search Data Conditioning Simple Aperture Photometry) light curves are preferred over the standard Simple Aperture Photometry (SAP) data, as they show significantly reduced scatter, minimize short-timescale flux variations, and effectively eliminate long-term non-astrophysical trends. Additionally, PDCSAP curves help mitigate systematic errors caused by poor instrumental performance. During the light curve analysis, transits that were incomplete or exhibited excessive noise were excluded. The absence of a complete transit impedes the precise determination of transit timings (\citealt{2013MNRAS.430.3032B}). To generate independent transit light curves, we divided the full dataset of detrended and normalized light curves into smaller segments centered around the predicted mid-transit times, with a window of $\pm ~$0.1 day. 

\subsection{Observational Data from ETD}
Since its launch in September 2008, the online repository, Exoplanet Transit Database (ETD) has been actively used by numerous amateur astronomers from observatories worldwide to upload thousands of their observed light curves. Till April 14, 2025, the database includes contributions from 1,717 observers and lists 62,424 objects. Given the extensive number of transits available for WASP-19b in the ETD, we selected only those with complete transit light curves and high-quality data flags (DQ $<3$) observed between 2010 and 2024. To ensure homogeneity and reliability in our analysis, we included only observations with a data quality index (DQ) of less than three. In total, we gathered 64 transit light curves from ETD (through 2024 May), which served as our secondary source for transit timing data and before doing light curve analysis, we normalised those light curves following the same procedure as mentioned in \citet{2024AJ....168..176B}. To account for atmospheric effects caused by interference from the Earth's atmosphere, we normalized the out-of-transit (OOT) flux values of the light curves by fitting a linear function to the OOT section, thereby bringing the values closer to unity. 
Although the Exoplanet Transit Database (ETD) is a valuable resource, it has some limitations, such as the use of JD/UTC or HJD/UTC instead of BJD/TDB for time measurements, and the lack of a feature for bulk downloading all transit light curves. To resolve these issues, we converted the time stamps of the 65 complete transits analyzed in this study from Julian Day (JD) or Heliocentric Julian Day (HJD) to Barycentric Julian Day (BJD) on the Barycentric Dynamical Time (TDB) timescale. This conversion corrects for relativistic time dilation and accounts for the Earth's motion. The details of all the ETD observations along with the observers name are shown in Table \ref{tab:6}. 

\subsection{Observational Data from Exoclock}
The Exoclock project, launched in September 2019, is an integrated, and interactive platform designed to facilitate the monitoring of transiting exoplanet timings through regular, real-time observations using small- and medium-scale telescopes. Additionally, it aims to monitor the ephemerides of targets selected for the Ariel Mission. It also promotes global collaboration among observatories to enhance the precision and coverage of exoplanet transit data. We collected a total of 24 complete light curves from the Exoclock Project and analysed those light curves to determine their mid-transit times. In this study, we integrated data from two networks: the Exoclock Project and the Exoplanet Transit Database (ETD). Although the number of Exoclock observations used in this analysis is limited, their contribution is valuable for extending the time baseline. We plan to incorporate additional data from the ETD and Exoclock platforms in future publications. Our derived mid-transit times are shown in Table \ref{tab:6}.

\subsection{Details of Literature Data}   \label{sec:literature observation}
In addition to the light curves acquired from TESS, ETD, and Exoclock, we have collected a total of 18 publicly available complete light curves and analyzed them, which include 12 from \citet{2013A&A...552A...2L}, 4 from \citet{2020A&A...636A..98C} and 2 from \citet{2024PSJ.....5..163A}. We compiled our derived 18 mid-transit times along with other 98 mid-transit times directly taken from the published literature to ensure a sufficiently long time span for detecting potential changes in the orbital period and to aid in refining the orbital ephemerides. These mid-transit times were sourced from various studies, with the following distribution: 1 from \citet{2010ApJ...708..224H}, 1 from \citet{2010A&A...513L...3A}, 1 from \citet{2011ApJ...730L..31H}, 1 from \citet{2013MNRAS.430.3422A}, 3 from \citet{2013MNRAS.428.3671T}, 1 from \citet{2013A&A...553A..49A}, 1 from \citet{2013A&A...552A...2L}, 23 from \citet{2013MNRAS.436....2M}, 2 from \citet{2011AJ....142..115D}, 2 from \citet{2016ApJ...823..122W}, 6 from \citet{2019MNRAS.482.2065E}, 3 from \citet{2020AJ....159..150P}, 1 from \citet{2015A&A...576L..11S}, a total of three mid-transit times are reported from \citet{2017Natur.549..238S}. However, two of these light curves were subsequently refitted by \citet{2020MNRAS.491.1243P}, and these re-analyzed light curves exhibit notably smaller error bars. Therefore, we have included only one mid-transit time from \citet{2017Natur.549..238S} in our analysis and 35 from \citet{2020MNRAS.491.1243P}, 7 from \citet{2024A&A...684A..78B}, and 9 from \citet{2024PSJ.....5..163A}. The majority of these 98 mid-transit times are taken directly from the databases of \citet{2024PSJ.....5..163A} and \citet{2024A&A...684A..78B}, as these studies provide detailed references to prior observations, clearly indicating the source for each reported mid-transit time. This comprehensive documentation facilitates the identification of the origin of each mid-transit measurement.

\section{\textbf{Modeling of the Transit Events}} \label{sec:light_curve_analysis}
To determine the individual mid-transit times (${T}_{m}$) and refine the stellar and planetary parameters of the WASP-19 system, we modeled a total of 223 transit light curves using the Transit Analysis Package (TAP; \citealt{2012AdAst2012E..30G}). These light curves include 116 transits from TESS, 65 from the Exoplanet Transit Database (ETD), and 24 from ExoClock, as well as 12 from \citet{2013A&A...552A...2L}, 4 from \citet{2020A&A...636A..98C} and 2 from \citet{2024PSJ.....5..163A}. This package allows us to fit the orbital and transit parameters using the analytic function of \citet{2002ApJ...580L.171M} through a Markov Chain Monte Carlo (MCMC) analysis.

For those light curves collected from ETD and Exoclock project, we considered only complete and high-quality transit light curves (LCs), which were publicly available to ensure the reliability of our results and for the sake of the homogeneity of the transit timing data sets, we redetermined all the mid-transit times of these LCs using TAP. This code accepts single or multiple light curves as input, but we loaded all the light curves of WASP-19b individually into TAP to derive the transit parameters.

To more robustly estimate parameter uncertainties and account for time-correlated noise in the light curves, TAP incorporates the wavelet-likelihood method developed by \citet{2009ApJ...704...51C}. To explore the space of the parameters for each light curve, we ran 5  Markov chain Monte Carlo (MCMC) chains of  $10^5$ steps each and removed the initial 10\% of each chain to account for burn-in.

To start the transit light-curve analysis, we ran TAP with the eccentricity (e) and the longitude of periastron ($\omega$) values fixed to zero, by following \citet{2009ApJ...704...51C}, and \citet{2010ApJ...711..374F}, and for the other parameters we followed the same approach as adopted by \citet{2024AJ....168..176B}, by varying these parameters, the orbital period (P), the semi-major axis scaled in stellar radii ($a/R_\ast$), orbital inclination (i), $u_1$ and $u_2$ under Gaussian penalties, and they are allowed to vary only between the $\pm 1\sigma$ values reported in the literature. The ratio of the planet-to-star radius (${R_p /R_\ast}$) and the midtransit time (${T}_{m}$ ) were allowed to float as completely free parameter. For all 89 transit light curves obtained from ETD and Exoclock, the linear ($u_1$) and quadratic ($u_2$) limb darkening coefficients (LDCs) were also fitted using the above mentioned method. The initial parameter setting is shown in Table \ref{tab:4} and the initial values of the LD coefficients calculated for different filters are listed in Table \ref{tab:3}.

\begin{table} [htbp]
\begin{center}
\caption{The Initial Values set for the Parameters}
\label{tab:4}
\begin{tabular}{ccc}
%\bottomrule
%\bottomrule
\hline
\hline
Parameter & Initial Value & During MCMC Chain \\
\hline
\hline
P (days) & 0.78883896$^{a}$ & A Gaussian Prior with $\sigma$ = 0.00000014$^{b}$ \\
i (deg) & 79.17$^{a}$ & A Gaussian Prior with $\sigma$ = 0.32$^{b}$\\
$a/R_\ast$ & 3.533$^{a}$ & A Gaussian Prior with $\sigma$ = 0.038$^{b}$\\
${R_p /R_\ast}$ & 0.14541$^{a}$ & Free\\
${T_m}$ & Set by eye & Free\\
$u_1$ & According to filter &   A Gaussian Prior with $\sigma$ = 0.028$^{c}$\\
$u_2$ & According to filter &   A Gaussian Prior with $\sigma$ = 0.024$^{c}$\\
\hline
\end{tabular}
\end{center}
{Notes:
$^a$ The initial values of the parameters P, i, $a/R_\ast$, and ${R_p /R_\ast}$ are directly adopted from \citet{2022A&A...668A.114R}.\\
$^b$ The priors of P, i, and $a/R_\ast$ taken from \citet{2022A&A...668A.114R}.\\
$^c$ The priors of $u_1$ and $u_2$ taken from \citet{2020A&A...636A..98C}.}\\
\end{table}

To calculate the initial values of the linear and quadratic LDCs ($u_1$ , $u_2$) for the TESS light curves of WASP-19b, we interpolated the coefficient effective temperature (${T}_{\rm eff}$), surface gravity ($log_{g}$), metallicity ([Fe/H]), and micro-turbulence velocity ($V_t$) from the tables of \citet{2017A&A...600A..30C}, where, ${T}_{\rm eff}=5500\,$ (\citealt{2010ApJ...708..224H}), $log_{g} = 4.421$ cm $s^{-2}$ (\citealt{2021MNRAS.505..435S}), $V_t$ = 1.1 km/s (\citealt{2010ApJ...708..224H}) and [Fe/H] = 0.02 (\citealt{2010ApJ...708..224H}).  The initial value of ${T}_{m}$ was set automatically for WASP-19b.

In ETD, all transit light curves were obtained using either V, R, I or clear filters, but in Exoclock Project some of them are observed using Luminance filter. To calculate the LD coefficients for the light curves observed in V, R, I filters, we followed \citet{2021AJ....161..108S} and linearly interpolated the values of $u_1$ and $u_2$ from the theoretical tables of \citet{2011A&A...529A..75C} using the EXOFAST package. (\citealt{2013PASP..125...83E}). As the clear filter covers V and R bands (\citealt{2013A&A...551A.108M}), the limb-darkening coefficients $u_1$ and $u_2$ for the clear filter are taken as the average of their value in V and R filters. However, the limb-darkening coefficients derived in V filter were taken for the Luminance filter.

To mitigate potential biases in the derived mid-transit times, we performed a simultaneous fit of the transit times along with the light curve normalization, following the methodology outlined in \citet{2013MNRAS.430.3032B}. This approach requires a well-constrained transit model, which we obtained through a global analysis of all available light curves using the Transit Analysis Package (TAP). In addition, we introduced two extra parameters — $F_{out}$, and $T_{grad}$, — to model the normalization. The priors on these normalization parameters were chosen to be uniform, reflecting a non-informative approach and allowing for equal weighting across a specified range during the fitting process.

\begin{table*} [htbp]
\begin{center}
\caption {The Calculated Values for Theoretical Limb-darkening Coefficients}
\label{tab:3}
\begin{tabular}{clcc}
%\bottomrule light curves
%\bottomrule
\hline
Object Name & Filter & $u_1$ & $u_2$\\
\hline
{WASP-19} & {\it V} & 0.5224 & 0.2193\\
& {\it R} & 0.4181 & 0.2510\\
& {\it I} & 0.3288 & 0.2552\\
& Clear & 0.4702 & 0.2352\\
& Luminance & 0.5224 & 0.2193\\
& TESS & 0.5353 & 0.3456\\
\hline

\end{tabular}\\
\label{tb:LD_coefficients}
\end{center}
\end{table*}

The 50.0th percentile is considered as the best-fit value, and the 15.9th and 84.1st percentile levels (i.e., 68\% credible intervals) of the posterior probability distribution for each model parameter are considered as its lower and upper $1\sigma$ uncertainties, respectively. The best-fit model parameter values P, ${T}_{m}$, i, $a/R_\ast$, ${R_p /R_\ast}$, $u_1$, and $u_2$, along with their $1\sigma$ uncertainties derived for each epoch from 116 TESS transit light curves as well as 65 ETD and 24 Exoclock light curves of WASP-19b, are presented in Table \ref {tb:5}. All of our derived mid-transit times are found to be consistent with the original values reported in the literature at the $1\sigma $ level. Furthermore, the graphical representations of all normalized light curves from TESS, ETD and Exoclock are shown in Figures \ref{fig:TESS1}-\ref{fig:TESS5},  \ref{fig:ETD1}-\ref{fig:ETD3} and \ref{fig:Exoclock}, respectively.

\begin{table} [htbp]
\begin{center}
\caption{The Best-ﬁt Values of Parameters $P$, T$_{m}$, i, a/R$_\ast$, R$_p$/R$_\ast$, $u_1$, and $u_2$ for 205 Transit Light Curves of WASP-19b using \texttt{TAP} \label{tb:5}}
\begin{tabular}{cllllllll}
%\bottomrule
%\bottomrule
\hline
\hline
{Epoch} & {Source} & \ \ \ \ \ \ \ \ \ \ \ {$P$} & \ \ \ \ \ \ \ \ {${T}_{m}$} & {$\it i$} & {$a/{R}_{\ast}$} & {R$_p$/R$_\ast$} & {u$_1$} & {u$_2$} \\
({E}) && \ \ \ \ \ \ \ \ (days) & \ \ \ \ (BJD$_{\rm TDB}$) & (deg) &  &  &  & \\
\hline
\hline
614 & TESS & $0.78883888^{+0.00000009}_{-0.00000011}$& $2455259.68419^{+0.00052}_{-0.00054}$ & $80.055^{+0.30}_{-0.31}$ & $3.578^{+0.030}_{-0.031}$ & $0.1392^{+0.0032}_{-0.0032}$ & $0.476^{+0.028}_{-0.028}$ & $0.239^{+0.024}_{-0.024}$\\
664 & TESS & $0.78883890^{+0.00000007}_{-0.00000009}$& $2455299.1280^{+0.00014}_{-0.00012}$ & $80.060^{+0.26}_{-0.27}$ & $3.557^{+0.0037}_{-0.0038}$ & $0.1409^{+0.0063}_{-0.0060}$ & $0.476^{+0.028}_{-0.028}$ & $0.238 ^{+0.024}_{-0.023}$\\
...&...&...&...&...&...&...&...&...\\
\hline
\end{tabular}

Note. This table is available in its entirety in machine-readable form.\\

\end{center}
\end{table}

\section{Stellar Activity}    \label{sec:Stellar Spots}
Stellar spots can significantly distort the transit light curves of planets orbiting active host stars (e.g., TrES-1b: \citealt{2009A&A...508.1011R}; CoRoT-2b: \citealt{2009A&A...504..561W}; WASP-10b: \citealt{2011MNRAS.411.1204M}; \citealt{2013MNRAS.430.3032B}). When a starspot is positioned along the planet's projected path on the visible portion of the stellar disk, it causes a reduction in the observed flux. WASP-19b orbits a highly active star, demonstrates photometric variability based on the findings by \citet{2010ApJ...708..224H} and \citet{2013MNRAS.434.3252H}. \citet{2013MNRAS.428.3671T} and \citet{2019MNRAS.482.2065E} suggested that the observed distortions in the light curves may be due to spurious effects arising from the starspot rotational modulation. In this study, we identified starspot-affected light curves through both manual inspection and visual analysis (\citealt{2011ApJ...743...61S}; \citealt{2011MNRAS.411.1204M}). A detailed examination of the 205 light curves obtained from ETD, Exoclock, and TESS revealed a positive "bump" near the midpoint of the transit (see, e.g., \citealt{2003ApJ...585L.147S, 2009A&A...508.1011R}) in some of the transit light curves. This feature, which distorts the shape of certain transit light curves, is likely attributable to the presence of starspots. Using this approach, we identified 16 starspot-affected light curves from ETD, 24 from TESS, 2 from \citet{2013MNRAS.428.3671T}, 8 from \citealt{2013MNRAS.436....2M} and 2 from \citealt{2019MNRAS.482.2065E}. Stellar spots can introduce systematic errors, leading to inaccurate estimations of mid-transit times, which may undermine the precision of the light curve measurements. Furthermore, they may generate spurious periodic Transit Timing Variation (TTV) signals, which may lead to misinterpretation of results and false-positive detections of non-transiting sub-stellar companions through the TTV method (e.g., \citealt{2009A&A...501L..23A, 2011ApJ...743...61S}). To accurately determine whether a true physical companion is present in the system, we excluded all 52 spot-affected light curves from TESS, ETD and literature. All spot-affected LCs from TESS and ETD and their corresponding residuals are presented in Figures \ref{fig:TESS_spot_affected_LCs} and \ref{fig:ETD_spot_affected_LCs}.

\section{\textbf{Transit Timing Analysis}}    \label{sec:timing_analysis}

After removing all spot-affected data from our dataset, and before starting the transit timing analysis, we made several corrections to our data: (1) We removed duplicate times that appeared in both the ETD and Exoclock databases, preferring the values from the ETD, which led to the removal of 16 transit light curves from Exoclock. We made this choice because the ETD data are more consistent and of higher quality. All the selected ETD light curves have a Data Quality (DQ) index below 3, which means they provide reliable measurements. Additionally, the ETD light curves have less noise and smaller uncertainties than the ones from ExoClock. To support our decision, we included a comparison table \ref{tab:Comparison_table} that shows the transit times reported by both ETD and ExoClock at same epoch, along with their uncertainties. This comparison confirms that the ETD data are consistent with the ExoClock data, further supporting the reliability of the ETD dataset, (2) We took special care with the timing corrections. For example, if a transit mid-time was reported in one timing system (e.g., HJD/UTC) in one study and later republished in another timing system (e.g., BJD/TDB), we standardized all mid-transit times to BJD/TDB to maintain homogeneity, converting other time stamps into BJD/TDB, (3) We removed some non-obvious duplicates. (4) If two or more groups used the same observations to generate different transit light curves or made different model fits, we only retained one mid-transit time value.

\begin{table} [htbp]
\begin{center}
\caption{Comparison of 16 mid-transit times obtained from ETD and ExoClock at identical epochs.}
\label{tab:Comparison_table}
\begin{tabular}{cc}
%\bottomrule
%\bottomrule
\hline
\hline
ETD Timings & Exoclock Timings \\
\hline
\hline
\(2458549.1452 \pm 0.00097\) & \(2458549.14511 \pm 0.00067\) \\
\(2458871.7780 \pm 0.0012\) & \(2458871.7779 \pm 0.0011\) \\
\(2458875.72274 \pm 0.00076\) & \(2458875.72268 \pm 0.00085\) \\
\(2458883.61074 \pm 0.0011\) & \(2458883.61105 \pm 0.00097\) \\
\(2458898.5975 \pm 0.0016\) & \(2458898.5997 \pm 0.0010\) \\
\(2458901.75633 \pm 0.00071\) & \(2458901.75627 \pm 0.00075\) \\
\(2458905.69992 \pm 0.00062\) & \(2458905.70004 \pm 0.00063\) \\
\(2458973.53994 \pm 0.00056\) & \(2458973.53989 \pm 0.00056\) \\
\(2459207.82674 \pm 0.00073\) & \(2459207.82673 \pm 0.00072\) \\
\(2459230.70137 \pm 0.00057\) & \(2459230.70140 \pm 0.00054\) \\
\(2459237.80108 \pm 0.00076\) & \(2459237.80121 \pm 0.00064\) \\
\(2459252.78896 \pm 0.00083\) & \(2459252.78916 \pm 0.00077\) \\
\(2459256.73251 \pm 0.00075\) & \(2459256.73254 \pm 0.00066\) \\
\(2459278.82091 \pm 0.00086\) & \(2459278.82087 \pm 0.00074\) \\
\(2459309.58449 \pm 0.00084\) & \(2459309.58450 \pm 0.00080\) \\
\(2459313.52941 \pm 0.00066\) & \(2459313.52932 \pm 0.00059\) \\
\hline
\end{tabular}
\end{center}
\end{table}

\subsection{The Linear Ephemeris}       \label{sec:linear}
Once we assembled all the mid-transit times, ${T}_{m}$, from TESS, ETD, Exoclock and literature, we adopted the methodology of transit timing analysis from \citet{2017AJ....154....4P}, \citet{2018AcA....68..371M}, and we fitted all the available mid-transit times using a linear ephemeris model (i.e., null-TTV model) to estimate new linear ephemerides for the orbital period P and mid-transit time $T_0$ of the hot Jupiter WASP-19b. The proposed equation for linear ephemeris, considering 327 mid-transit times is:
\begin{equation}
T_{c} (E) = T_0 + EP, \label{1}
\end{equation}
For the linear model, the planet was assumed to have a circular orbit and a constant orbital period P. In equation \eqref{1}, $T_0$ is the reference mid-transit time at zero epoch (The first transit of WASP-19b observed by \citet{2010ApJ...708..224H} is considered as the zeroth transit), P is the orbital period of the planet, E is the number of orbits counted from the designated reference transit $T_0$ and $T_c$ is the calculated mid-transit times.

The free parameters $T_0$ and P are to be fitted in this model, with initial guesses of values taken from the discovery paper (\citealt{2010ApJ...708..224H}) and by assuming a Gaussian likelihood function by imposing uniform priors on P and $T_0$ and by following the algorithm proposed by \citet{2010CAMCS...5...65G}. Here, we used EMCEE, an affine invariant MCMC ensemble sampler implementation method (\citealt{2006ApJ...642..505F, 2011MNRAS.410...94G, 2013PASP..125..306F}) to perform an MCMC sampling of the posterior probability distribution for all the free parameters in parameter space, given broad uniform priors on all the parameters. For linear model, the number of degrees of freedom, N$_{dof}$ = 250 which is determined by taking difference between the number of data points and the number of fit parameters.

To initiate the Markov Chain Monte Carlo (MCMC) process, we employed 100 walkers, each undergoing an initial burn-in phase consisting of 300 steps. Upon completion of these 300 steps, the final positions of the walkers were taken as the starting points for the main sampling phase. In this phase, each walker performed an additional 20,000 steps to thoroughly explore the posterior probability distributions of the model parameters.
To evaluate the performance of the MCMC chains in terms of convergence and sampling efficiency, we computed several diagnostic measures, including the mean acceptance fraction ($a_f$), the integrated autocorrelation time ($\tau$), and the effective number of independent samples (N$_{eff}$). These diagnostic parameters, summarized in Table \ref{tb:timing_models}, provide insight into the quality of the MCMC chains. The mean acceptance fraction, which indicates the proportion of accepted proposals during the sampling process, was approximately 0.44. This value lies within the optimal range of 0.2–0.5, as recommended for efficient sampling (\citealt{2010CAMCS...5...65G, 2013PASP..125..306F}). 

The integrated autocorrelation time ($\tau$), which quantifies the correlation between successive samples and indicates the number of steps required for the chain to achieve independence, was estimated to be approximately 19 steps across all parameters. To further assess the sampling efficiency, we computed the effective number of independent samples (N$_{eff}$), which is derived by dividing the total number of steps per walker (20000) by the integrated autocorrelation time ($\tau$). This calculation yielded an effective sample size of approximately 1053 independent samples per walker. This value exceeds the minimum threshold of 50 independent samples per walker, which is typically considered adequate for robust statistical inference (\citealt{2017AJ....153..135M}). 

After verifying the convergence and sampling efficiency of the MCMC chains, we proceeded with the Bayesian parameter extraction. However, prior to extracting the model parameters, we applied an additional burn-in step by discarding the first 37 steps (approximately 2$\tau$) from each walker's total of 20000 steps. This removal of initial steps was essential to eliminate the strongly correlated parameters typically present in the early stages of MCMC chains (\citealt{2016A&A...595L...5A, 2018ApJS..236...11H}). The remaining samples, which were deemed independent and well-converged, were then used for the final Bayesian parameter extraction of the model parameters, including the parameters of interest, P and $T_0$. 

Using the procedure described above, we derived a new linear ephemeris for the orbital period P and mid-transit time $T_0$.
We represented the best fit values of all parameters along with their lower and upper $1\sigma$ uncertainties adopted from the median value, 16 and 84 percentile values of the drawn posterior distributions and the uniform prior used for each parameter in Table \ref{tb:timing_models}.

Our refined orbital period, P = ($0.7888390520 \pm 0.000000010$) days, is found to be more precise and lies within
$1\sigma$ confidence interval of the value reported in the discovery paper (\citealt{2010ApJ...708..224H}), due to the extended observation time span of 14 years (2010–2024). Our derived mid-transit time is also $5.6\sigma$ times more precise than the ephemeris published by \citet{2024A&A...692A..35M} under the constant-period assumption. 

Next we calculated the timing residuals, O-C (pronounced ``O minus C''; \citealt{2005ASPC..335....3S}), where O denotes the observed mid-transit times and C is the calculated mid-transit times. In the absence of transit timing variation, we would anticipate no significant deviations of the derived O–C (observed-minus-calculated) values from zero. However, we observed a significant deviation on both sides of zero and this deviation from the zero value could be the ﬁrst indication of timing anomalies caused by additional planets or moons (\citealt{2009AN....330..475R}), which we have explained in Section \ref{sec:additional}. The estimated timing residuals (O-C) along with their corresponding epochs and original mid-transit times ($T_m$) are shown in Table \ref{tab:6} and we have also  constructed an O-C diagram to analyze this possible shift in transit times (see Figure \ref{fig:O-C_diagram}) by plotting the timing residuals, i.e., O-C values as a function of epoch, E.

To explore this further and to check whether the ephemeris is up-to-date as well as to evaluate how well the observed data fit the linear model, a chi-squared fitting statistic was performed using the observed transit mid-point and the expected transit mid-point. Previous studies on TTV analysis (e.g., \citealt{2017AJ....153..191S, 2024AJ....168..176B}) have shown that $\chi^{2}_{red}$ values typically exceed 1.0 in such analyses. In our case, the calculated $\chi^{2}_{red}$ value is also significantly greater than 1, indicating that the linear model does not adequately fit the transit time data. Consequently, this suggests the possibility of TTV in the WASP-19 system.

In addition to examining the reduced chi-squared ($\chi^{2}_{red}$) statistics, we also calculated two commonly used model selection criteria: the Bayesian Information Criterion (BIC; \citealt{1978AnSta...6..461S}) and the Akaike Information Criterion (AIC; \citealt{1974ITAC...19..716A}). These metrics are used to determine which model best represents the data through model comparison. The formulas used for calculating BIC and AIC are as follows: BIC= ${\chi}^{2} + k_F\log{N_P}$ and AIC = ${\chi}^{2} $ + 2$k_F$, where $N_P$ is the total number of data points (252 in this analysis), and $k_F$ represents the number of free parameters ($k_F$ = 2 for a linear fit). For the constant period model, the minimum chi-square, degrees of freedom, the resulting reduced chi-squared ($\chi^{2}_{red}$) value, and the calculated AIC, BIC values are given in Table \ref{tb:timing_models}. A lower BIC value generally indicates a better model, and the difference in BIC values ($\bigtriangleup\mathrm{BIC}$) between models is typically used to determine model preference, with a $\bigtriangleup\mathrm{BIC} > 10$ indicating very strong evidence in favor of the model with the smaller BIC. Similarly, AIC assists in model selection by balancing the goodness-of-fit with model complexity. The calculated values for $\chi^{2}$, $\chi^{2}_{red}$, BIC, and AIC corresponding to the linear ephemeris model are presented in Table \ref{tb:timing_models}.

\begin{center}
\small\addtolength{\tabcolsep}{-3pt}
\begin{longtable*}{cp{3.5cm}cp{1.5cm}cp{3.5cm}cp{0.5cm}cp{3.5cm}cp{3.5cm}cp{3.5cm}cccclcc}
\caption{Mid-transit Times $({T}_{m})$ and Timing Residuals (O-C) for all 252 Transit Light Curves of WASP-19b} 
\label{tab:6}\\
\hline
\hline 
Transit Number & \ \ \ \ ${T}_{m}$  & Uncertainty & $O-C$ & TRESCA ID of ETD & \ Transit Source & Timing Source\\
({E}) & \ \ \ \ (BJD$_{\rm TDB}$) &  (days) & \ \ (days)  & light curves \\
\hline
\endfirsthead
0 & 2454775.33720 & 0.0015 & -0.0008519 & ... & \citet{2010ApJ...708..224H} & \citet{2024PSJ.....5..163A}\\
2 & 2454776.91566 & 0.00019 & -0.0000582 & ... & \citet{Anderson_2010} & \citet{Bernab__2024}\\
 ...&...&...&...&...&...&...&\\
\hline
\end{longtable*}
Note. This table is available in its entirety in machine-readable form.\\
References. \citet{2010ApJ...708..224H, 2010A&A...513L...3A, 2011AJ....142..115D, 2011ApJ...730L..31H, 2013A&A...552A...2L, 2013MNRAS.430.3422A, 2013MNRAS.428.3671T, 2013A&A...553A..49A, 2013MNRAS.436....2M, 2015A&A...576L..11S, 2016ApJ...823..122W, 2017Natur.549..238S, 2019MNRAS.482.2065E, 2020A&A...636A..98C, 2020AJ....159..150P, 2020MNRAS.491.1243P, 2024PSJ.....5..163A, 2024A&A...684A..78B}.
\end{center}

\subsection{A Search for Orbital Decay Phenomenon} \label{sec:decay}
Given the ongoing debate regarding the detection of orbital decay-i.e., a shortening of the orbital period over time-in WASP-19b over the past several years, and the availability of transit observational baselines spanning more than a decade, we identified this as an ideal opportunity to detect long-term TTV, particularly to investigate the orbital decay phenomenon in WASP-19b. It is one of the best candidates for observing decay in its orbit due to its ultra-short orbital period, which is less than a day, and its high mass, which is slightly less than 2.0 M$_J$. These characteristics make it an excellent target for detecting potential signs of orbital decay. 

Ultra-short period hot Jupiters, due to their relatively large mass and short orbital periods, are expected to experience strong tidal interactions with their host stars (e.g., \citealt{1999ssd..book.....M, 2008ApJ...681.1631J, 2010exop.book..239C, 2010ApJ...723..285H, 2012ApJ...757....6H, 2012ApJ...751...96P, 2014ARA&A..52..171O, 2019EAS....82....5M}). The tidal forces exerted by a hot Jupiter induce large-amplitude waves within its host star. Over time, this leads to the exchange of the planet's orbital angular momentum and energy with the host star's spin (\citealt{1996ApJ...470.1187R, 2003ApJ...596.1327S}), facilitated by the tidal bulge raised in the star by the close-in planet. As a result, the planet gradually spirals inward, causing a contraction of the orbit's semi-major axis, eventually leading to the engulfment of the planet by its host star. This process, in which the orbital period progressively decreases over time, is known as tidal orbital decay. Consequently, the transits in the system occur increasingly earlier (\citealt{2010ApJ...725.1995M}). In order to confirm or rule out the orbital decay phenomenon in the system, we employed the orbital decay ephemeris model as a function of epoch outlined by \citet{2014ApJ...781..116B} and \citealt{2016AJ....151...17J}, incorporating a uniform period change over time:
\begin{equation}
T_{q} (E) = {T}_{0} + P E + \ \frac{1}{2} \ \frac{dP}{dE} \ {E}^2,
\label{lab:equation}
\end{equation}
      The quadratic model represented above also assumes a circular orbit, but incorporates a steady rate of change in the orbital period between successive transits, represented by an additional term​, $\frac{dP}{dE}$. In this model, ${T}_{q} (E)$  denotes the calculated mid-transit time. In this analysis, we treated $T_0$, P, and $\frac{dP}{dE}$​ as free parameters and applied the EMCEE sampling method to determine their posterior distributions, as described in the linear ephemeris model. The initial guesses for these parameters P, $T_0$ and $\frac{dP}{dE}$ taken from \citealt{2022A&A...668A.114R}. The prior for the quadratic model allowed the period derivative to have any sign ranging from negative to positive values.

We derive the quadratic coefficient of $\frac{dP}{dE}$ = (- $0.28 \pm 0.10$)$\times 10^{-10}$ for WASP-19b using transit data from TESS, ETD, Exoclock, and archival sources. The result is statistically significant at the $2.8\sigma$ level. Our result is consistent with that of \citet{2022A&A...668A.114R}, who found no significant evidence for orbital decay, reporting $\frac{dP}{dE}$ = (- $0.35 \pm 0.22$)$\times 10^{-10}$. Additionally, it aligns with the findings of \citet{2020AJ....159..150P} ($\frac{dP}{dE}$= (- $2.06 \pm 0.42$)$\times 10^{-10}$) and \citet{2023ApJS..265....4K} ($\frac{dP}{dE}$ = (- $0.87 \pm 0.13$)$\times 10^{-10}$), with significance levels of $4.1\sigma$ and $3.6\sigma$, respectively. Furthermore, our value is within the $6.5\sigma$ confidence interval of the period decrease rate reported by \citet{2023Univ...10...12K}, $\frac{dP}{dE}$= (- $1.2 \pm 0.1$)$\times 10^{-10}$.

Similar to the linear ephemeris model, the degrees of freedom, $\chi^{2}$, $\chi^{2}_{red}$, AIC, and BIC values for the orbital decay model are shown in Table \ref{tb:timing_models}. To assess the significance of period change detection and to evaluate whether adding additional free parameters improves the model fit, we compared the $\chi^{2}_{red}$, AIC, and BIC values of the two models. Although the smaller values of these three statistical quantities—$\chi^{2}_{red}$, AIC, and BIC—suggest that the orbital decay model provides a better fit than the linear model, indicating a preference for the inclusion of these additional parameters, the $\bigtriangleup\mathrm{AIC} = 5.64$ and $\bigtriangleup\mathrm{BIC} = 2.11$ values, being less than 10, do not provide strong evidence in favor of the orbital decay model.

Now to determine the rate of this decay, we substituted the derived values of P and $\frac{dP}{dE}$​ into equation (4) from \citet{2017AJ....154....4P}, and the change in the orbital period as a function of epoch, $\frac{dP}{dE}$, is translated into the rate of change of the orbital period over time, $\dot {P}$= $\frac{dP}{dt}$, typically expressed in milliseconds per year (ms/yr). This allows for the straightforward calculation of $\dot {P}$ for WASP-19b. The observational signature of a tidal orbital decay is a decrease of the orbital period. The negative sign in our calculated value of $\dot {P}$, $\sim - 1.1 \pm 0.40$ \textbf{ms/yr}, indicates that the orbital period of WASP-19b is decreasing, as inferred from the full set of available transit timing data spanning 15 years.

\subsubsection{Comparison with Previous Studies} \label{sec:comparison}

In past few years, several authors (\citealt{2013MNRAS.436....2M, 2019MNRAS.482.2065E}) have performed the empirical studies of WASP-19b using transit observations and they provided hints of transit timing variations in WASP-19b by studying the rate of decay. In our study, we observed a period change rate of $\sim - 1.1 \pm 0.40$ ms/yr, based on the 252 mid-transit times collected from TESS, ETD, Exoclock and literature, which is slower than some previous reports but consistent with others. \citet{2013MNRAS.436....2M} and \citealt{2019MNRAS.482.2065E} identified an anomaly in the light curve and attributed the observed timing inconsistencies to the effects of starspot activity. \citet{2020AJ....159..150P}, upon incorporating additional transit data to those previously reported, found strong evidence of orbital decay with a decay rate of $\sim - 6.50 \pm 1.33$ ms/yr. This rate is approximately six times faster than the value we derived, likely due to the inclusion of the ETD and Exoclock data in our analysis. However, they suggested that this result might be misleading, attributing potential systematic errors to the influence of stellar activity. Alongside \citet{2020AJ....159..150P}, \citet{2023ApJS..265....4K}, and \citet{2023Univ...10...12K} also presented strong evidence for orbital decay. \citet{2023Univ...10...12K} reported a period change rate of $\sim - 3.7 \pm 0.5$ ms/yr, and our derived $\dot {P}$ value is in good agreement with their result, with a significance of $4\sigma$. Additionally, our value agrees to within $1\sigma$ of the value reported by \citet{Shen_2024}. In contrast, \citet{2020MNRAS.491.1243P}, \citet{2020AJ....159..104W}, and \citet{2022A&A...668A.114R} dismissed the evidence of orbital decay in their analyses. \citet{2020MNRAS.491.1243P}, after including 74 transits from both existing literature and their own observations spanning a period of 10 years, determined that the orbital decay rate was consistent with zero. This result supports the constant orbital period model, indicating the absence of a long-term trend. Nonetheless, they established an upper bound for the period variation of $\sim - 2.294 $ ms/yr, a value that is in agreement with the result we have derived. \citet{2022A&A...668A.114R} incorporated observational data from sectors 9 and 36 of the Transiting Exoplanet Survey Satellite (TESS) and arrived at a similar conclusion. More recently, \citet{2024PSJ.....5..163A} extended the analysis by including data from sectors 62 and 63, also finding no evidence of orbital decay. While this paper was being written, two new studies on WASP-19b were published: one by \citet{2024A&A...692A..35M}, and another by \citet{2025arXiv250105704M}. \citet{2024A&A...692A..35M} investigated long-term orbital period variations for six hot Jupiters, including WASP-19b, using all available TESS data. \citet{2025arXiv250105704M} compiled spectroscopic transit observations for 37 short-period exoplanets, including WASP-19b, from the Hubble Space Telescope (HST), along with previously published TESS data from \citet{2024ApJS..270...14W}. Both studies reported results consistent with prior research, finding no evidence of orbital decay. In contrast, \citet{2024ApJS..270...14W} reported a marginal detection of orbital decay. We have compared our estimated value of the period change rate for WASP-19b, with the values reported in various studies, as summarized in Table \ref {tab:7}.

\begin{table*}[h]
    \centering\small\centering\renewcommand{\arraystretch}{1.4}
        \caption{Comparison of the values of period change rate and orbital decay timescale of WASP-19b as estimated by previous studies.}
        \label{tab:7}
    \begin{tabular}{l c}
    \hline\hline
         Reference & Period change rate, {$\dot {P}$} \\
         & [ms/yr] \\
         \hline
         This work & $-$1.1 $\pm$ 0.40 \\
         \citet{2025arXiv250105704M} & $-$0.26 $\pm$ 0.30 \\
           \citet{2024ApJS..270...14W} &  $-$1.64 $\pm$ 0.29 \\
         \citet{Shen_2024} &  $-$1.153 $\pm$ 0.539 \\
         \citet{2024PSJ.....5..163A} & 0.6 $\pm$ 0.8 \\
         \citet{2023Univ...10...12K} & $-$3.7 $\pm$ 0.5 \\
         \citet{2023ApJS..265....4K} & $-$3.47 $\pm$ 0.47 \\
         \citet{2022ApJS..259...62I} & $-$3.54 $\pm$ 1.18  \\
         \citet{Ros_rio_2022} & $-$1.40 $\pm$ 0.88  \\       
         \citet{2020MNRAS.491.1243P} & 0.036 $\pm$ 2.32  \\
         \citet{2020AJ....159..150P} & $-$6.50 $\pm$ 1.33 \\
         \hline
    \end{tabular}
    
    \label{tab:comparison_of_orbital_decay_rate}

\end{table*}

%-------------------------------------------------------------------

\subsubsection{Calculation of Inspiral Timescale of WASP-19b } \label{sec:Inspiral_timescale}

The Inspiral Timescale or the potential characteristic decay timescale ($T_d$) of WASP-19b can be calculated using the formula $T_d$ = {P}/{$\dot {P}$}, where P and $\dot {P}$ are the period and period change rate, respectively, as discussed above. In our analysis, this timescale represents how long it would take for a hot Jupiter to be engulfed by its host star and the timescale over which the orbit is shrinking is, $T_d$ = 61.40 Myr, which represents approximately 0.62\% of the derived age of the host star, WASP-19 (Age = 9.95 Gyr; \citealt{2010ApJ...708..224H}). 

Similar to the linear model, here we also calculated the timing residuals, i.e., O-C values, by subtracting the mid-transit times calculated using the linear ephemeris, ${T}_{q} (E)$ from those derived after ﬁtting the orbital decay model, ${T}_{q} (E)$; to analyze the possible shift in transit times and the TTV diagram (O-C diagram), we represented these timing residuals as a function of epoch in the O-C diagram, with the red dashed curve showing the quadratic ephemeris model (see Figure \ref{fig:O-C_diagram}). To explore the future trend of the orbital decay scenario, we sampled 100 random realizations from the posterior distributions of the orbital decay model. These are represented by the brown solid lines in the O-C diagrams, which are extrapolated to the next 10 years to illustrate the projected evolution of the decay.

The O-C diagram indicates that the orbital decay model does not show a significant deviation from the linear ephemeris. This suggests that the available observational data do not provide strong evidence for orbital decay, implying that any potential period change is too small to be detected given the current measurement uncertainties. Long-term high-precision follow-up transit observations of these exoplanets are needed to reveal the true nature of their O-C variations.       

      For planets with eccentric orbits in close proximity to their host star, the gravitational interaction between the planet and the star can induce a tidal bulge on the star. This interaction may result in the planet undergoing inward migration (\citealt{1996ApJ...470.1187R, 2009ApJ...692L...9L, 2010ApJ...725.1995M}), particularly when the star's rotation is slower than the planet's orbital motion. Such migration has important implications for the planet's long-term stability and evolution. If a planet's orbital period is shorter than its rotation period, it typically indicates that the planet is tidally locked or that its rotation is slowing relative to its orbit (\citealt{2024ApJS..270...14W}) and the orbit is gradually shrinking. For WASP-19b, where the stellar rotation period ($P_{rot}$​= $10.5\pm0.2 $days) is longer than the planet's orbital period (P = 0.788839 days; see Table \ref{tb:timing_models}), tidal friction may cause a gradual slowdown of the planet's rotation over time. Although the period change is significant at the $2.6\sigma$ level (see Section \ref{sec:decay}), the lack of a significant deviation between the orbital decay model and the linear model prevents us from definitively concluding whether tidal interactions are responsible for any notable changes in the planet's orbital period. To refine and confirm our findings, continuous and regular transit observations over an extended timescale are required.

\subsection{Apsidal Precession Study for WASP-19 System}      \label{sec:apsidal}
A potential mechanism that could induce long-term transit timing variation (TTV) is apsidal precession, which refers to the gradual rotation of the orbit's elliptical shape. Specifically, this process involves the movement of the periapsis, the point of closest approach to the star, around the star-planet orbital path. Apsidal precession arises due to the non-point-mass component of the gravitational field (\citealt{1928MNRAS..88..641R, 1938MNRAS..98..734C, 1939MNRAS..99..451S, 1939MNRAS..99..662S}). According to the theoretical framework developed by \citet{2009ApJ...698.1778R}, hot Jupiters with slightly eccentric orbits (eccentricity, $e > 0.003$), which induce sinusoidal variations in transit times, are considered ideal candidates for investigating apsidal precession. This phenomenon has been proposed as a potential explanation for long-term TTVs observed in numerous hot Jupiter systems (\citealt{2009ApJ...698.1778R, 2017AJ....154....4P, 2019AJ....157..217B, 2020ApJ...888L...5Y, 2022AJ....163...77A}.

In the case of the exoplanet WASP-19b, \citet{2011ApJ...730L..31H} reported an eccentricity value of 0.0046, which exceeds the threshold value. As a result, we explored the possibility of apsidal precession in the WASP-19 system by applying the apsidal precession model outlined in Equation (5) of \citet{2024AJ....168..176B}, which was derived based on the precession model introduced by \citet{1995Ap&SS.226...99G}:
\begin{equation}
T_{ap} (E) = T_{ap0} + P_{s} E -  \frac{e {P_s} \cos{({\omega}_0 + E \frac{d\omega}{dE})}}{\pi(1-\frac{\frac{d\omega}{dE}}{2 \pi})},
\label{eq:1}
\end{equation}
In this equation, a planet has a constant orbital period, but the orbit is precessing. \( E \) represents the epoch, \( T_{\mathrm{ap}0} \) is the calculated mid-transit time, and \( \omega_0 \) is the argument of periastron. The five free parameters in the model are: \( T_{\mathrm{ap}0} \), the mid-transit time at \( E = 0 \); \( P_s \), the sidereal period; \( e \), the orbital eccentricity; \( \omega_0 \), the argument of periastron at \( E = 0 \), which describes the angle between the ascending node and the periastron; and \( \frac{d\omega}{dE} \), the precession rate of the argument of periastron.

To determine the best-fit ephemeris of the apsidal precession model, we adopted the same technique as \citet{2024AJ....168..176B}. To account for additional noise sources—particularly those arising from instrumental calibration—we introduced a jitter term, \( \mathrm{jit}_{\mathrm{tr}} \). Initially, following \citet{2024A&A...684A..78B}, we fixed the jitter value at 0.00005 days. In a second approach, this value was used as an initial guess, and the jitter term was treated as a free parameter during the fitting process. The latter approach resulted in lower values of the reduced chi-squared (\( \chi^2_{\mathrm{red}} \)), the Akaike Information Criterion (AIC), and the Bayesian Information Criterion (BIC). Therefore, we adopted the model with the jitter term as a free parameter.

In Table~\ref{tb:timing_models}, the best-fitting parameters from the MCMC posterior probability distribution are presented.

The best-fit parameter values of e, $\omega_0$, and $\frac{d\omega}{dE}$, obtained from the MCMC posterior probability distribution, are $e = 0.0058^{+0.00013}_{-0.00021}$, $\omega_0$ = $-0.72^{+1.17}_{-0.62}$,  $\frac{d\omega}{dE}$ = $0.0008^{+0.000181}_{-0.000185}$. The derived value of the model parameter $\frac{d\omega}{dE}$ is found to be statistically significant for WASP-19b, whereas the value of $\omega_0$​ is statistically insignificant, as indicated by the large $1\sigma$ uncertainties associated with this parameter. These unusual results may arise from strong correlations between the model parameters or from the adoption of an incorrect model for the timing data.

The eccentricity of WASP-19b, as determined in this study, consistent with the upper limit (e $\sim 0.0046$) reported by \citet{2011ApJ...730L..31H}. However, this value deviates from the eccentricity of $0.001780 \pm 0.000248$ reported by \citet{Shen_2024} at a $13.4\sigma$ significance level. Additionally, our apsidal precession analysis yields a precession rate ($\frac{d\omega}{dt}$) of $0.05^\circ/\text{day}$, which is consistent with the $0.03623^\circ/\text{day}$ value reported by \citet{Shen_2024} and with the $0.065^\circ/\text{day}$ value reported by \citet{2024A&A...684A..78B} at a $1\sigma$ significance level. The slower precession rate observed in this study implies a longer apsidal motion period of about 18.64 years, which reflects a gradual and long-term change in the orbital orientation of the planet. As the observational window is not long enough, that's why to accurately measure this effect and distinguish it from other possible influences on the system, extended and continuous observation over many years is required as well as whether our calculated value of WASP-19b's eccentricity is sufficiently large to induce apsidal precession can be addressed through future observations of eclipse center times. These upcoming eclipse timing measurements will be crucial for distinguishing between the orbital decay and apsidal precession models. While orbital decay results in a gradual decrease in both transit and eclipse periods, apsidal precession produces an anti-correlated period trend between the transit and eclipse center times.

As shown in Table \ref{tb:timing_models}, with significantly lower values of $\chi^{2}_{red}$, AIC, and BIC compared to the other two timing models, the apsidal precession model provides a better explanation of transit timing variations. Moreover, the $\bigtriangleup\mathrm{BIC} > 10$ suggests a strong preference for the apsidal precession model, which more effectively accounts for model complexity, consistent with the explanation provided by \citet{2024A&A...684A..78B}. Furthermore, the superiority of the apsidal precession model is supported by the fact that, while the orbital decay model has only three free parameters, the apsidal precession model incorporates five free parameters.

\begin{figure}
    \centering
    \includegraphics[width=1.1\linewidth]{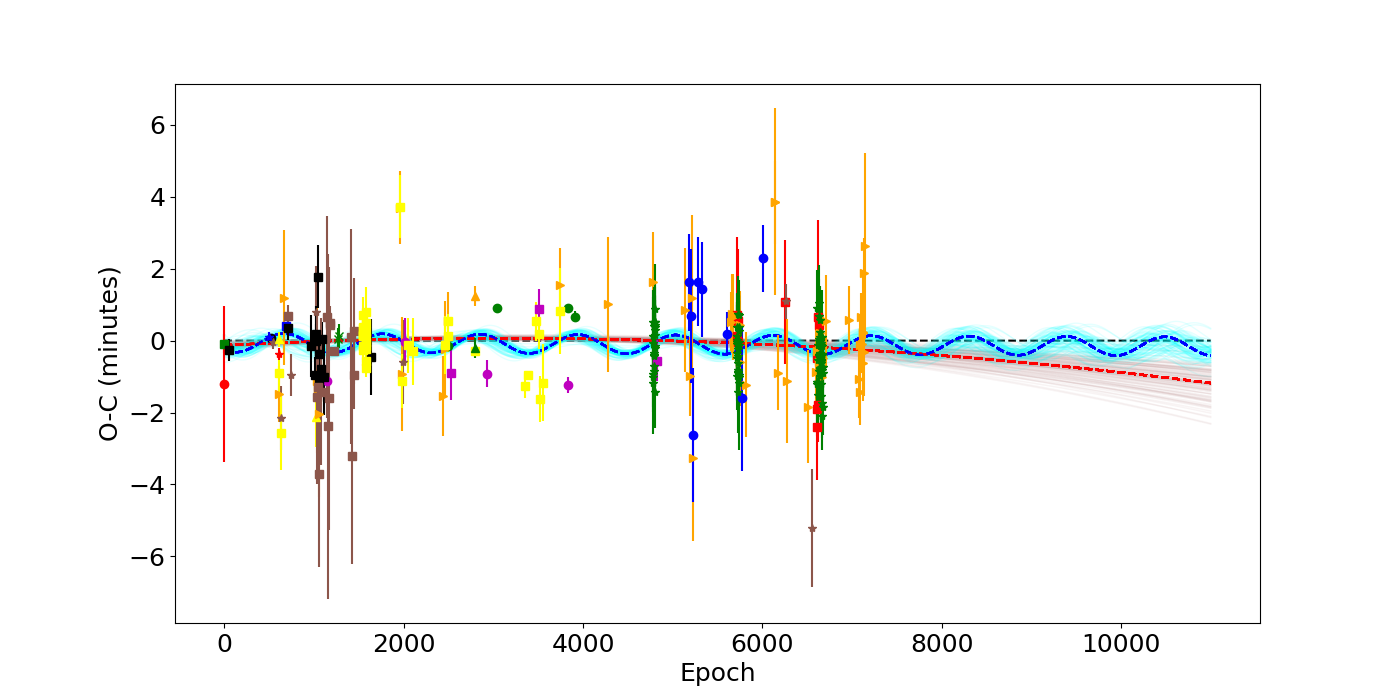}
    \caption{O-C diagram for analysing 252 mid-transit times of WASP-19b. The blue filled square show the data from \citet{2013A&A...553A..49A}, the red filled squares are from \citet{2024PSJ.....5..163A}, the green filled square is from \citet{2010A&A...513L...3A},  the black filled circle is from \citet{2013MNRAS.430.3422A}, the brown filled asterisks are from \citet{2024A&A...684A..78B}, the magenta filled circles are from \citet{2020A&A...636A..98C}, the yellow filled upward triangles are from \citet{2011AJ....142..115D}, the green filled circles are from \citet{2019MNRAS.482.2065E}, the orange filled right triangles are from the quality 1 and quality 2 data of ETD, the blue filled circles are from Exoclock, the red  filled circle is from \citet{2010ApJ...708..224H}, the green asterisk is from \citet{2011ApJ...730L..31H}, the black filled squares are from \citet{2013A&A...552A...2L}, the brown filled squares are from \citet{2013MNRAS.436....2M}, the magenta filled squares are from \citet{2020AJ....159..150P}, the yellow filled squares are from \citet{2020MNRAS.491.1243P}, the green filled upward triangle is from \citet{2015A&A...576L..11S},  the orange filled upward triangle is from \citet{2017AJ....153..191S},  the blue filled asterisks is from TESS-data, the red filled asterisks is from \citet{2013MNRAS.428.3671T}, the green x-shaped markers are from \citet{2016ApJ...823..122W}. The dashed black line, red and blue curves represent the linear, orbital decay and apsidal precession models. The lines are drawn for 100 randomly chosen sets of parameters from the Markov chains of posteriors of the orbital decay (brown) and apsidal precession (cyan) models. The models are extrapolated for the next $\sim 10 $ years to illustrate the broad spectrum of possible solutions.}
    \label{fig:O-C_diagram}
\end{figure}

We again plotted the O-C diagram with the timing residuals $T_{ap}(E)$ - $T_{c}(E)$ as a function of the epoch, where the $T_{ap}(E)$ and $T_{c}(E)$ values are derived using the best-ﬁt ephemeris for the apsidal precession and linear models. In the O-C diagram, the blue dashed curve depicts the apsidal precession model. To see the future trend for WASP-19b, we randomly drew a sample of 100 parameter sets from the posteriors of the apsidal precession model (shown by a cyan colored line; see Figure \ref{fig:O-C_diagram}) and extrapolated them for the next $\sim  10 $yr. Investigating the intrinsic nature of the O-C diagram will be valuable through the acquisition of additional long-term, high-precision follow-up observations of transits and occultations. As noted by \citet{2009ApJ...698.1778R}, a longer observational baseline is essential to accurately measure the timing variation caused by apsidal precession in many systems hosting extremely hot Jupiters. 

The fitting of the apsidal precession model (Figure \ref{fig:O-C_diagram}) reveals a sinusoidal variation when compared to the transit time data. This variation may be influenced by the light-time effect (LiTE), which could be induced by the gravitational perturbation by another planet in the WASP-19 system. Furthermore, the observed decrease in orbital period (see Section \ref{sec:decay}) may be attributed to mechanisms other than orbital decay, such as the $R\phi$mer effect or light-time effect (LiTE) caused by a distant companion in the system (\citealt{2023Univ...10...12K}). The phenomenon of apsidal precession has been ruled out for certain systems, such as WASP-12b (e.g., \citealt{2017AJ....154....4P, 2020ApJ...888L...5Y, 2021AJ....161...72T}), likely due to the very low eccentricities observed in many hot Jupiters.

Tidal interactions typically act to circularize elliptical orbits of hot Jupiters on short timescales ($\sim1$ Myr), but in the case of WASP-19b, we inferred a notable increase in the eccentricity value (e = 0.0058) based on our model fit, compared to previous findings, such as e = 0.00172 (\citealt{2024A&A...684A..78B}) and e = 0.001780 (\citealt{Shen_2024}). This excitement in the eccentricity of its orbit suggests a strong evidence for apsidal precession. However, since the apsidal precession model produces a sinusoidal variation in the O–C diagram (Figure \ref{fig:O-C_diagram}), which could also arise from the gravitational influence of a third body in the system, we further investigated this possibility in Section~\ref{sec:additional} by fitting a sinusoidal model to the timing residuals.

\begin{table*}[!htb]
\caption{Best-Fit Model Parameters for WASP-19b}
    \centering
    \begin{tabular}{lcccc}
     \hline
     Parameter                          &  Symbol &   units    & Posterior value  & 1 $\sigma$ uncertainty  \\
     \hline
     \multicolumn{5}{c}{\textbf{Constant Period Model}} \\
        Period                          & P$_{\text{orb}}$          & days         & 0.7888390520 & $^{+0.000000010}_{-0.000000010}$    \\
        Mid-transit time                & T$_{0}$   &BJD$_{\rm TDB}$     & 2454775.33804012  &$^{+0.0000266}_{-0.0000266}$ \\
        \hline                                                              
        $a_f$, $\tau$, $(N_{\rm eff})^{a}$  &                 &           &$\sim 0.44, \sim 19, \sim 1053$ &\\
        N$_{dof}$                         &            &              &250         & \\
        $\chi^{2}$, $\chi^{2}_{red}$	                    &            &              & 685, 2.74       &       \\
        AIC                             &           &               & 689.72 \\
        BIC                             &           &               & 696.78 \\
     \hline
     \hline 
    \multicolumn{5}{c}{ \textbf{Orbital Decay Model}}\\  
        Period                         & P$_{\text{orb}}$        &  days         & 0.7888391381 & $^{+0.0000000325}_{-0.0000000327}$ \\
        Mid-transit time               & T$_{q0}$    &BJD$_{\rm TDB}$      & 2454775.33795158 & $^{+0.0000416}_{-0.0000415}$ \\
        Decay Rate                      & dP/dE     &days/epoch     & -0.28 $\times$ 10$^{-10}$ &  $^{+0.10\times10^{-10}}_{-0.0997\times10^{-10}}$  \\
        Decay Rate                      & dP/dt     & ms/yr       & -1.1                   & 0.40      \\
        \hline
        $a_f$, $\tau$, $(N_{\rm eff})$  &                 &           &$\sim 0.35, \sim 30, \sim 1067$ &\\
        N$_{dof}$                         &            &              &249         & \\
        $\chi^{2}$, $\chi^{2}_{red}$	                    &            &              & 677.28, 2.72       &       \\
        AIC                             &           &               & 684.08 \\
        BIC                             &           &               & 694.67 \\
        \hline 
        \hline
       \multicolumn{5}{c}{  \textbf{Apsidal Precession Model}} \\
        Sidereal Period                  & P$_{s}$    & days          &  0.7888390453      &  $^{+0.0000000154}_{-0.0000000156}$             \\
        Mid-transit time                 & T$_{ap0}$    & BJD$_{\rm TDB}$     & 2454775.338000036    & $^{+0.0000592}_{-0.0000597}$                  \\
        Eccentricity                    & e         &               &  0.0058         & $^{+0.00013}_{-0.00021}$ \\
        Argument of Periastron          & $\omega$$_{0}$& rad           &    -0.71          &$^{+1.17}_{-0.62}$               \\
        Precession Rate                 & d$\omega$/dE& rad/epoch     &  0.000725          & $^{+0.000181}_{-0.000185}$                 \\
       Jitter & $\mathrm{jit}_{\mathrm{tr}}$ & days & 0.0002 & $^{+0.00000298}_{-0.00000618}$ \\
       \hline                                                          
        $a_f$, $\tau$, $(N_{\rm eff})$  &                 &           &$\sim 0.23, \sim 255, \sim 392$ &\\
        N$_{dof}$                         &            &              &246         & \\
        $\chi^{2}$, $\chi^{2}_{red}$	                    &            &              & 292.74, 1.19       &       \\
        AIC                             &           &               & 303.65 \\
        BIC                             &           &               & 324.83 \\
        \hline 
        \hline
    \end{tabular}
   
    \label{tb:timing_models}
     {Note: $^a$ $N_{\rm eff}$ is the effective number of independent samples.
}
\end{table*}

\subsection{Searching for nearby, possible companions} \label{sec:additional}
By employing the TTV method (\citealt{2002ApJ...564.1019M, 2005MNRAS.359..567A, 2005Sci...307.1288H}), we can detect additional low-mass planets in hot Jupiter systems, which are typically difficult to identify using other techniques (\citealt{2013MNRAS.433.3246S}). Numerous studies have focused on this area (\citealt{2011MNRAS.413L..43P, Hoyer_2012, 2013MNRAS.434...46H, 2013A&A...553A..17S}), yet only a few hot Jupiters have been found to host additional close-in planets, such as WASP-47b (\citealt{2015ApJ...812L..18B}), Kepler-730b (\citealt{2019ApJ...870L..17C}), and TOI-1130b (\citealt{2023Univ...10...12K}). The Kepler mission, launched in 2009, played a pivotal role in detecting multiple transiting exoplanets (e.g., \citealt{2013MNRAS.433.3246S, 2014ApJ...787...80H}) by studying planet-planet interactions induced by TTVs through continuous photometric monitoring of thousands of stars. By combining the new time-series data from TESS with recent ground-based observations, we can now confirm the presence of short-term TTVs in short-period exoplanets, such as hot Jupiters. Since the discovery of WASP-19b, several scientists tried to find whether  there is any presence of short-term TTVs, caused by an additional planet, in this system. \citet{2020A&A...636A..98C} using both photometry and radial velocity data, neither found any additional companion nor any evidence for long-term trends in the data. \citet{2013MNRAS.436....2M} and \citet{2019MNRAS.482.2065E} found that a linear ephemeris poorly fits the data for WASP-19b. To explore if this is due to perturbations from another body, \citet{2020MNRAS.491.1243P} analyzed sinusoidal variations in the mid-transit times and reported a null detection of periodic variations, ultimately ruling out the existence of planetary companions in the system.

Our O-C diagram (Figure \ref{fig:O-C_diagram}) indicates that the fitting of the apsidal precession model exhibits a sinusoidal variation in the transit time data, potentially influenced by periodic variations resulting from the light-time effect (LiTE) caused by a third component in the WASP-19 system. To further explore this phenomenon, we probed periodicity in the timing residuals given in Table~\ref{tab:6}, by computing the generalized Lomb–Scargle periodogram (GLS; \citealt{2009A&A...496..577Z}) in the frequency domain. We also calculated the false alarm probability (FAP) associated with the highest power peak by randomly permuting the timing residuals to the observing epochs using a bootstrap resampling method, finding that the FAP value for the peak power of 0.311776, at a frequency of \( 0.076375 \pm 0.000140 \) cycles/epoch, was 22.5\%. The amplitude and phase values are \( A_{\mathrm{TTV}} = 0.280291 \pm 0.107079 \) minutes and \( \phi = 1.071963 \pm 0.236848 \), respectively.

\begin{figure}
    \centering
    \includegraphics[width=1.1\linewidth]{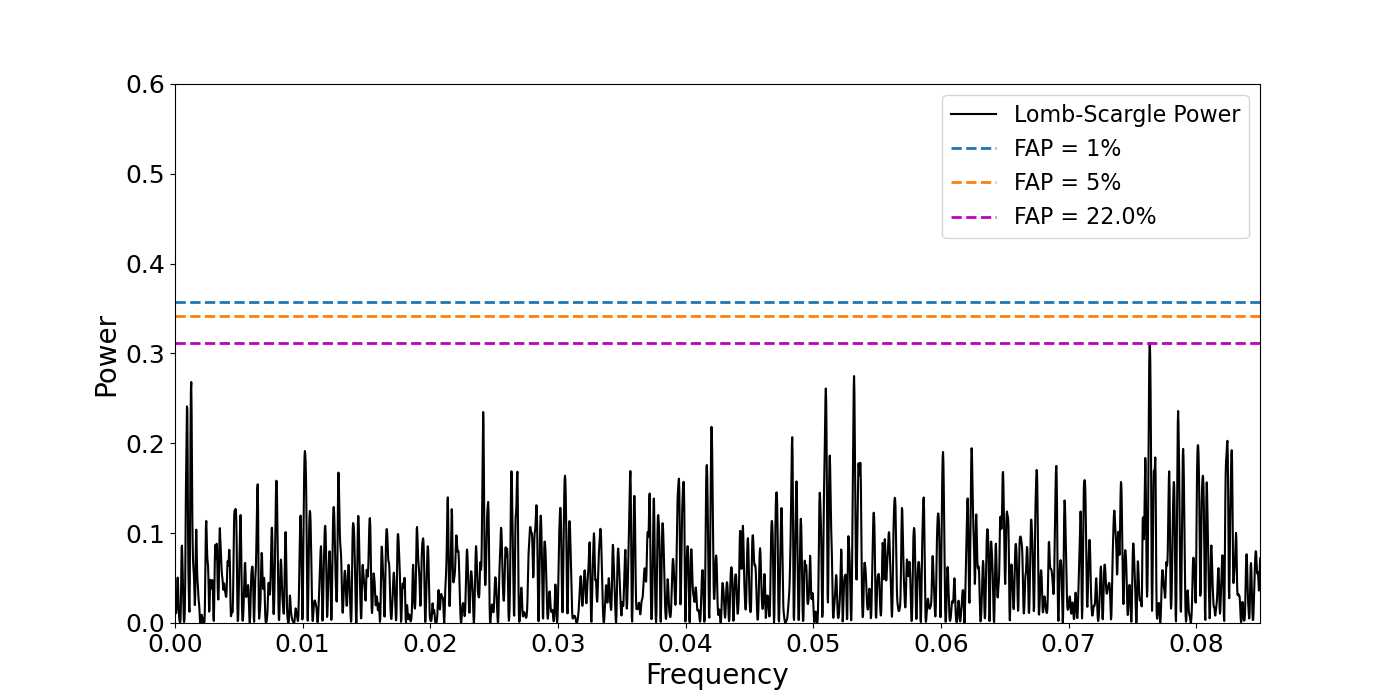}
    \caption{Generalized Lomb-Scargle periodogram computed for the 252 timing residuals of WASP-19b. The purple dashed line indicates the FAP level of the highest peak. The other dashed lines from top to bottom indicate the threshold levels of FAP = 1\% and 5\%, respectively..}
    \label{fig:periodogram}
\end{figure}

The results of the generalized Lomb-Scargle analysis are presented in the periodogram of the power spectrum as a function of frequency in Figure \ref{fig:periodogram}. As shown in Figure \ref{fig:periodogram}, the false alarm probability (FAP) associated with the highest power peak is significantly below the commonly adopted thresholds of 5\% and 1\%. This result suggests that the potential transit timing variations (TTVs) in the WASP-19 system do not exhibit any detectable periodic behavior. 

Nonetheless, under the assumption that the TTVs follow a sinusoidal modulation, we derived the amplitude of this cyclic variation by applying the same methodology described in \citet{2019A&A...622A..71V} and \citet{2022AJ....163...77A, 2023AJ....166..223A}. The timing residuals (as listed in Table \ref{tab:6}) were fitted using the following function:
\begin{equation}
\label{eq:frequency}
TTV(E) = A_{TTV}\sin(2{\pi}fE - \phi),
\end{equation}

where \( A_{\mathrm{TTV}} \) is the amplitude (in minutes) of the timing residuals, \( f \) is the frequency at the highest peak in the power periodogram, and \( \phi \) is the orbital phase. The initial estimates of the amplitude, frequency, and phase were obtained using the bootstrap analysis of the highest peak in the periodogram as explained above. These values served as starting points for model fitting, during which the frequency, amplitude, and phase were allowed to vary freely. 

We adopted uniform prior distributions for the parameters within the following bounds: amplitude \( A_{\mathrm{TTV}} \in (0, 0.8) \), frequency \( f \in (0.02, 0.10) \), and phase \( \phi \in (0, 2\pi) \), i.e.,
\[
A_{\mathrm{TTV}} \sim \mathcal{U}(0, 0.8), \quad f \sim \mathcal{U}(0.02, 0.10), \quad \phi \sim \mathcal{U}(0, 2\pi).
\]

The best-fit values for the highest peak (hereafter, Peak 1) are: amplitude 
\( A_{\mathrm{TTV}} = 0.475892^{+0.034750}_{-0.034826} \), frequency 
\( f = 0.076354^{+0.000011}_{-0.000015} \), and phase 
\( \phi = 0.643350^{+0.174561}_{-0.233109} \).

We also calculated the $\chi^{2}_{red}$,, Bayesian Information Criterion (BIC), and Akaike Information Criterion (AIC) for this model. While these statistical metrics are significantly improved compared to the three alternative timing models, the fitted frequency is high enough to cause the TTV model to oscillate more rapidly than the typical gap between transit epochs. As a result, we conclude that this model is not physically viable and that is consistent with the outcome of the frequency analysis performed using the bootstrap method.

\begin{figure}
    \centering
    \includegraphics[width=1.1\linewidth]{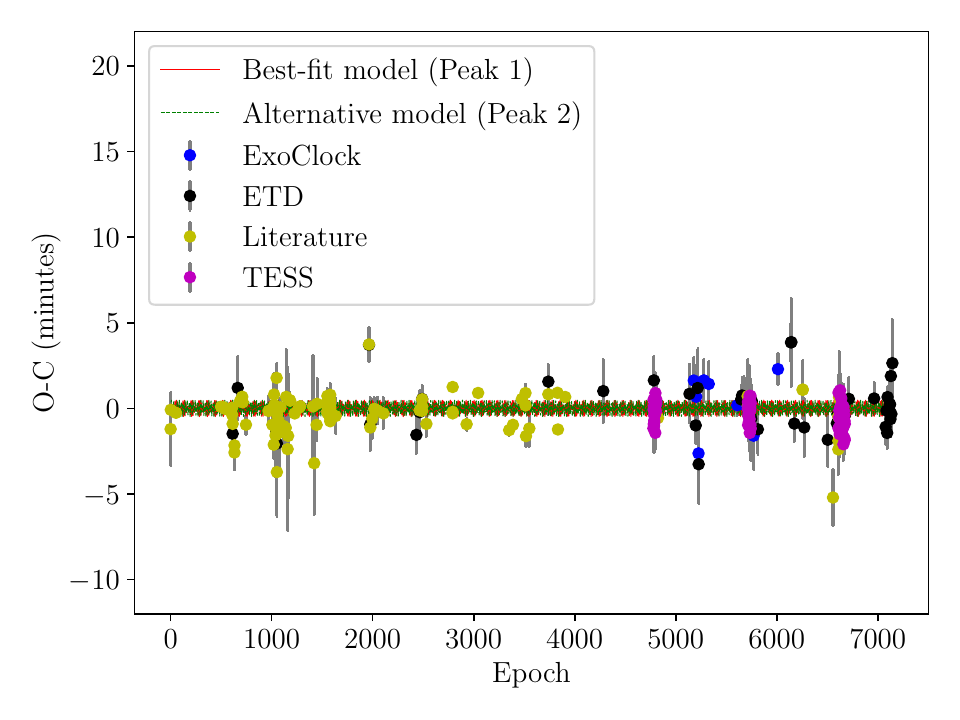}
    \caption{The O−C diagram as a function of epoch along with the best-fit sinusoidal variability.}
    \label{fig:sinusoidal_model}
\end{figure}

\begin{figure}
    \centering
    \includegraphics[width=1.1\linewidth]{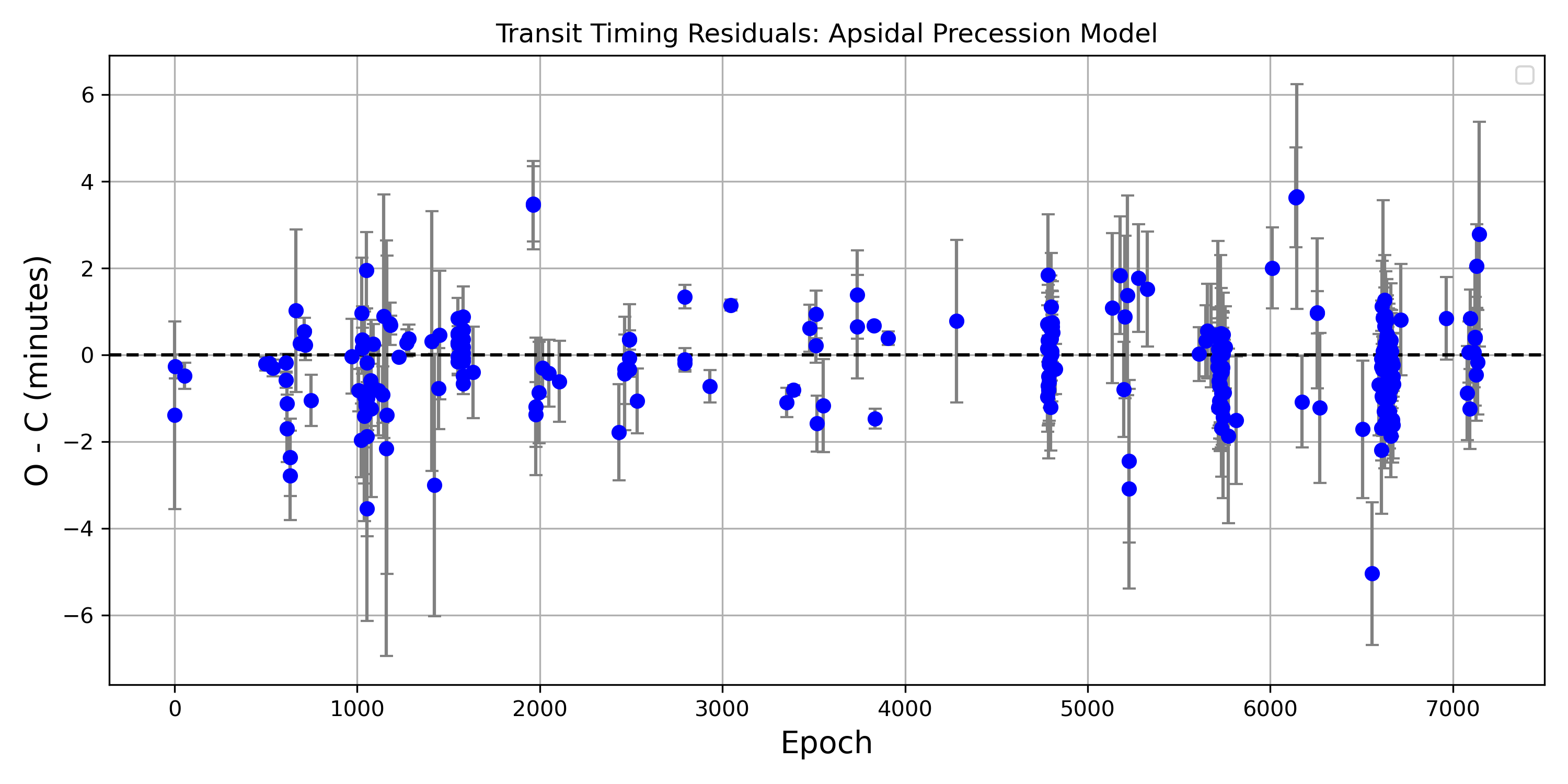}
    \caption{Residual Plot for Apsidal Precession Model}
    \label{fig:Residual_plot}
\end{figure}

To further investigate, we repeated the fitting procedure for the second-highest peak (Peak 2) and generated residuals versus epoch plots for both cases (see Figure \ref{fig:sinusoidal_model}). In both instances, the results were qualitatively similar, leading us to rule out the presence of an additional planet based on the current dataset. Therefore, we conclude that the apsidal precession model provides a better explanation for the observed TTVs in our system. The residuals corresponding to this preferred model are presented in Figure \ref{fig:Residual_plot}.

\section{Discussion}     \label{sec:results}

\subsection{Derivation of the Stellar Tidal Quality Factor} \label{sec:tidal quality factor}
The discovery of hot Jupiters with tight orbits has reignited interest in understanding energy dissipation in stars through tidal interactions (\citealt{2018ARA&A..56..175D}), particularly with regard to the tidal quality factor, ${Q}_{\ast}$, which quantifies the efficiency with which a celestial body dissipates tidal energy. The tidal quality factor ${Q}_{\ast}$ is also defined as the ratio of the maximum energy stored in the tidal deformation of a star over an orbital period to the total energy dissipated through frictional forces during the same period (see, e.g. Eq. (2.19.) of \citealt{2008EAS....29...67Z}). This parameter plays a crucial role in characterizing the timescales over which tidal interactions between a planet and its host star influence the star's rotational dynamics and the evolution of its orbit. Although we do not find significant evidence for orbital decay, we can explore the implications under the assumption that the model is valid and that the observed reduction in orbital period is predominantly caused by tidal dissipation within the host star.

Empirically, the efficiency of tidal dissipation is parametrized using a dimensionless parameter: modified tidal quality factor (${Q}^{'}_{\ast}$). This dimensionless parameter quantifies the efficiency of tidal kinetic energy dissipation within the star. It is defined as the ratio of the maximum tidal energy stored in the system to the energy dissipated over one tidal cycle (\citealt{1963MNRAS.126..257G}). This parameter is crucial for characterizing the system's evolution under the influence of tidal forces. It can be represented in combination with the potential Love number of the second order: (\citealt{2007ApJ...661.1180O}), which takes into account the internal density stratification of the object, and is defined from the tidal interaction model by \citet{1977A&A....57..383Z} as ${Q}^{'}_{\ast} = \frac{3}{2} \left(\frac{Q_\ast}{K_\ast}\right)$ (\citealt{2010ApJ...725.1995M, 2016AJ....151..137H, 2018AJ....155..165P, 2022AJ....163..281T}), where ${K}_{\ast}$ is the stellar Love number, which quantifies the concentration of the star's density in its core and characterizes its response to tidal forces and ${Q}_{\ast}$ is the tidal quality factor. The 1/${Q}_{\ast}$ is the energy lost over one tidal period normalized by the energy stored in the tidal deformation; k* is the ratio between the linear perturbation of the self-gravitation potential induced by the presence of the companion and the perturbing tidal potential evaluated at the stellar surface. By definition, the modified tidal quality factor, ${Q}^{'}_{\ast}$, is the inverse of the phase lag angle between the tidal forcing potential and the resulting tidal bulge (\citealt{1966Icar....5..375G}). To calculate ${Q}^{'}_{\ast}$ for the WASP-19 system, we employed the approach outlined by \citet{2017AJ....154....4P} and \citet{2018AcA....68..371M}, utilizing Kepler's third law in conjunction with Equation (20) from the modified ``constant phase lag (CPL'' model proposed by \citet{1966Icar....5..375G}:
\begin{equation}
{Q}^{'}_{\ast} = -\frac{27}{2}{\pi}\left(\frac{M_p}{M_\ast}\right)\left(\frac{a}{R_\ast}\right)^{-5}\left(\frac{1}{\dot{P}}\right),
\label{eqn:quality_factor}
\end{equation}
which is based on the traditional equilibrium tide model where the tidal bulge on the star lags behind the line joining the star and the planet by a constant angle. Where P is the orbital period derived using the orbital decay model, $\dot{P}$ is the decay rate of the orbital period, $\frac{M_p}{M_\ast}$ is the planet-to-star mass ratio, and $\frac{a}{R_\ast}$ is the ratio of the semi-major axis to the stellar radius. From the expression, it is evident that the time derivative is inversely proportional to the modified stellar tidal quality factor, ${Q}^{'}_{\ast}$. This expression is based on the assumption that the host star's rotational frequency is much smaller than the planet's orbital frequency.

In this study, the values of the planetary and stellar masses, M$_P$ = 1.168\, M$_J$ and M$_\ast$ = 0.965\ M$_{\sun}$, as well as the  semi-major axis relative to the stellar radius, $\frac{a}{R_\ast}$ = 3.533, were adopted from \citet{2020A&A...636A..98C}. By substituting the derived value of $\dot{P}$ (see Section \ref{sec:comparison}) and the above values in Equation \ref{eqn:quality_factor}, we have inferred the modiﬁed stellar tidal quality factor to be 
$\sim 2.6 \times 10^{6}$ for WASP-19b; where the higher value corresponds to relatively less efficient tidal dissipation (\citealt{1966Icar....5..375G, 2017ApJ...836L..24W}), resulting in weaker tidal forces and slower changes in the system over time. Our calculated value of ${Q}^{'}_{\ast}$ falls within the $1\sigma$ confidence interval of the lower limit reported by \citet{2022A&A...668A.114R}, specifically ${Q}^{'}_{\ast}$ = $ 1.26 \times 10^{6}$.

This value also agrees well with the values reported for binary star systems (${10}^{5}-{10}^{7}$; \citealt{2005ApJ...620..970M, 2007ApJ...661.1180O, 2010A&A...512A..77L, 2015Natur.517..589M}), for the stars hosting the hot Jupiters (${10}^{5}–{10}^{6.5}$; e.g., \citealt{2008ApJ...681.1631J, 2012MNRAS.422.3151H, 2020MNRAS.498.2270B}), for transiting giant planets (${10}^{4}–{10}^{8}$; \citealt{2017A&A...602A.107B}), for the stars hosting gas giant planets with 0.5 days $< P < 2 $ days (${10}^{5}–{10}^{7}$; \citealt{2018AJ....155..165P}) and for short-period ($P \leq 2 $ days) and massive ( $M_{P} \geq 0.5\ M_{J}$) hot Jupiters (${10}^{5}–{10}^{6}$; a theoretical study by \citealt{2016ApJ...816...18E}). Our value is fully consistent with the ${Q}^{'}_{\ast}$ reported by \citet{2006ApJ...653..621M} (${Q}^{'}_{\ast}$ $\sim 10^{6}$, solar-type binaries in open clusters). Additionally, \citet{2024ApJ...960...50W} estimated ${Q}^{'}_{\ast} < {10}^{6}$ for systems similar to WASP-19 b with M$_\ast < 1.1\ M_{\sun}$ and ${P} < 1$ day, in agreement with our ${Q}^{'}_{\ast}$. However, our finding is also in agreement with the upper limit of the typical modiﬁed stellar tidal quality factor value of hot-Jupiter host stars (${Q}^{'}_{\ast} \leq {10}^{7}$), reported by \citet{2019AJ....158..190H}, for the period range 2 days$ < P < 5\ $days and the mass range $0.5\ M_{Jup} < M_{P} < 2\ M_{Jup}$. The comparison of our derived ${Q}^{'}_{\ast}$ value with those from previous studies is shown in Table \ref{tab:comparison_of_tidal_quality_factor}.

\begin{table*}[htbp]
    \centering
    \small
    \renewcommand{\arraystretch}{1.4}
        \caption{Comparison of the values of Stellar Tidal Quality Factor of WASP-19b as estimated by previous works.}
    \begin{tabular}{l c}
    \hline\hline
         Stellar tidal quality factor, ${Q}^{'}_{\ast}$ & Reference \\
        
         \hline
         $\sim$$ 2.6 \times 10^{6}$ & This Work\\
        $  \gtrapprox 2.5 \times 10^{6}$ & \citet{2024PSJ.....5..163A}\\
         $~7 \times 10^{5}$ & \citet{2023Univ...10...12K} \\
           $4.6 \times 10^{6}$ &  \citet{2023RAA....23i5014G} \\
           $  > 1.26 \times 10^{6}$ &  \citet{Ros_rio_2022} \\
          $(6.5-8.1) \times 10^{6} $& \citet{2018AJ....155..165P}  \\
           5 $\times 10^{5}$ & \citet{2020AJ....159..150P} \\
         $  > 1.23 \times 10^{6}$ &  \citet{2020MNRAS.491.1243P} \\
         
         \hline
    \end{tabular}
    
    \label{tab:comparison_of_tidal_quality_factor}
\end{table*}

%-------------------------------------------------------------------
While the derived value of ${Q}^{'}_{\ast}$ is consistent with the previously mentioned results, it is at least 1-2 orders of magnitude smaller than the typical values reported for Sun-like primary stars in eclipsing binaries (${10}^{7.8}$; \citealt{2022MNRAS.512.3651P}); for a simplified tidal evolution model (${10}^{7.5}$-${10}^{8.5}$; \citealt{2010ApJ...723..285H}), for hot Jupiters in dynamical and equilibrium tide regimes (${10}^{7.3}$-${10}^{8.3}$; \citet{2018MNRAS.476.2542C} and nearly 1 order of magnitude higher than the value, ${Q}^{'}_{\ast}$ $\sim {10}^{5.5}$ reported by \citet{2008ApJ...681.1631J}.

\citet{2022A&A...668A.114R} determined new lower limits for ${Q}^{'}_{\ast}$ of $(1.26 \pm 0.10) \times 10^{6}$ for WASP-19, and our derived ${Q}^{'}_{\ast}$ value for WASP-19b is statistically consistent with their results. Furthermore, our ${Q}^{'}_{\ast}$ value for WASP-19b is at least one order of magnitude higher than that of WASP-12b (\citealt{2024A&A...686A..84L}) and two orders of magnitude higher than that of WASP-4b (\citealt{2020ApJ...893L..29B}), suggesting that the orbit of WASP-19b may be decaying gradually, which could result in a slow orbital evolution.

\subsubsection{Calculation of the remaining lifetime}   \label{sec:remaining lifetime}
The remaining lifetime of a hot Jupiter is the time until its orbit decays sufficiently for the planet to spiral inward and ultimately collide with its host star. This orbital decay is primarily driven by stellar tidal dissipation, which is typically quantified by the tidal quality factor ${Q}^{'}_{\ast}$, a measure of the efficiency of energy dissipation in the star. The timescale for the orbital evolution of hot Jupiters is therefore influenced by the value of ${Q}^{'}_{\ast}$ (\citealt{2016A&A...588L...6M, 2017AJ....154....4P}). By substituting the appropriate value of ${Q}^{'}_{\ast}$ (as detailed in Section \ref{sec:tidal quality factor}) and other relevant parameters from \citet{2020A&A...636A..98C} into Equation (5) from \citet{2009ApJ...692L...9L}:
\begin{equation}
T_{\text{remain}} = \frac{1}{48}\frac{Q'_{\star}}{n}\,\left(\frac{a}{R_{\star}}\right)^5\,\left(\frac{M_{\star}}{M_p}\right) \,,
\label{timing_a}
\end{equation}
we calculated the value of $T_{remain}$ of WASP-19b to be $\sim 8.8 $Myr, where n = $\frac{2\pi}{P}$ is the frequency of the mean orbital motion of the planet.

\subsubsection{Calculation of the shift in transit time}   \label{sec:shift in transit time}
The expected shift in the transit arrival time of an exoplanet, commonly referred to as $T_{shift}$, represents the predicted variation in the timing of the planet's transits as observed from Earth, resulting from various influences on the planet's orbital dynamics.
To estimate the expected shift in the transit arrival time of WASP-19b due to its decaying orbit, we used Equation (7) of \citet{2014MNRAS.440.1470B}:
\begin{equation}
T_{\rm shift}=\frac{1}{2}T^{2}\left(\frac{dn}{dT}\right)\left(\frac{P}{2\pi}\right),
\label{eqn:shift}
\end{equation}
where $\frac{dn}{dT}$ is current rate of orbital frequency change of the planet. The P and T values are adopted from Table \ref{tb:timing_models} derived for Orbital Decay Model. For the calculated value of $\frac{dn}{dT}$ = $\sim 4.66 \times 10^{-20} rad\, s^{-2} $corresponding to ${Q}^{'}_{\ast} = 2.6 \times 10^{6}$, the expected shift in the transit arrival time of WASP-19b after 15 yr (T = 15 yr) is found to be $T_{shift} \sim56.56 $s. This value of $T_{shift}$ is consistent with the rms of the obtained timing residuals (see, Section \ref{sec:additional}). Since ${Q}^{'}_{\ast}$ is inversely proportional to $\dot{P}$, any uncertainties or errors in $\dot{P}$ will have a significant impact on the uncertainties in ${Q}^{'}_{\ast}$. Therefore, even small errors in P can lead to large deviations in the value of ${Q}^{'}_{\ast}$. Given the uncertainty in our estimate of $\dot{P}$ ($\pm0.00040 $ s/yr), we calculate ${Q}^{'}_{\ast} \approx 6.9 \times 10^{6}$ and if ${Q}^{'}_{\ast} = 2.6 \times 10^{6}$ measured from our timing analysis is maintained for another five years (i.e., in a total of
20 yr monitoring), one can expect $T_{shift} \sim62.64 $s. This conclusion can be further confirmed through subsequent follow-up observations.

\subsection{Planetary Love Number \texorpdfstring{$k_p$}{kp}} 
\label{sec:Planetery Love Number}
According to tidal evolution theory, tidal forces are expected to lead to the circularization of hot Jupiter orbits on a timescale shorter than the ages of their host systems (\citealt{2007A&A...462L...5L, 2018ARA&A..56..175D}). The system in question, WASP-19, is relatively old, yet it continues to exhibit orbital eccentricity. This observation contradicts the predictions of tidal theory, suggesting the presence of additional dynamical processes that may be influencing the system's behavior. Our findings indicate that a companion may contribute to the excitation of eccentricity, thereby preventing circularization (see Section \ref{sec:additional}). By rederiving Equation (25) from \citet{1966Icar....5..375G} and incorporating Equation (17) from \citet{2017AJ....154....4P}, as outlined below:
\begin{equation}
\tau_e = \frac{e}{|de/dt|} =
\frac{2Q_{\rm p}}{63\pi} \left(\frac{M_{\rm p}}{M_\star}\right)
\left( \frac{a}{R_{\rm p}} \right)^5 P_{\rm orb}.
\end{equation}
and by assuming the value of the planetary tidal quality factor to be $Q_{\rm p}\sim 10^6$, the expected timescale for the tidal orbital circularization of WASP-19b, $\tau_e$ = {e}/{$|\dot{e}|$} $\sim50.36 $ Myr, which is several orders of magnitude shorter than the age of the host star (WASP-19: 9.95 Gyr; \citealt{2010ApJ...708..224H}). Given that the system exhibits observable orbital eccentricity (see Section \ref{sec:apsidal}) and is much older than the estimated timescale for tidal circularization, this indicates that tidal forces have not had enough time to circularize the orbit. Therefore, the presence of an eccentric orbit is inconsistent with tidal theory, which anticipates a rapid circularization of orbits due to tidal interactions. Therefore, alongside the existence of a companion body, there must be some other dynamic process that can excite and maintain a nonzero eccentricity (see \citealt{2019AJ....157..217B, 2021A&A...656A..88M, 2022ApJ...941L..31V}).

In this context, \citet{2009ApJ...698.1778R} introduced the concept of planetary love number. The planetary Love number ($k_{2p}$ , twice the apsidal motion constant, hereafter, $k_{p}$), named after mathematician A. E. H. Love, quantifies a planet's elastic response to tidal forces, characterizing its deformation due to the gravitational influence of external perturbing body. This dimensionless parameter is a measure of the degree of central concentration of the planet's interior density distribution, which has a direct effect on the star-planet orbit through the gravitational quadrupole of rotational and tidal bulges. Thus, $k_{p}$ is a crucial tool for studying the relationship between the interior density of hot Jupiters and their interactions with host stars, enhancing our understanding of their complex behaviors. To estimate the value of planetary love number, $k_{p}$, we adopted the following equation of \citet{2017AJ....154....4P}:
\begin{equation}
\frac{d\omega}{dE}={15}{\pi}{k_p}\left(\frac{M_\ast}{M_p}\right)\left(\frac{R_p}{a}\right)^5,
\end{equation}
where the value $\frac{d\omega}{dE}$ is taken from Table \ref{tb:timing_models}, and the other relevant parameters are taken from \citet{2020A&A...636A..98C}). The higher value of $k_{p}$ leads to a higher precession rate with a smaller centrally condensed interior structure.

Our estimated value of the planetary Love number, $k_{p}$ = $1.21\pm0.56$ is approximately one order of magnitude higher than that of Jupiter, $k_{p}$ = 0.59 (\citealt{2016ApJ...831...14W}), suggesting that the planet possesses a less centrally condensed internal structure. A higher Love number reflects a greater susceptibility to tidal deformation. Since the apsidal precession rate is directly proportional to the Love number $k_{p}$ (\citealt{2009ApJ...698.1778R, 2017AJ....154....4P}), this enhanced deformability results in a faster apsidal precession, thereby influencing the orientation of the orbital ellipse over time.

Furthermore, a relatively high Love number (around 0.6 or greater) implies that the planet is more responsive to tidal forces, exhibiting a more deformable and less rigid structure. This is characteristic of gas giants—especially hot Jupiters—which often have substantial gaseous envelopes and complex internal density distributions. A higher degree of tidal deformation can induce significant shifts in the orbital orientation, with potential implications for the long-term dynamical stability and evolution of the planetary orbit. The inferred Love number for this planet aligns with similarly elevated values reported in other systems, such as WASP-4b ($1.20\pm0.23$; \citealt{2019AJ....157..217B}), HAT-P-13b ($0.81\pm0.10$; \citealt{2017ApJ...836..143H}), and WASP-18b ($0.62\pm0.37$; \citealt{Csizmadia_2019}). Even higher values have been derived from tidal deformation modeling for planets like WASP-103b ($1.590\pm0.49$; \citealt{Barros_2022}) and WASP-12b ($1.550\pm0.47$; \citealt{2024A&A...685A..63A}). These values exceed what is expected from a purely static tidal response and may necessitate invoking additional mechanisms, such as rapid planetary rotation. However, such rapid rotation appears inconsistent with the presumed tidally locked states of these close-in exoplanets.

\subsection{Other Possible Causes for the Long-term Variation of the Orbital Period} \label{sec: other causes}
There are several other possible scenarios to explain the long-term variation of the orbital period of exoplanets. Here, we provide a brief overview of some of these, including the Applegate mechanism, and Shklovskii effect.
\subsubsection{Applegate mechanism} \label{sec: applegate mechanism}
The observed variations in the orbital period of close stellar binary systems may be explained by the Applegate mechanism (\citealt{1987ApJ...322L..99A, 1992ApJ...385..621A}). This mechanism not only influences the orbital motion of binary stars but has also been shown to affect the dynamics of ultra-short-period (USP) gas giants, as demonstrated by Watson and Marsh, who calculated that the Applegate effect can have a measurable impact that depends on the stellar activity cycle. The underlying cause of this effect is the magnetic activity cycles in the low-mass components of binary systems, which redistribute angular momentum within the stellar interior. This redistribution results in fluctuations in the stellar quadrupole moment of the low-mass primary star, induced by spin–orbit coupling (\citealt{2002ApJ...564.1019M}), ultimately leading to variations in the orbital period of the binary components. However, these variations are not perfectly periodic; instead, they occur on quasiperiodic timescales that are linked to the stellar activity cycle, introducing irregularities in the transit times.

\citet{2010MNRAS.405.2037W} estimated that the TTV amplitudes induced by the Applegate effect in several exoplanet systems, including WASP-19, range from a few seconds for an orbital period modulation timescale of approximately 11 years, to several minutes for a modulation timescale of around 50 years.
The magnitude of TTV driven by the Applegate mechanism is proportional to $a^{-2} T^{-\frac{3}{2}}$ modulation (\citealt{2010MNRAS.405.2037W}), which results in stronger effects for hot Jupiters due to their small star–planet separation. For WASP-19b, the largest TTV amplitude (or O-C variation) predicted by \citet{2010MNRAS.405.2037W} for the Applegate mechanism was $ \delta{t} \sim 27.5$~s over a 50-year activity cycle, which is in good agreement with the $\sim 16.82$~s TTV amplitude calculated in our study (see Section \ref{sec:linear}). This suggests that the variation in transit times is most likely caused by Applegate mechanism  or the magnetic activity of the host star, WASP-19.

\subsubsection{Shklovskii effect} \label{sec: Shklovskii effect}
Proper motion can exert a significant (and in some cases, dominant) influence on transit timing, particularly in short-period systems. In this context, \citet{Rafikov_2009} identified two additional phenomena that also contribute to changes in the apparent orbital period of a transiting planet. One such phenomenon is the Shklovskii effect (\citealt{1970SvA....13..562S}), first described by Russian astrophysicist I. S. Shklovskii in 1970. This effect arises from the relative motion between the star (or planet) and the observer, causing an apparent variation in the orbital period of the exoplanet. It arises from relativistic effects, including the Doppler shift and the star's motion, and can cause small periodic variations in the observed period of exoplanet orbits. While this effect is typically small, it becomes important when making highly precise measurements of orbital parameters, especially in systems with long orbital periods. Here, we note that the observed distance, d and proper motion, $\mu$ imply a period derivative of $P\mu^2 d/c \sim 6.9\times 10^{-14}$, too small to explain the data.

The second phenomenon, also influenced by proper motion, is the apparent apsidal precession. This effect arises not from any intrinsic changes in the orbit, but due to the alteration in the observer's viewing angle as a result of the system's proper motion. Consequently, the period derivative is of the order of approximately $(P\mu)^2/2\pi$, which in this case is on the order of $10^{-20}$. This value is so minuscule that it has a negligible impact on the observed orbital period over any practical timescale, particularly in systems with long orbital periods.

\section{Concluding Remarks}         \label{sec:remarks}
According to the theoretical predictions of \citet{2014ApJ...787L...9V}; \citet{2016ApJ...816...18E}, being an UHJ, the orbit of WASP-19b should rapidly decay with time, but some previous photometric studies of WASP-19b in last few years (\citealt{2013A&A...552A...2L, 2016ApJ...823..122W, 2020MNRAS.491.1243P}) observed lack of TTV in WASP-19b. \citet{2020MNRAS.491.1243P} also found no evidence of orbital decay or any periodic variations in transit timings that may be caused by additional bodies present in the system, thus contradicting the theoretical predictions. These contradictory findings and due to the availability of the data more than a decade, we seek this an ideal opportunity to study the exoplanet, WASP-19b by adding more data to the previous analysis. In this work, we have incorporated the ground-based photometric data along with space-based telescope photometry in order to perform the transit timing analysis. For this purpose we have taken total 116 complete transit light curves of WASP-19b observed by TESS in four it's sectors 9, 35, 62 and 63. To extend the existing observational time baseline, we incorporated additional high-quality transit light curves sourced from public databases, including the Exoplanet Transit Database (ETD) and ExoClock, along with 18 publicly available complete transit observations. In total, 222 light curves were modeled to refine the system’s physical and orbital parameters. Given the high stellar activity of WASP-19, particularly the presence of starspots, which can significantly distort mid-transit time measurements, all light curves exhibiting spot-induced anomalies were excluded from the analysis. The remaining mid-transit times were then combined with those from archival data, yielding a final dataset of 252 transit times. A detailed transit timing analysis was subsequently conducted using three distinct timing models, employing the emcee Markov Chain Monte Carlo (MCMC) sampling algorithm to derive robust parameter estimates. For the first model, linear model, we assume a circular orbit with a constant period. The second one, the orbital decay model, also assumes a circular orbit, with an exception of a steady change in the orbital period, denoted by $\frac{dP}{dE}$. The third model is the apsidal precession model, which assumes non-eccentric planetary orbit, with the argument of periastron, $\omega_0$, uniformly precessing. By fitting these three models, we have derived new ephemerides for the free parameters (see Table \ref{tb:timing_models}), these values are consistent with those values derived by others, but more precised than their derived values, because of the inclusion of more data from ETD and Exoclock. The FAP value of 22.5\%, calculated using the timing residuals obtained after linear model fitting, suggests that no additional planet may be present in the system. The determined value of the orbital decay rate, $\dot {P}$, for WASP-19b, is approximately $\sim - 1.1 \pm 0.40$~ms~yr$^{-1}$, exhibiting a declining trend. We have also calculated the $\chi^{2}_{red}$ value and the AIC, BIC values to check the goodness of the fit and to check statistically which model better fits our data. When the timing residuals (O–C) were plotted as a function of epoch (E), the apsidal precession model revealed a clear sinusoidal variation. This variation may be attributed to the light-time effect (LiTE) caused by gravitational perturbations from an additional planetary companion in the WASP-19 system. These findings potentially contradict the previously derived higher false alarm probability (FAP) value, suggesting that the presence of a second planet cannot be ruled out. To explore it further, we fitted a sinusoidal model to the mid-transit time data. Although the sinusoidal model provided a statistically improved fit compared to other timing models, the resulting frequency was unphysically high. This caused the model to exhibit oscillations on timescales shorter than the typical intervals between observed transits, leading to poor consistency with the temporal sampling of the dataset. Consequently, we reject the sinusoidal model as a viable explanation and do not propose the presence of an additional planetary companion. Instead, we conclude that the apsidal precession model remains the most plausible explanation for the observed TTVs.

Based on the negative value of $\dot {P}$, we calculated the modified stellar tidal quality factor ($ {Q}^{'}_{\ast}$) which is approximately $\sim 2.6 \times {10}^{6}$, and is consistent with theoretical predictions. Our derived values of the model parameter e and $\frac{d\omega}{dE}$ (section \ref{sec:apsidal}) are found to be statistically significant for WASP-19b, whereas the value of $\omega_0$​ is statistically insignificant. Moreover, our estimated value of planetary Love number, $k_{p}$ = $1.21\pm0.56$ is one order of magnitude greater than the $k_{p}$ value derived by \citet{2016ApJ...831...14W}, i.e., 0.59, indicating  it is more deformable and has a faster apsidal precession. In addition to tidal interaction, apsidal precession and Light-time effect, we also explored alternative mechanisms such as the Applegate effect and Shklovskii effect. And we have found the magnetic activity in the star caused by the changes in the quadrupole moment or Applegate mechanism could be a possible explanation for the observed TTV in WASP-19b, however the Shklovskii effect  fails to explain the data due to the very small value of period derivative. 

In future research, we intend to expand this study by incorporating additional space-based transit observation data from the PLATO mission (\citealt{2014ExA....38..249R}), and other relevant sources. Extending the observational time baseline will facilitate more accurate estimates of the orbital ephemerides and enable the investigation of transit timing variations (TTVs) within the system. This approach is also applicable to the study of other short-period, close-in transiting hot Jupiter systems. Moreover, the data presented in this study, when complemented by additional high-precision photometric follow-up observations of primary and secondary eclipses, along with radial velocity measurements from larger aperture ground-based telescopes, will enable further confirmation of our findings. Additionally, the ephemerides derived in this study will also aid in scheduling future follow-up observations of this system.

\section*{Acknowledgements}\label{sec:Acknowledgments}

We thank the anonymous referee and the editor for their valuable suggestions and constructive feedback, which have substantially enhanced the quality of this paper. This research is supported in part by the National Science and Technology Council (NSTC), Taiwan, under the Grant NSTC 113-2112-M-007-030 and the Grant NSTC 113-2115-M-007-008. We thank \citet{2024PSJ.....5..163A} and \citet{2024A&A...684A..78B} for compiling the TTV database, which served as a valuable resource for this study. This work also incorporates data collected from the Transiting Exoplanet Survey Satellite (TESS) mission, publicly available through the Mikulski Archive for Space Telescopes (MAST), and we extend our gratitude to the TESS mission for its significant contribution to exoplanet science by providing high-precision transit photometric data. The specific observations analyzed are retrievable at the TESS Light Curves—All Sectors page (\citealt{10.17909/t9-nmc8-f686,2021ApJS..254...39G}). The funding for the TESS mission is provided by NASA's Explorer Program. Additionally, this study utilizes publicly available transit light curves from the Exoplanet Transit Database (ETD) and the Exoclock Project, and we thank the contributors of these initiatives for making the data publicly accessible. Currently, all ETD light curves are available on VarAstro server, which was launched last year. We gratefully acknowledge the valuable suggestions and insightful discussions shared by Susana Barros, Martin Schlecker, Javier López Santiago, Paul Robertson, and Nhat Quang Hoang Tran, during preparing this manuscript. Furthermore, this work made use of the VizieR catalog access tool, operated by the Centre de Données astronomiques de Strasbourg (CDS), France; the SIMBAD database; NASA's Astrophysics Data System (ADS) Bibliographic Services; and the NASA Exoplanet Archive, operated by the California Institute of Technology under contract with NASA as part of the Exoplanet Exploration Program.

\appendix
\restartappendixnumbering

\section{Graphical Representation of individual transit events taken from TESS, ETD and Exoclock} \label{app:individual_transits}
\begin{figure} [h]
    \centering
    \includegraphics[width=1.1\linewidth]{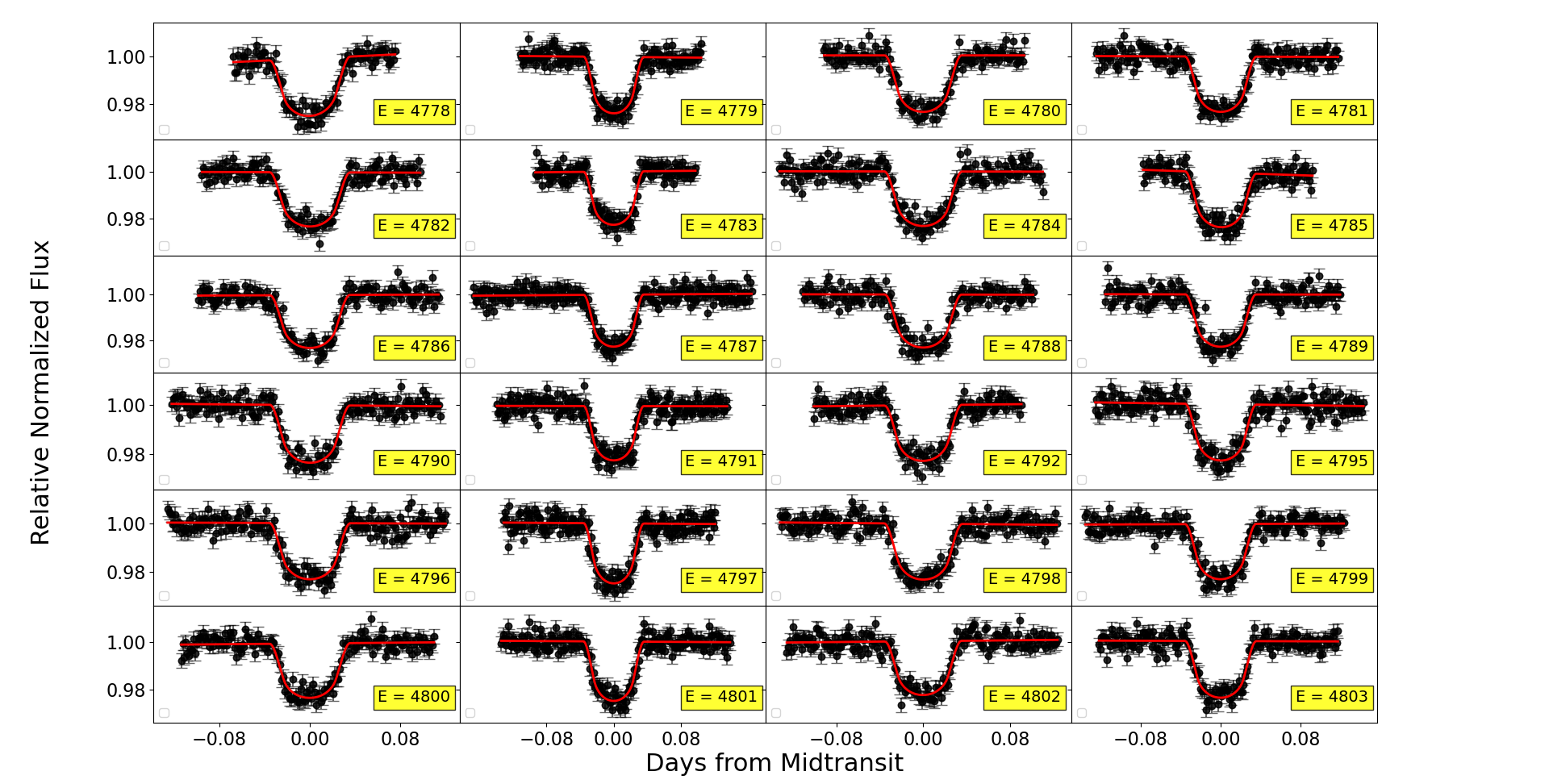}
    \caption{The normalized relative flux of WASP-19b as a function of the time (the offset from mid-transit time and in TDB-based BJD) of individual transit observed by TESS between epochs (4778 - 4803) : the points are the data of raw flux, solid red lines are best-fit models for model flux, and E is the calculated epoch number.}
    \label{fig:TESS1}
\end{figure}

\begin{figure}
    \centering
    \includegraphics[width=1.1\linewidth]{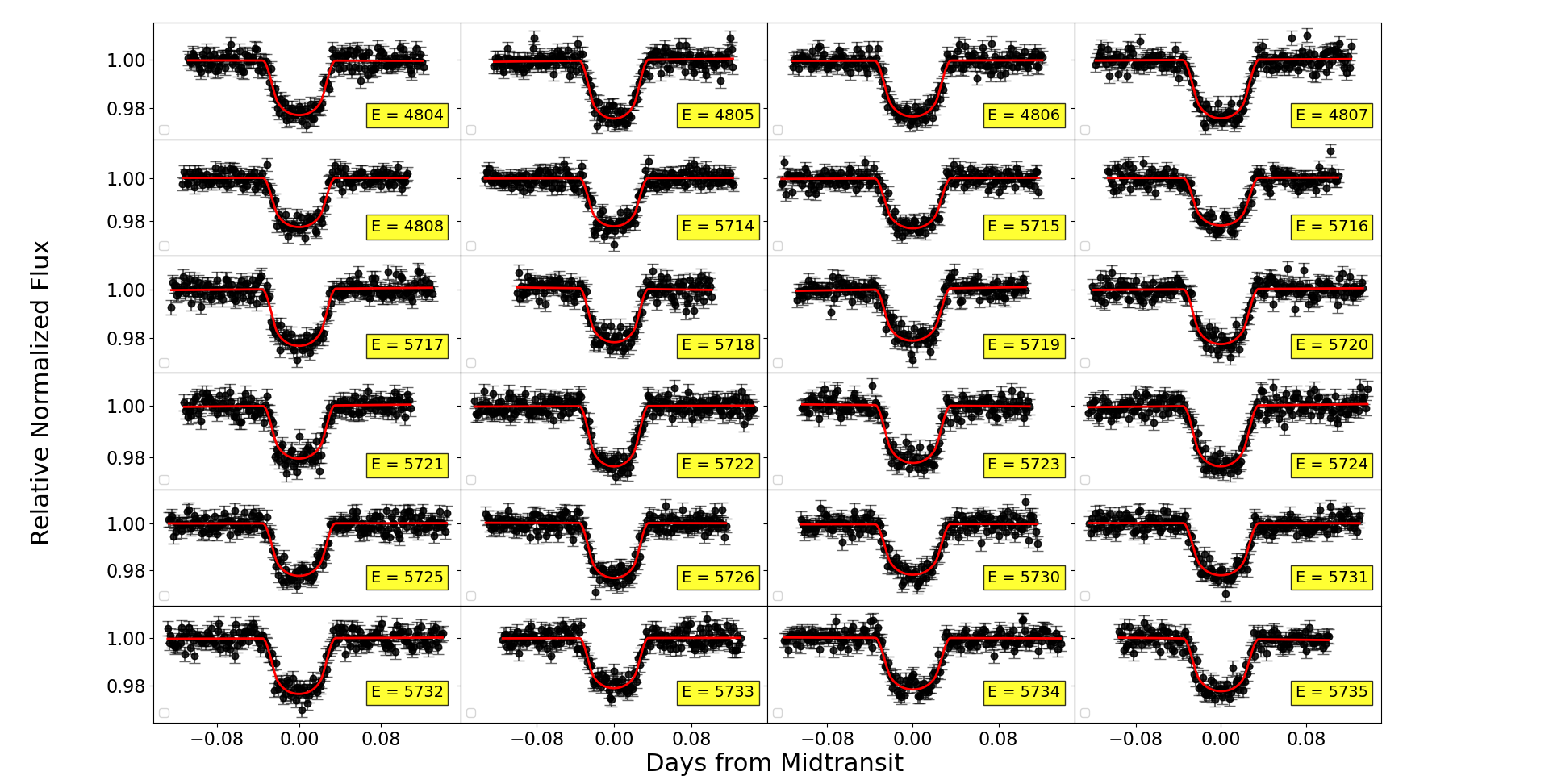}
    \caption{Same as the Figure \ref{fig:TESS1} but for epochs (4804 - 5735)}
    \label{fig:TESS2}
\end{figure}

\begin{figure}
    \centering
    \includegraphics[width=1.1\linewidth]{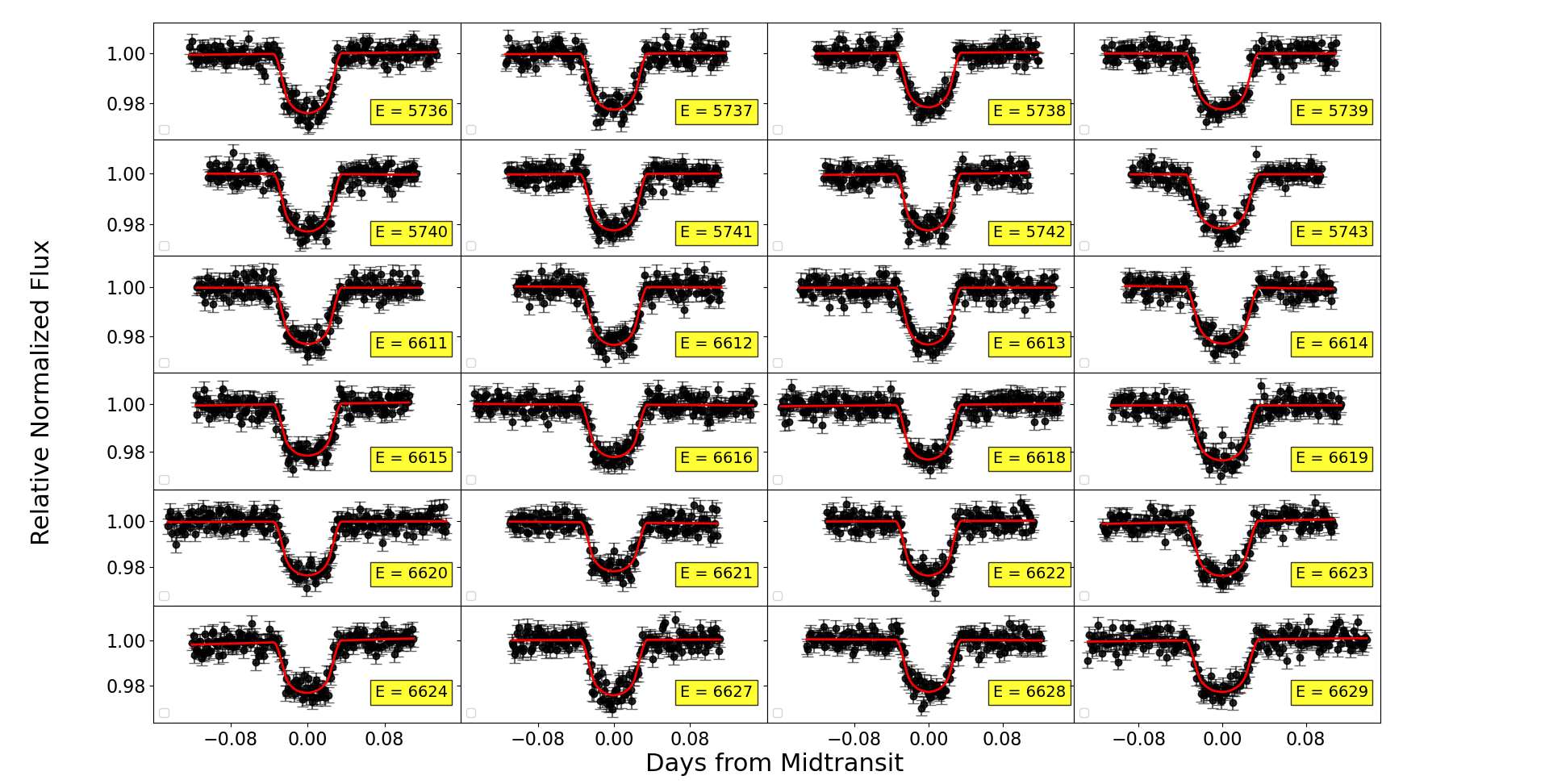}
    \caption{Same as the Figure \ref{fig:TESS1} but for epochs (5736 - 6629)}
    \label{fig:TESS3}
\end{figure}

\begin{figure}
    \centering
    \includegraphics[width=1.1\linewidth]{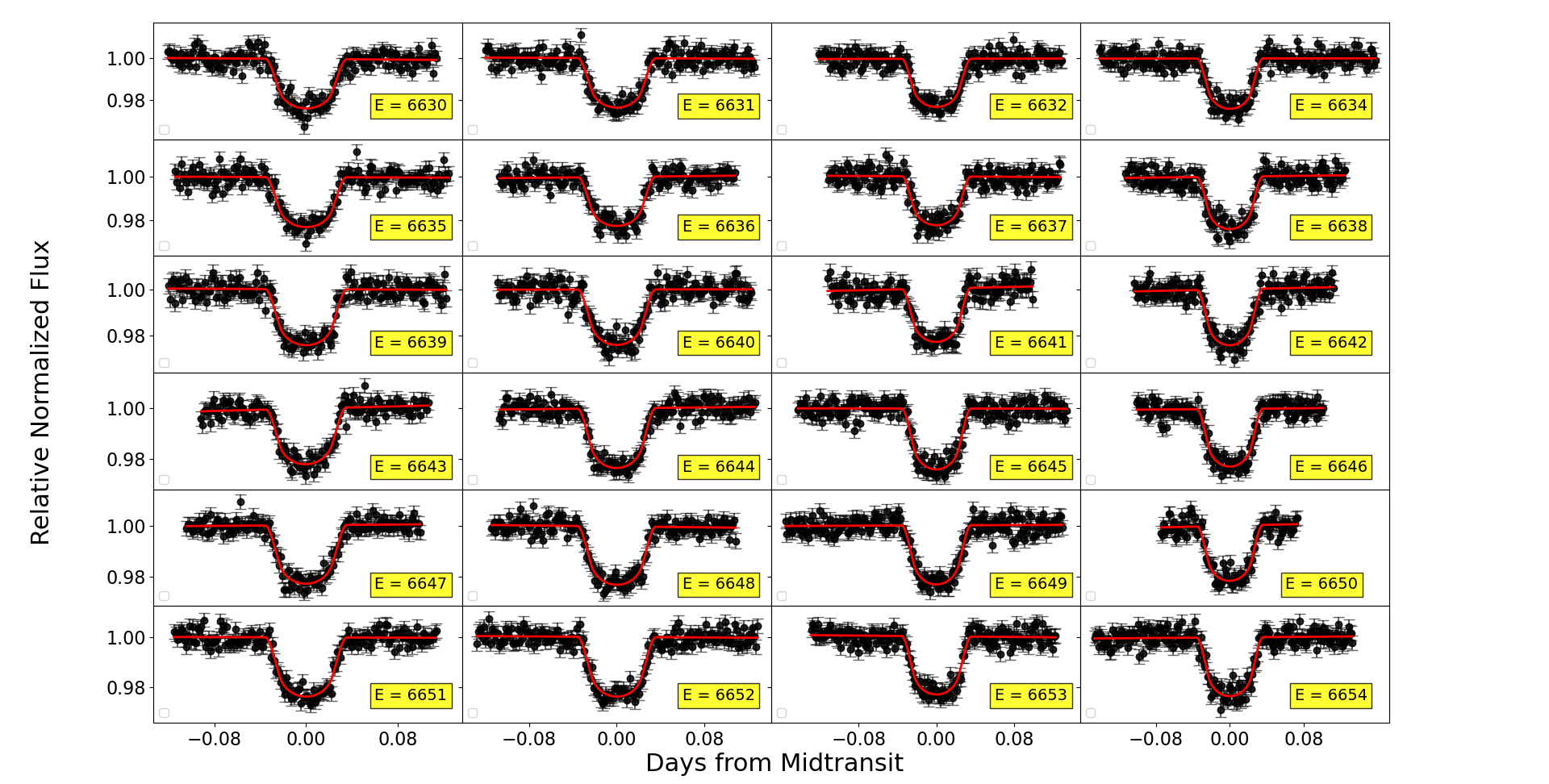}
    \caption{Same as the Figure \ref{fig:TESS1} but for epochs (6630 - 6654)}
    \label{fig:TESS4}
\end{figure}

\begin{figure}
    \centering
    \includegraphics[width=1.1\linewidth]{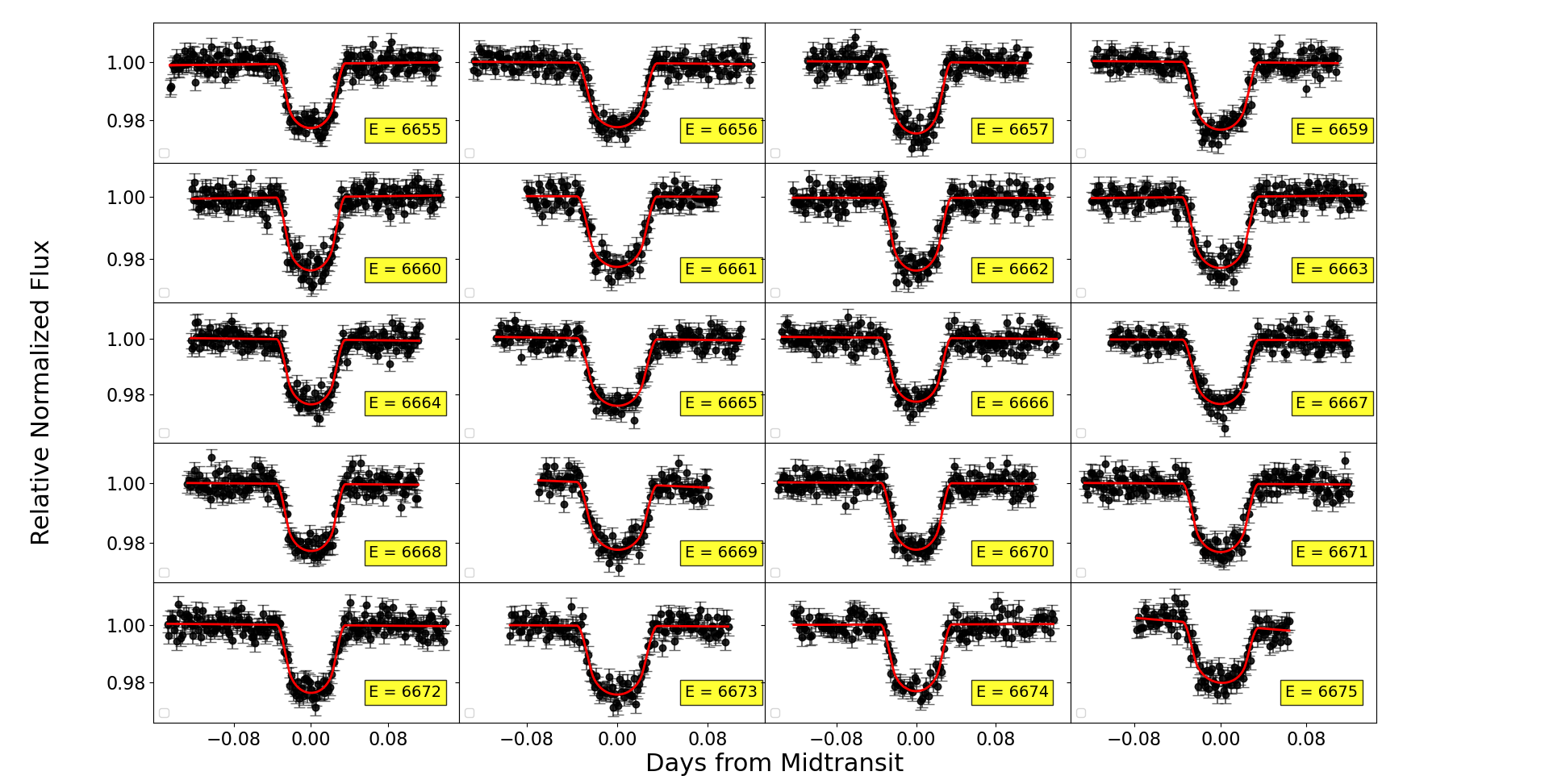}
    \caption{Same as the Figure \ref{fig:TESS1} but for epochs (6655 - 6675)}
    \label{fig:TESS5}
\end{figure}

\begin{figure}
    \centering
    \includegraphics[width=1.1\linewidth]{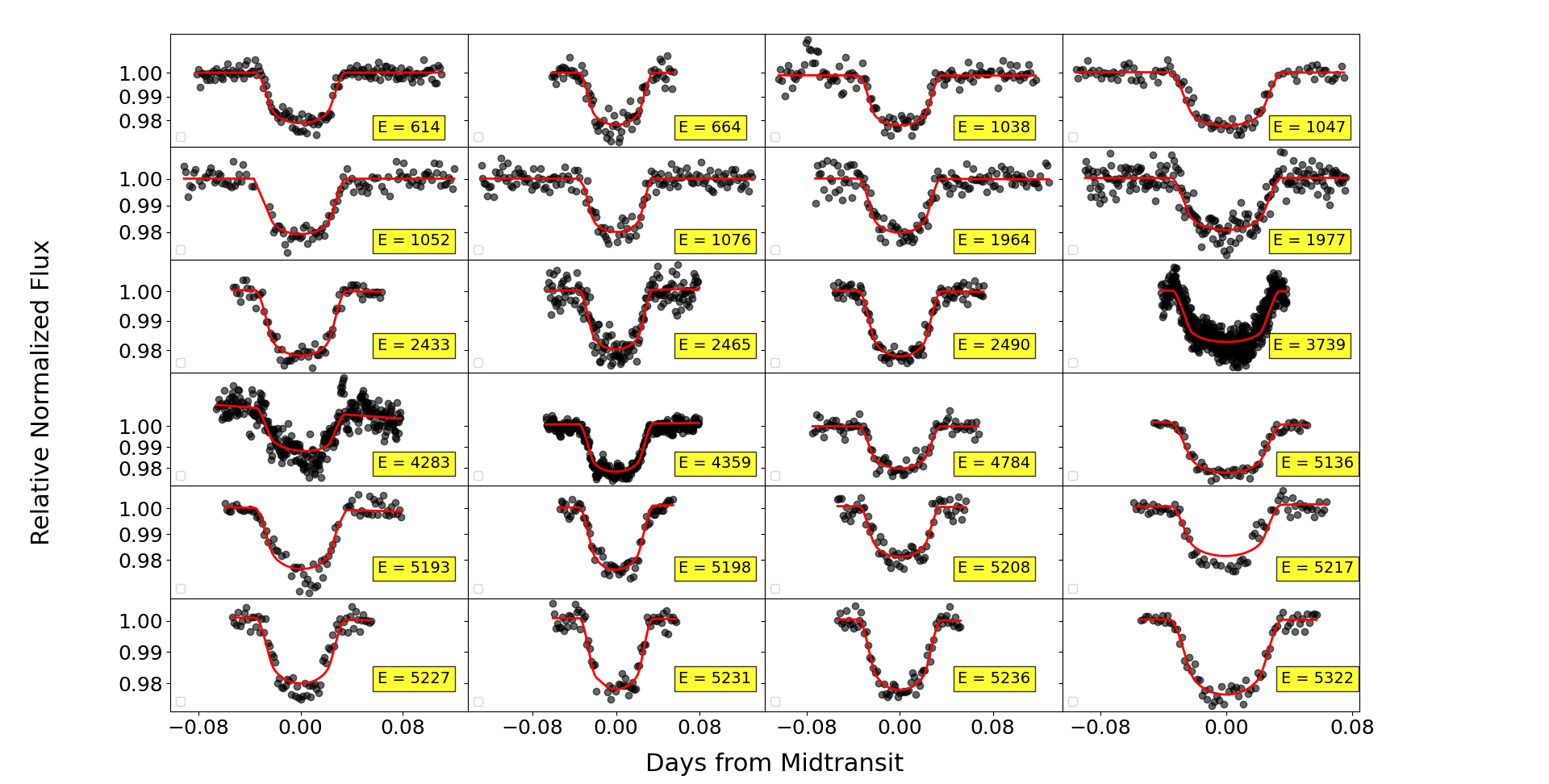}
    \caption{The normalized relative flux of WASP-19b as a function of the time (the offset from mid-transit time and in TDB-based BJD) of individual transits taken from ETD  between epochs (614 - 5322) : here the points, solid red lines, and E represent the same elements as described in Figure \ref{fig:TESS1}.}
    \label{fig:ETD1}
\end{figure}

\begin{figure}
    \centering
    \includegraphics[width=1.1\linewidth]{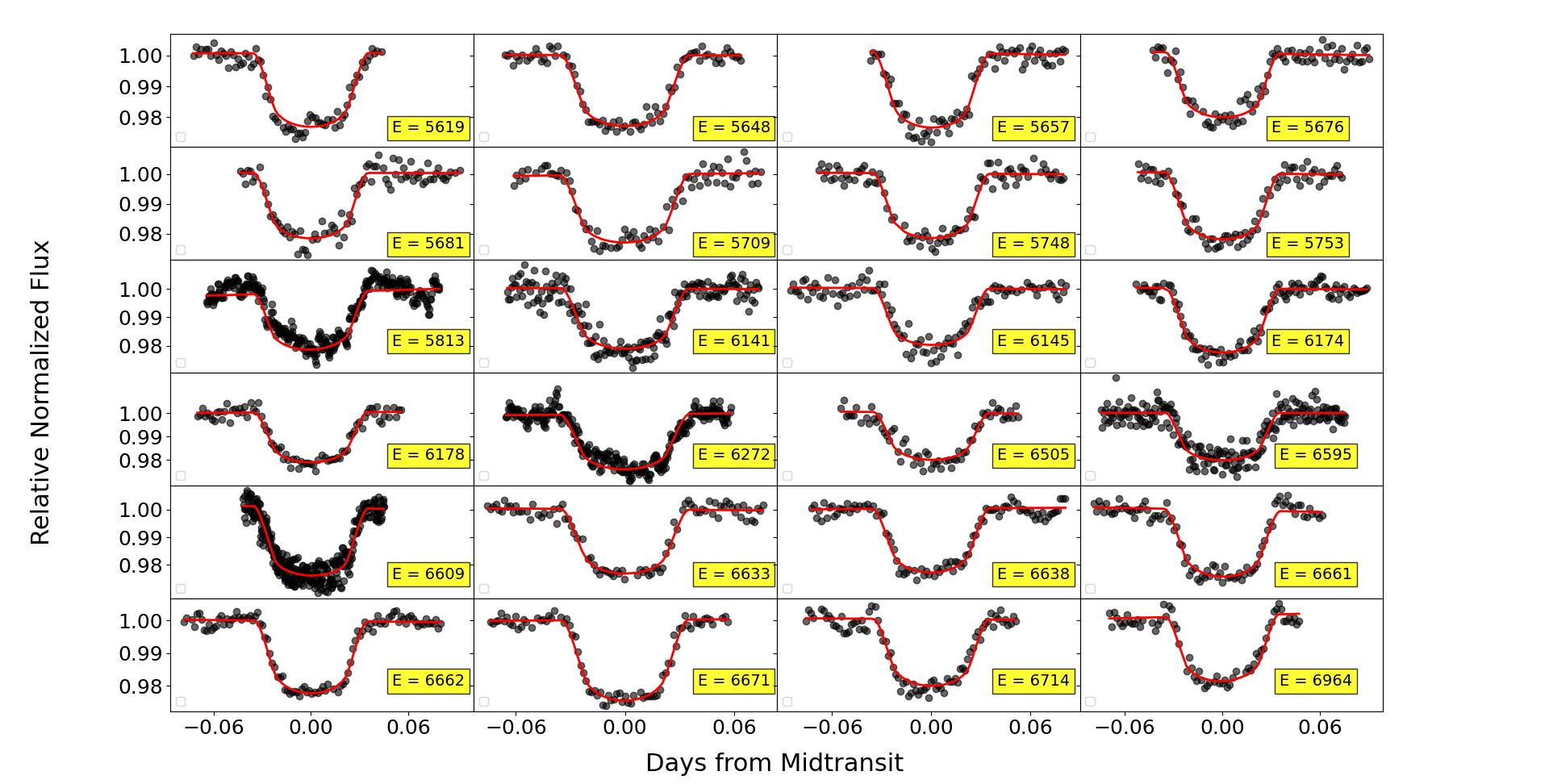}
    \caption{Same as the Figure \ref{fig:ETD1} but for epochs (5619 - 6964)}
    \label{fig:ETD2}
\end{figure}

\begin{figure}
    \centering
    \includegraphics[width=1.1\linewidth]{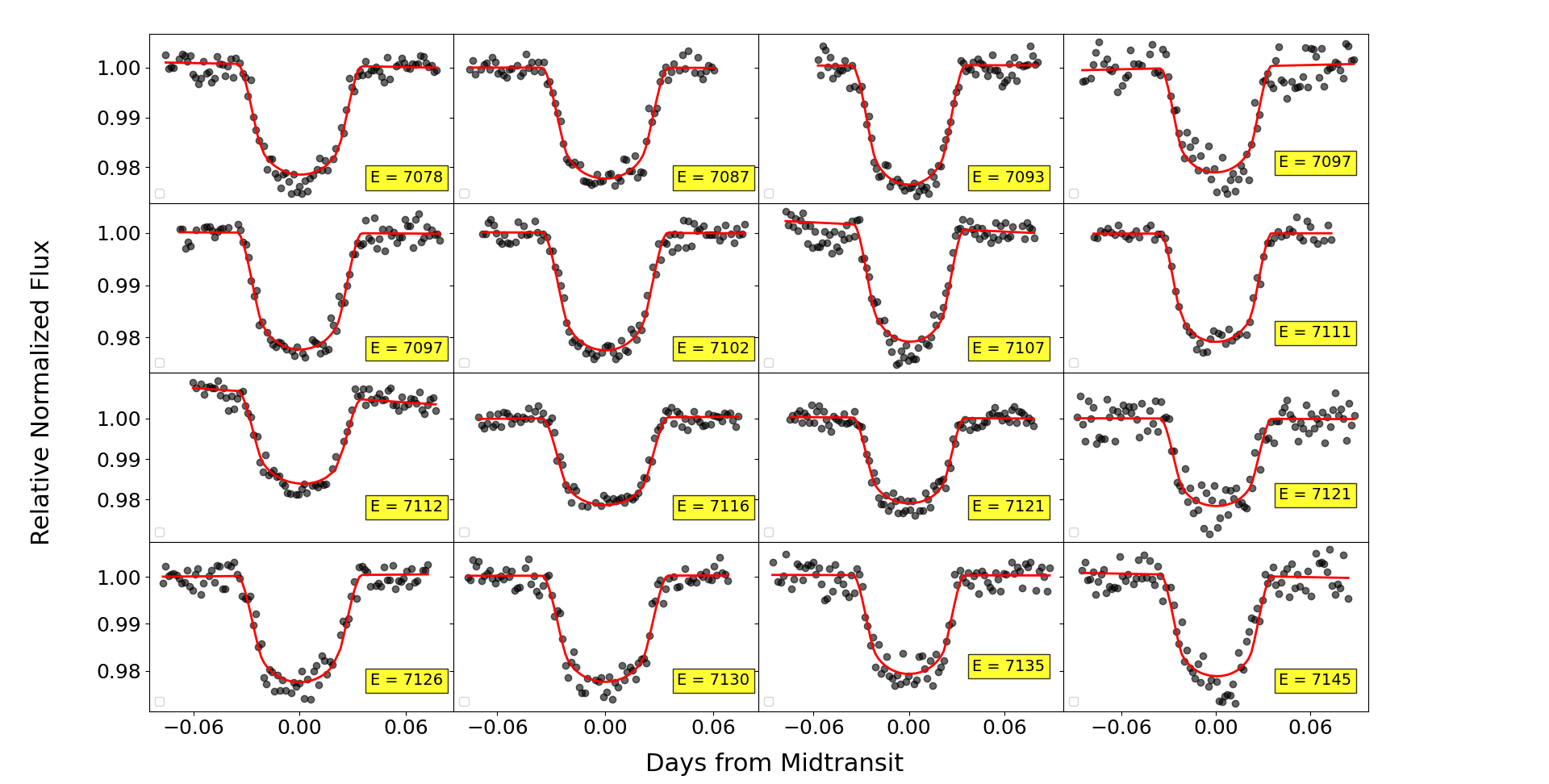}
    \caption{Same as the Figure \ref{fig:ETD1} but for epochs (7078 - 7145)}
    \label{fig:ETD3}
\end{figure}

\begin{figure}
    \centering
    \includegraphics[width=1.1\linewidth]{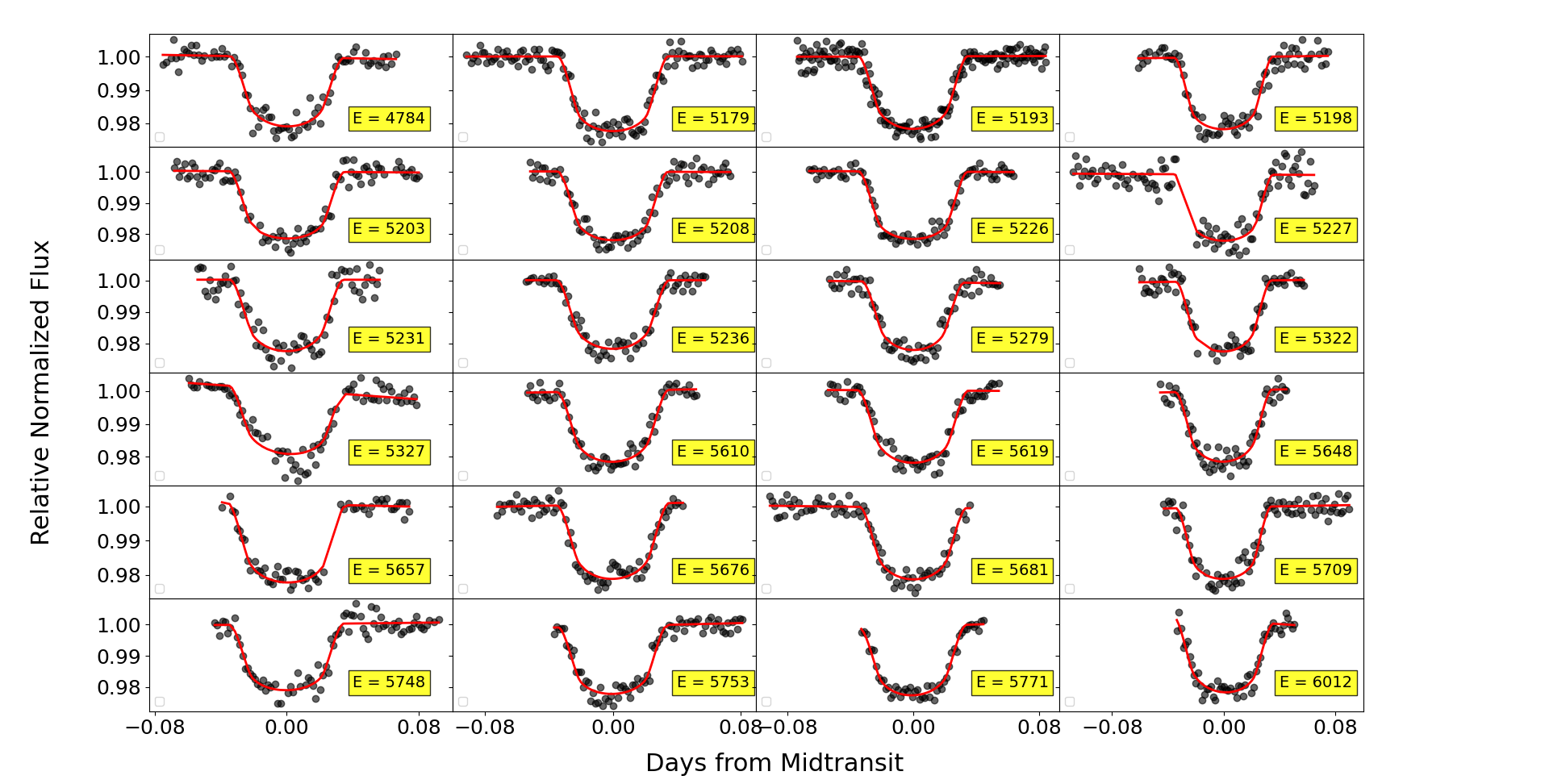}
    \caption{The normalized relative flux of WASP-19b as a function of the time (the offset from mid-transit time and in TDB-based BJD) of individual transits taken from Exoclock  between epochs (4784 - 6012) : here the points, solid red lines, and E represent the same elements as described in Figure \ref{fig:TESS1}.}
    \label{fig:Exoclock}
\end{figure}

\begin{figure}
    \centering
    \includegraphics[width=0.23\textwidth,
    height=3.5cm]{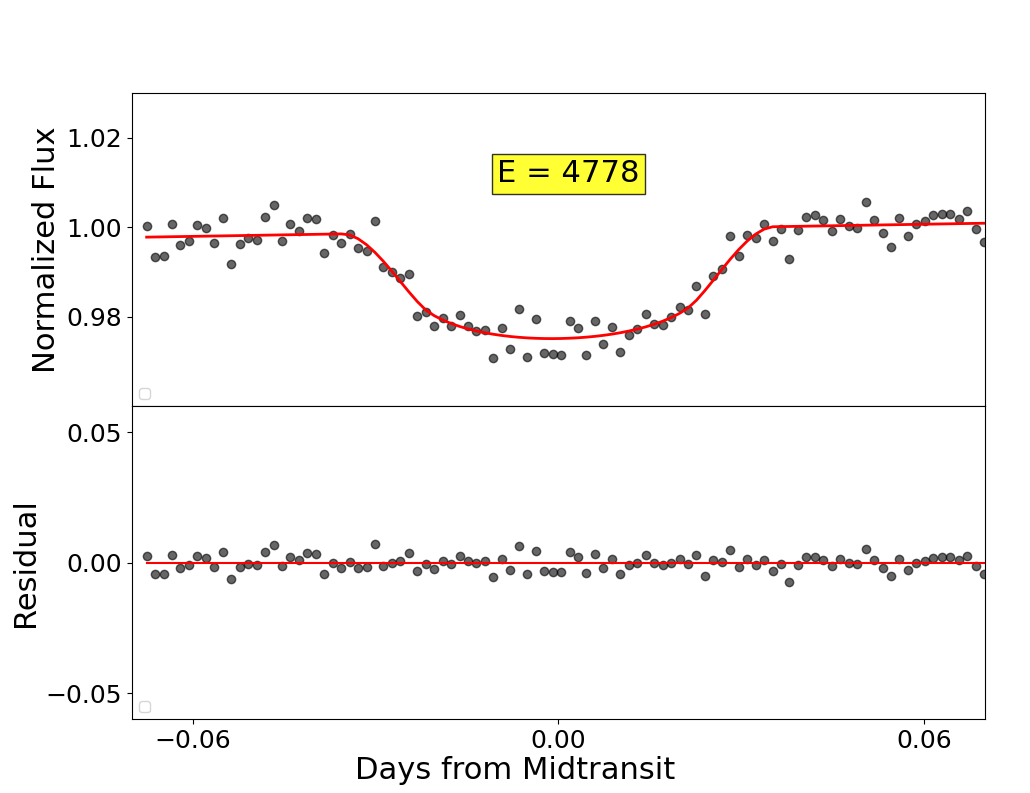}
    \hspace{0.0\textwidth} % Space between figures
    \includegraphics[width=0.23\textwidth,
    height=3.5cm]{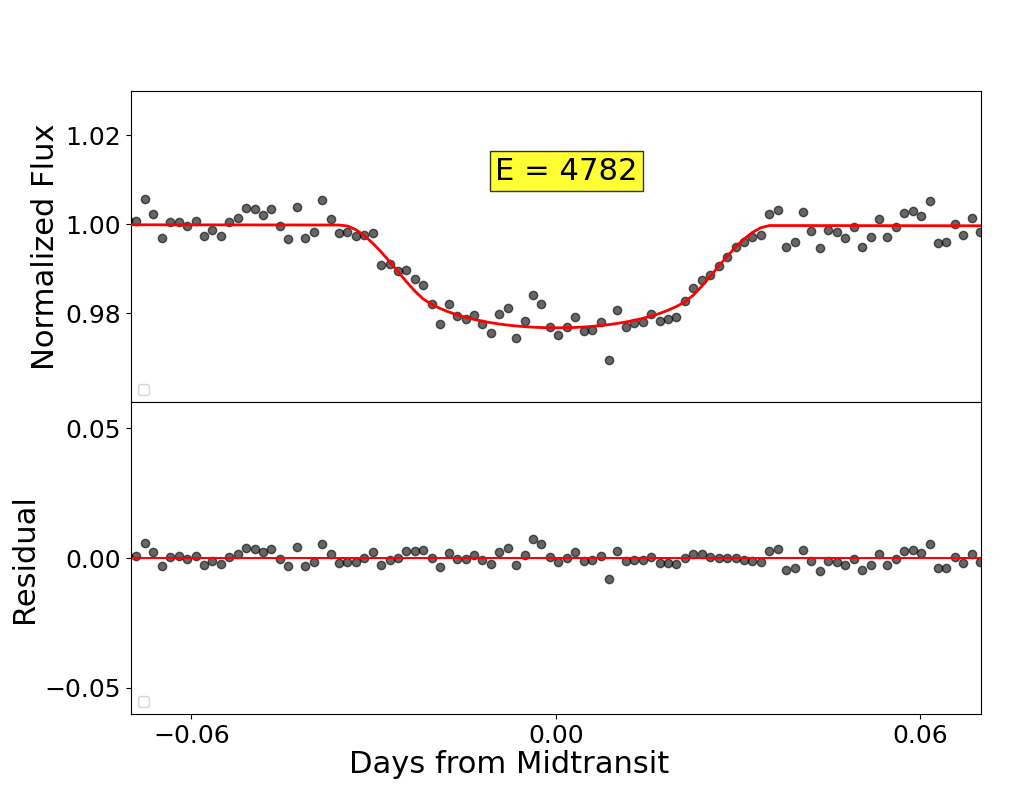}
    \hspace{0.0\textwidth}
\includegraphics[width=0.23\textwidth,
    height=3.5cm]{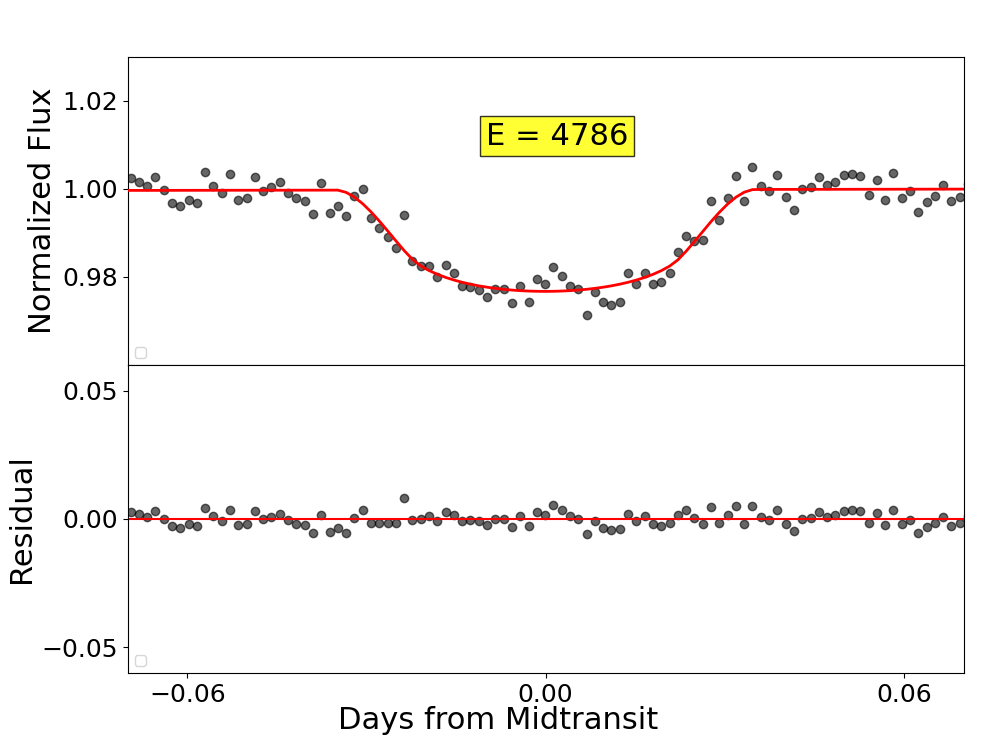}
    \hspace{0.0\textwidth}
    \includegraphics[width=0.23\textwidth,
    height=3.5cm]{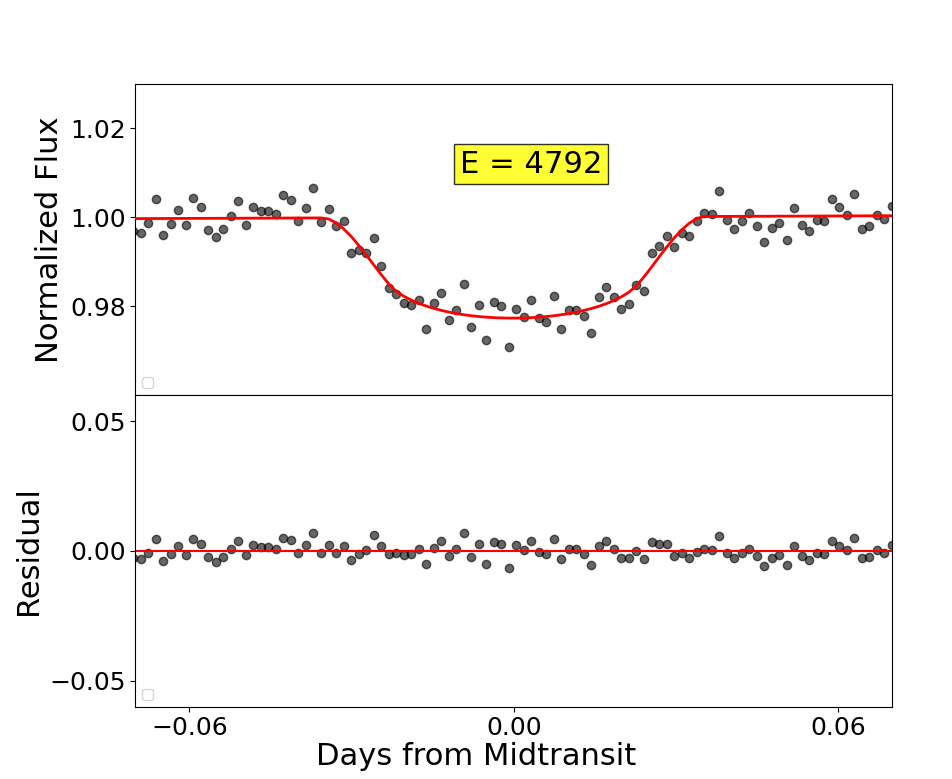}    

 \vspace{0.5cm} % Space between rows

    \includegraphics[width=0.23\textwidth, height=3.5cm]{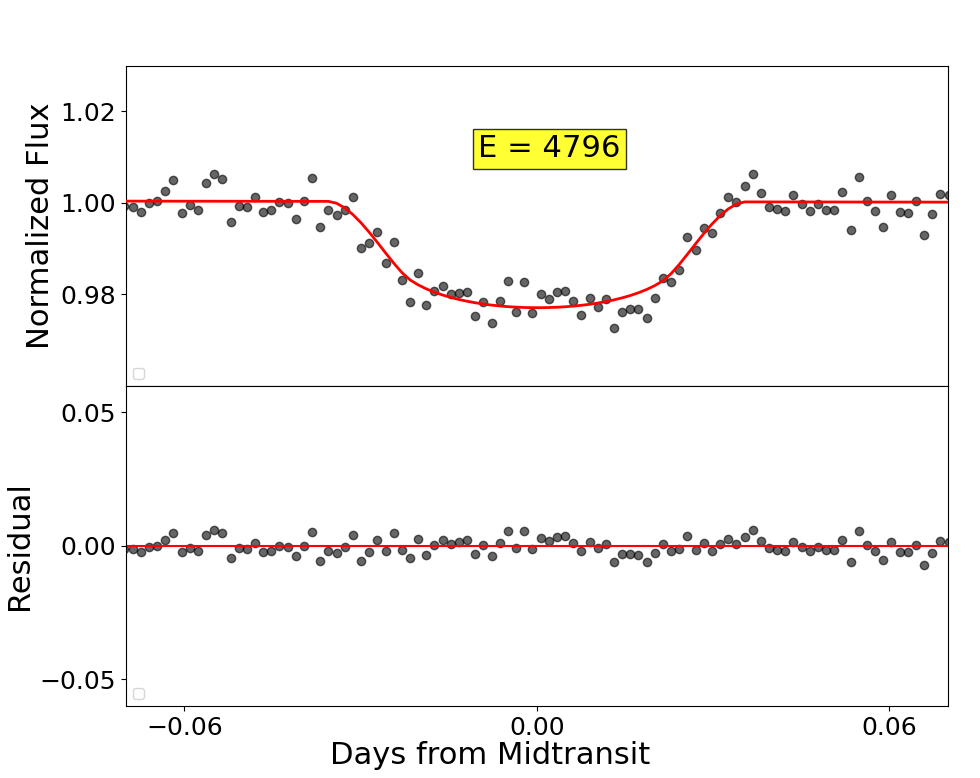}
    \hspace{0.0\textwidth}
    \includegraphics[width=0.23\textwidth, height=3.5cm]{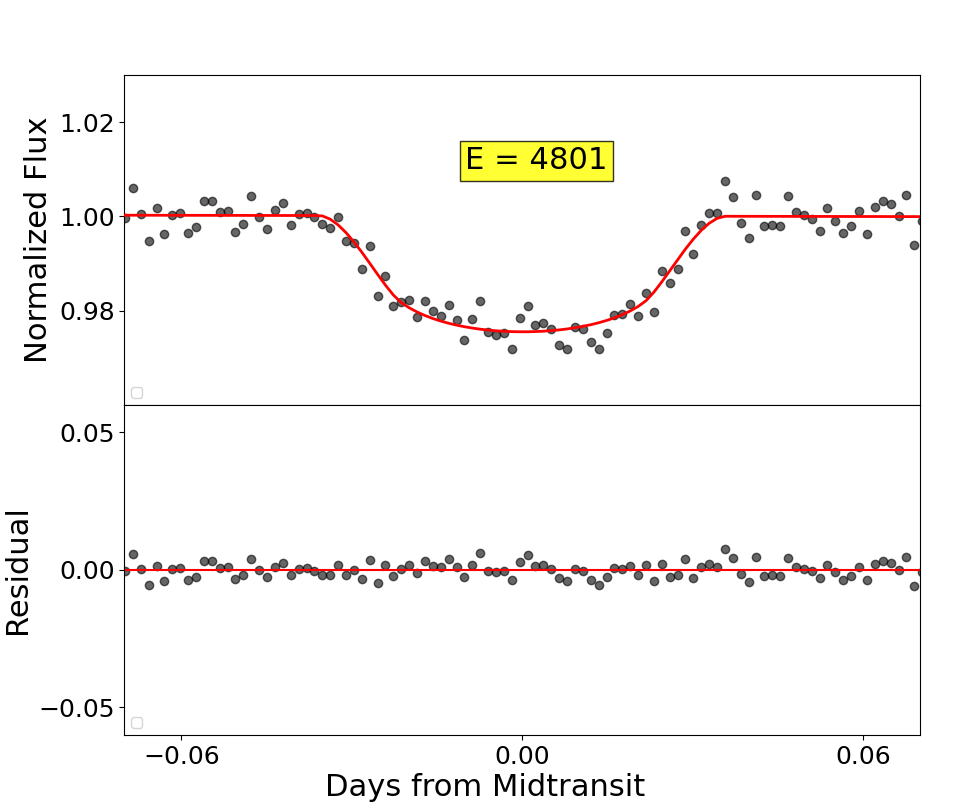}
    \hspace{0.0\textwidth}
    \includegraphics[width=0.23\textwidth, height=3.5cm]{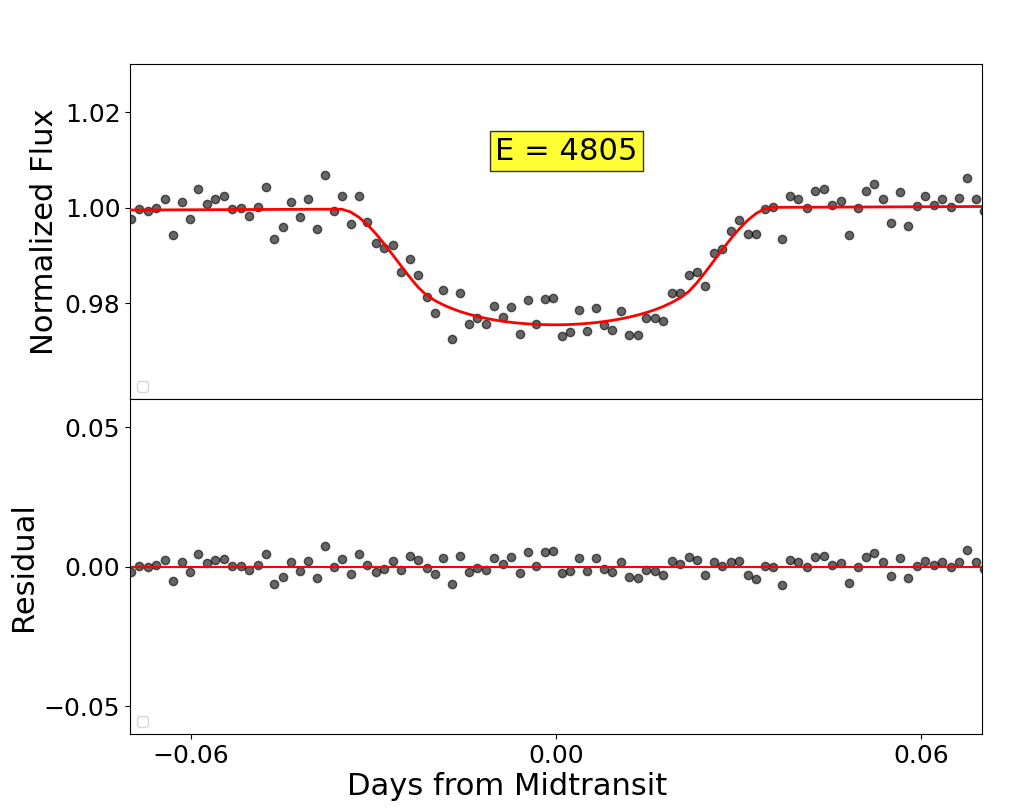}
    \hspace{0.0\textwidth}
    \includegraphics[width=0.23\textwidth, height=3.5cm]{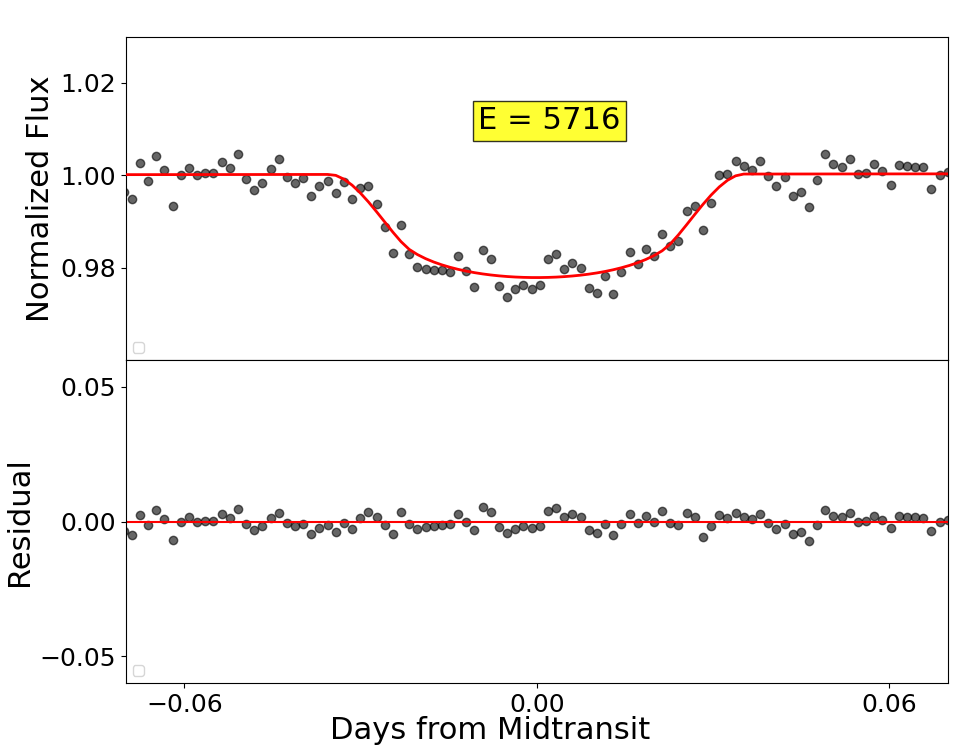}

     \vspace{0.5cm}
    
    \includegraphics[width=0.23\textwidth, height=3.5cm]{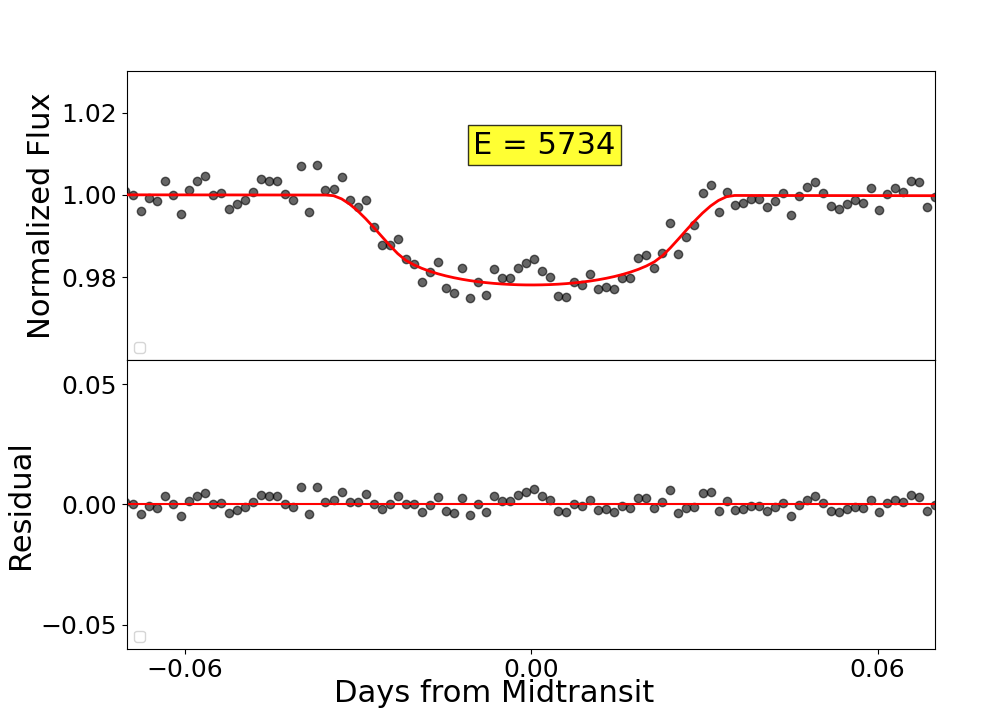}
     \hspace{0.0\textwidth}
    \includegraphics[width=0.23\textwidth, height=3.5cm]{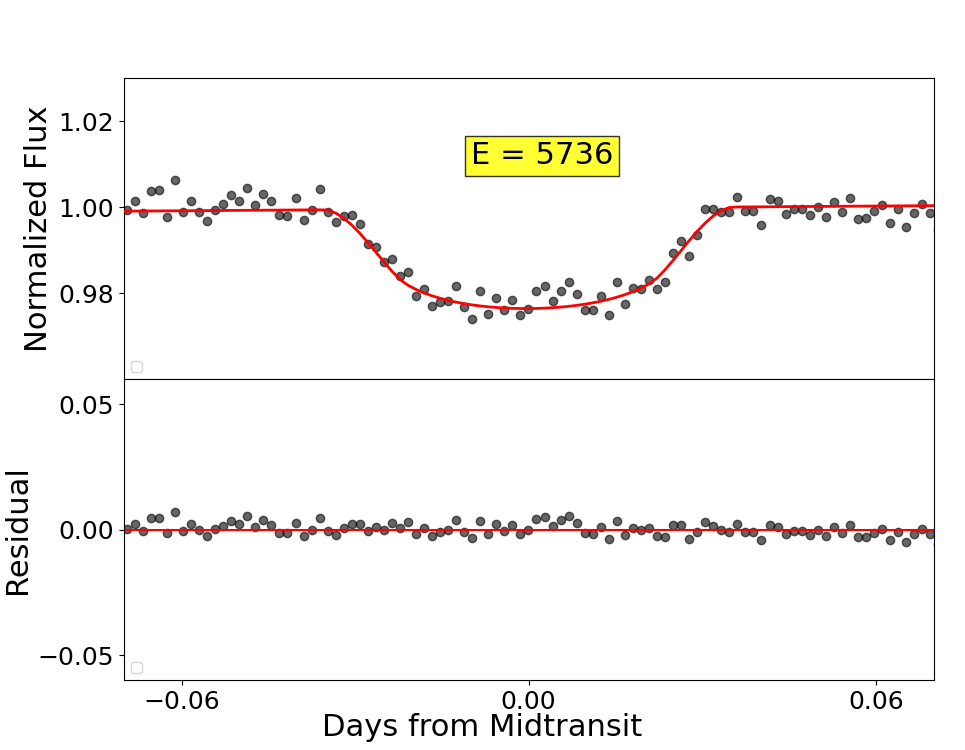}
    \hspace{0.0\textwidth}
    \includegraphics[width=0.23\textwidth, height=3.5cm]{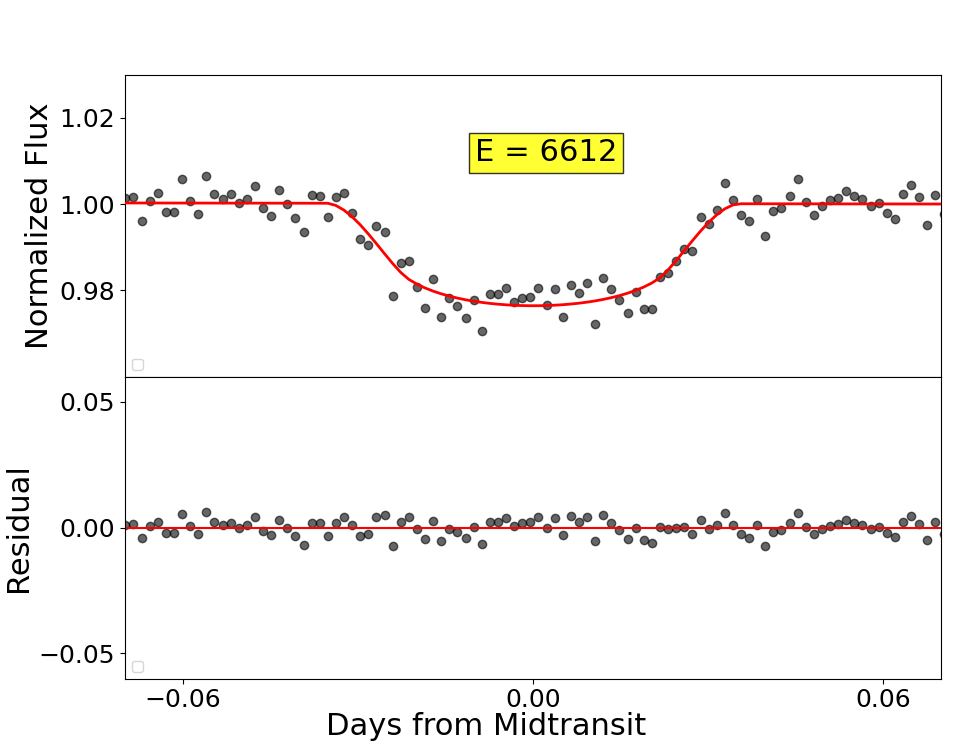}
     \hspace{0.0\textwidth}
    \includegraphics[width=0.23\textwidth, height=3.5cm]{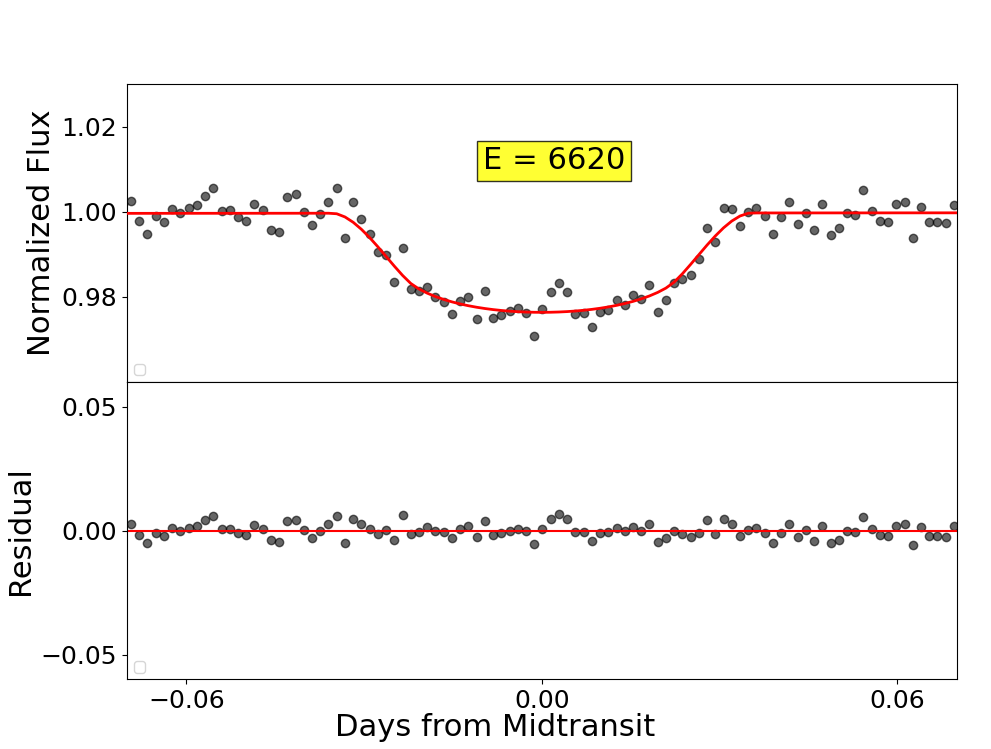}
    
    \vspace{0.5cm}
    
    \includegraphics[width=0.23\textwidth, height=3.5cm]{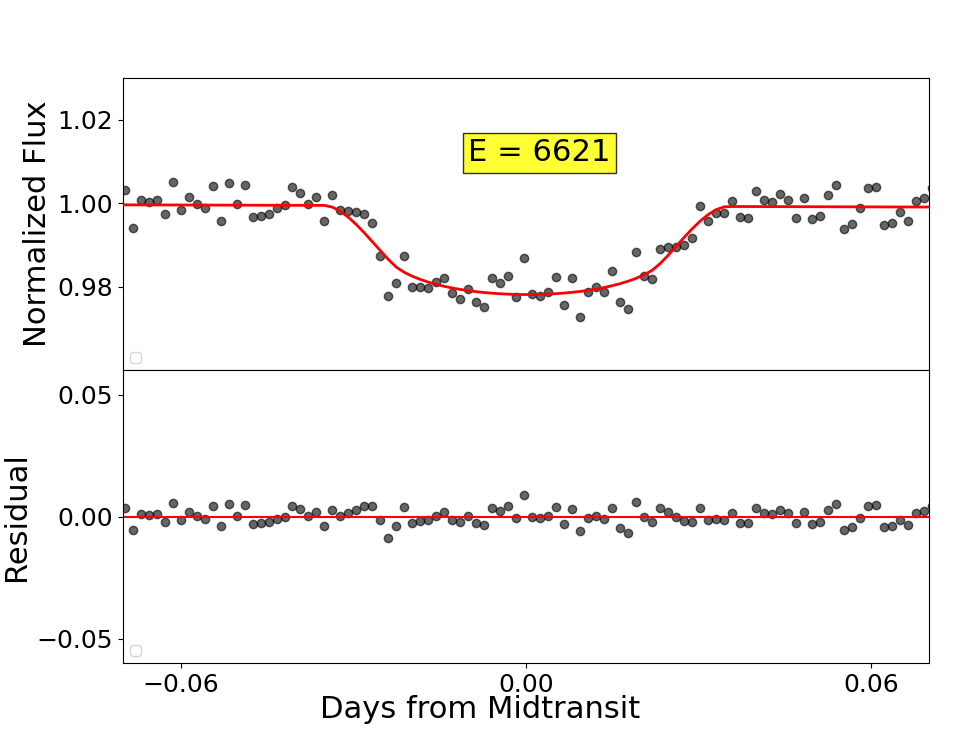}
    \hspace{0.0\textwidth} % Space between figures
    \includegraphics[width=0.23\textwidth, height=3.5cm]{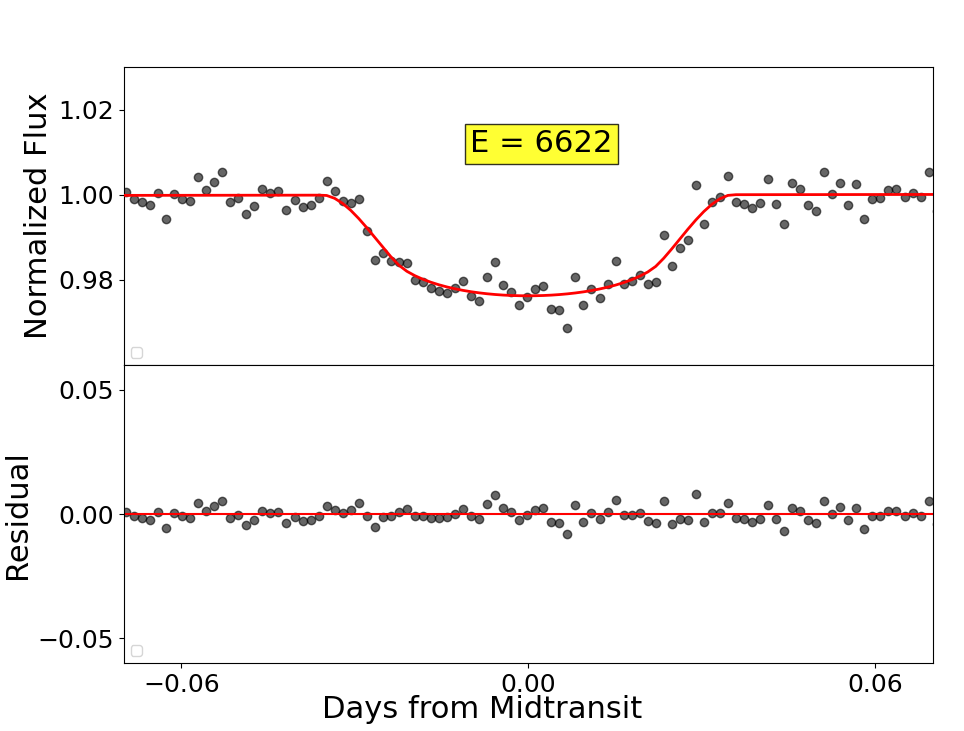}
    \hspace{0.0\textwidth}
    \includegraphics[width=0.23\textwidth, height=3.5cm]{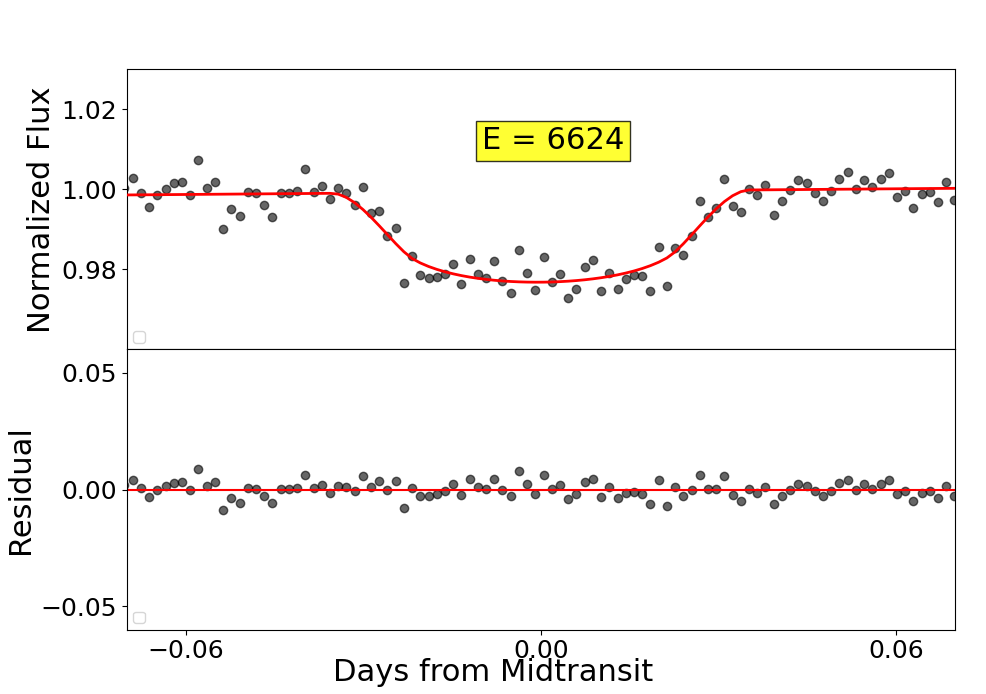}
    \hspace{0.0\textwidth}
    \includegraphics[width=0.23\textwidth, height=3.5cm]{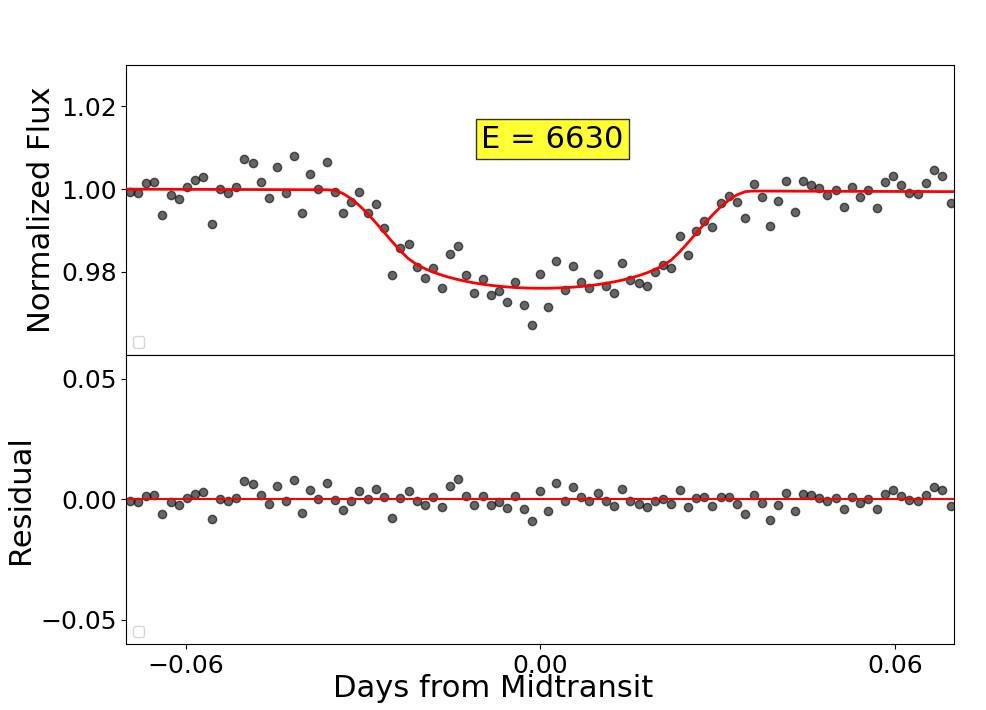}

    \vspace{0.5cm}
    
    \includegraphics[width=0.23\textwidth, height=3.5cm]{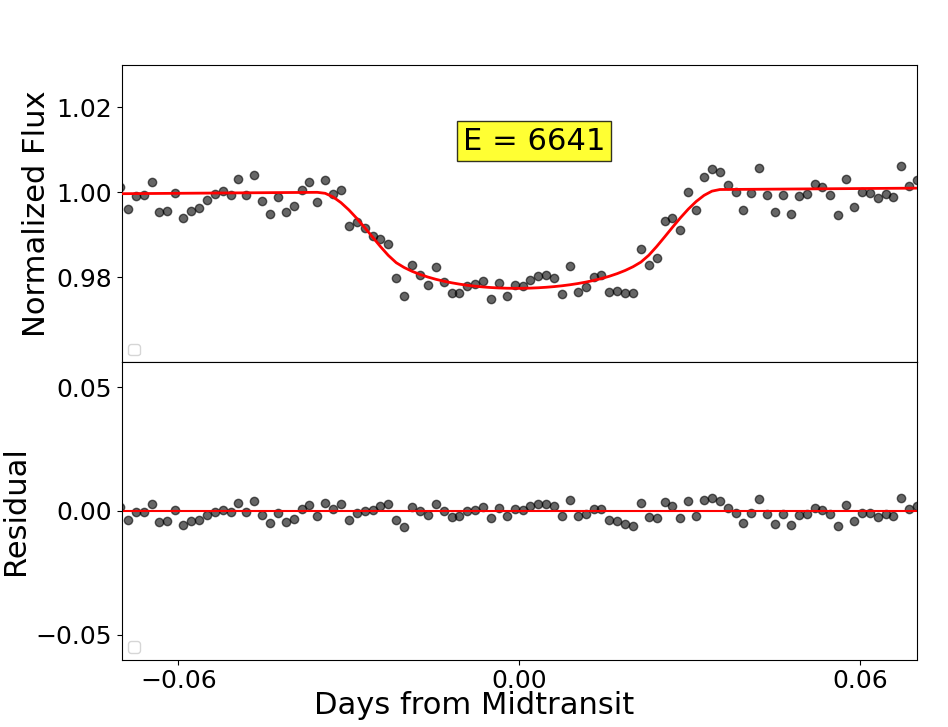}
    \hspace{0.0\textwidth} % Space between figures
    \includegraphics[width=0.23\textwidth, height=3.5cm]{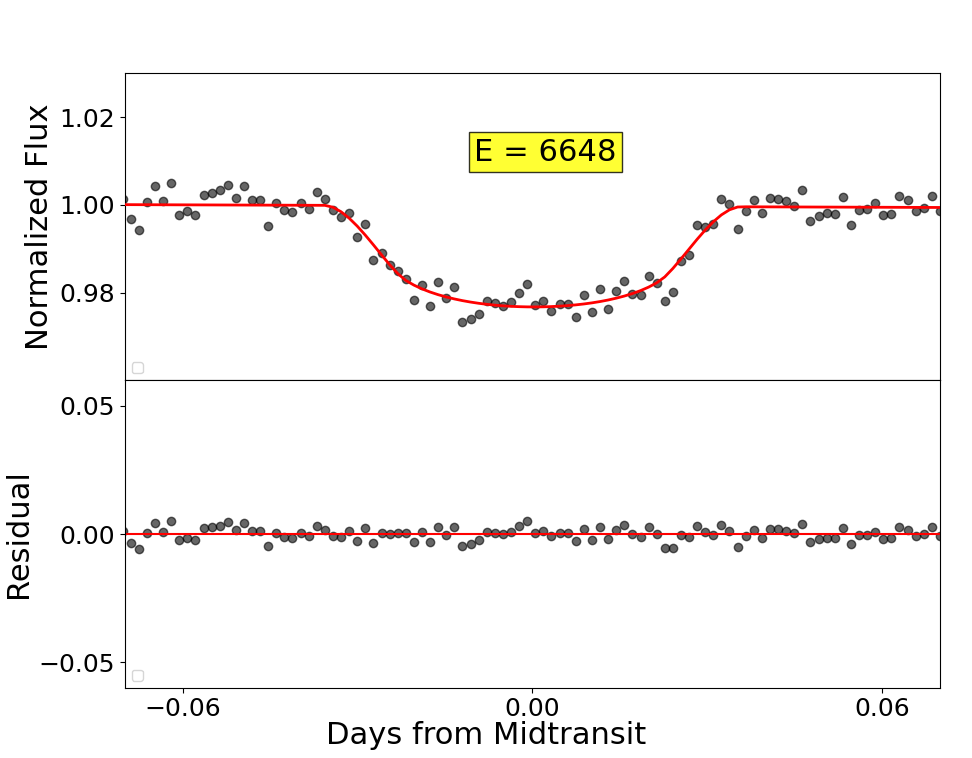}
    \hspace{0.0\textwidth}
    \includegraphics[width=0.23\textwidth, height=3.5cm]{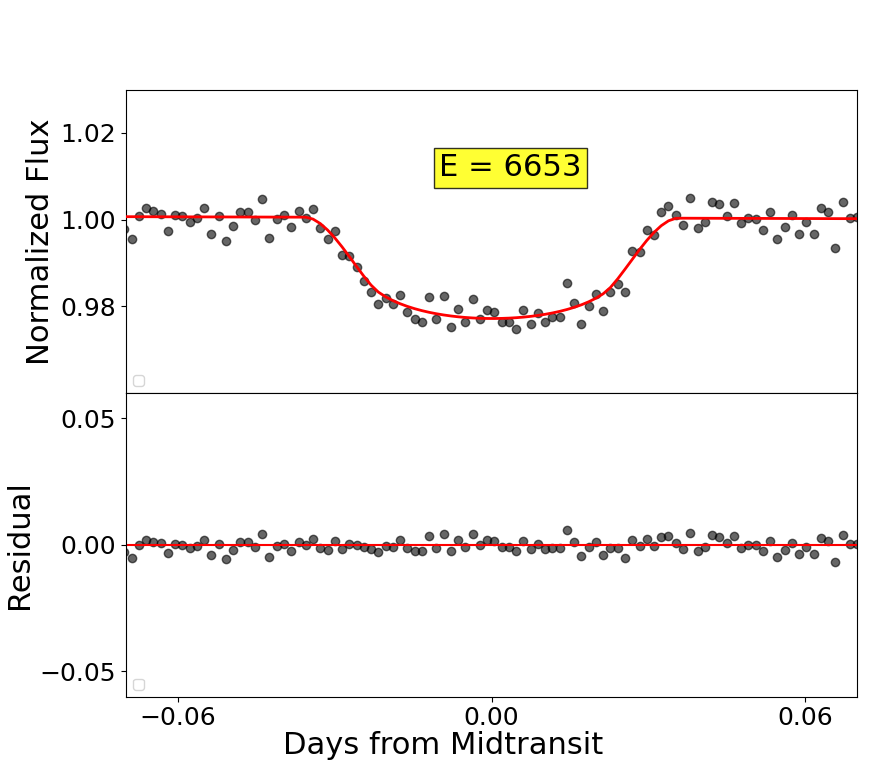}
    \hspace{0.0\textwidth}
    \includegraphics[width=0.23\textwidth, height=3.5cm]{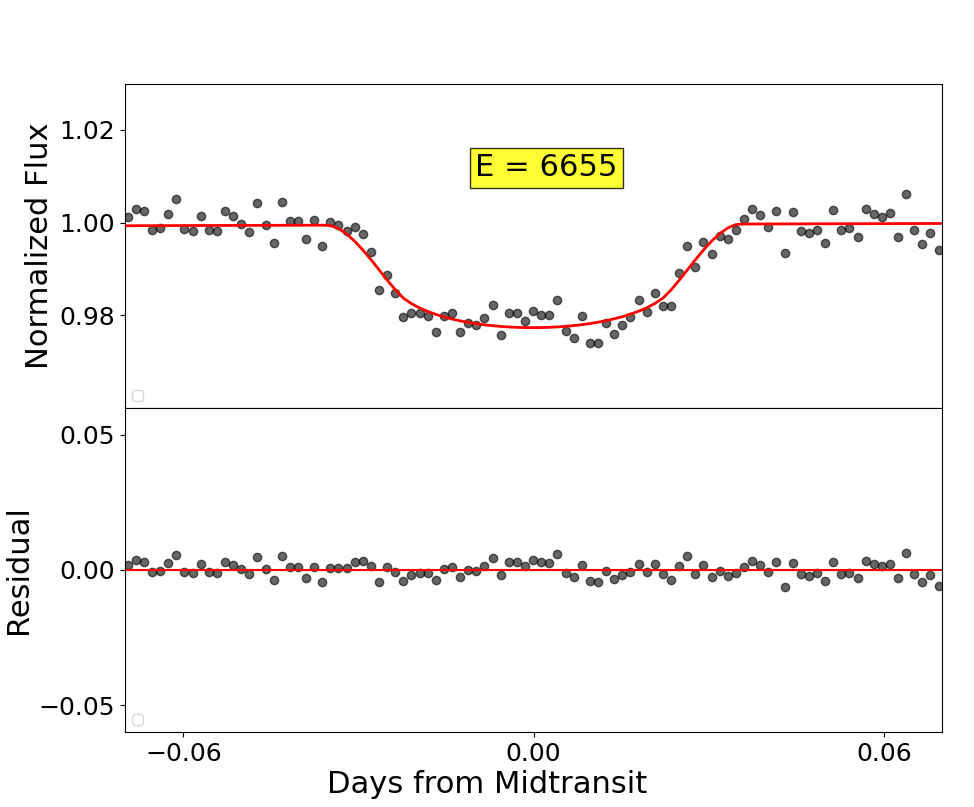}

    \vspace{0.5cm}
    
    \includegraphics[width=0.23\textwidth, height=3.5cm]{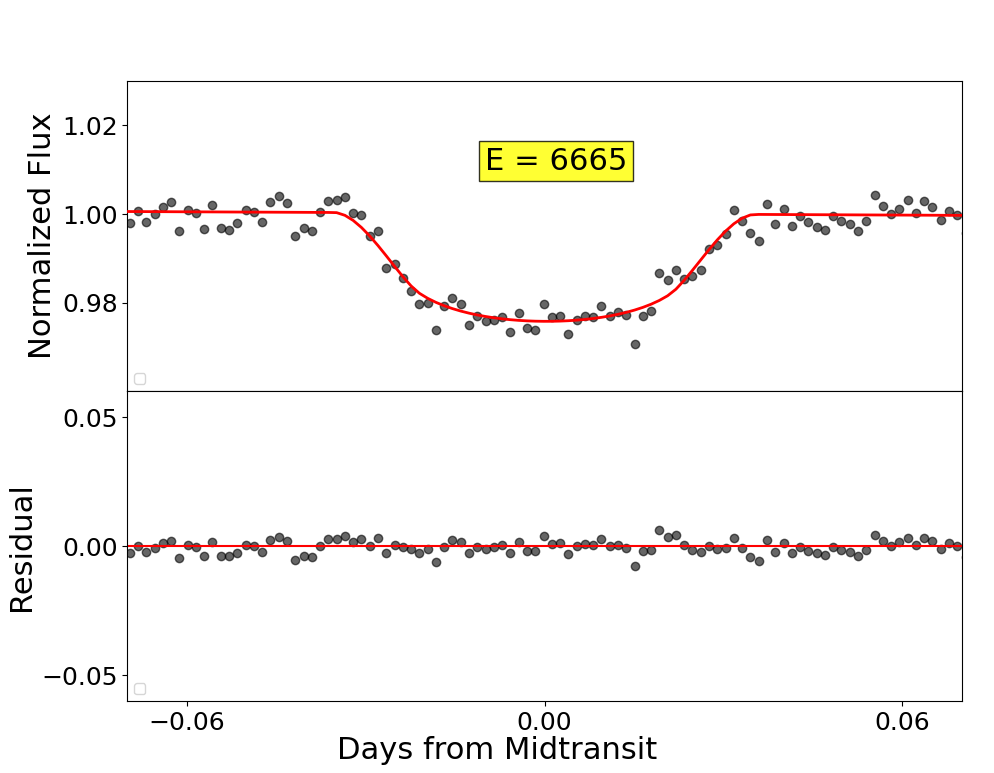}
    \hspace{0.0\textwidth} % Space between figures
    \includegraphics[width=0.23\textwidth, height=3.5cm]{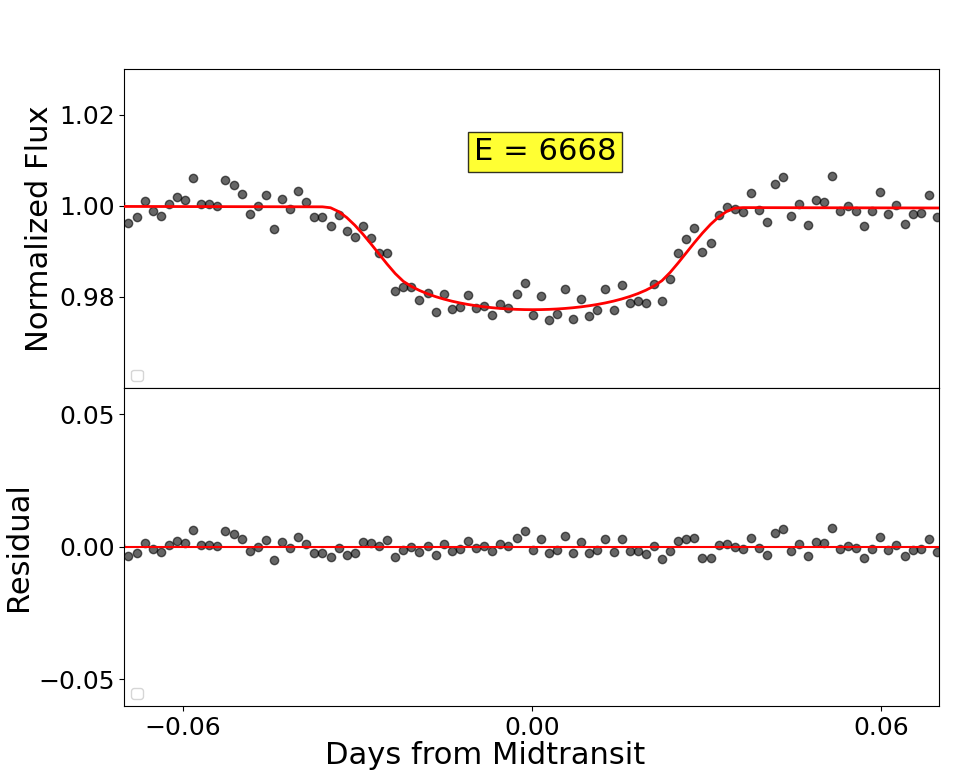}
    \hspace{0.0\textwidth}
    \includegraphics[width=0.23\textwidth, height=3.5cm]{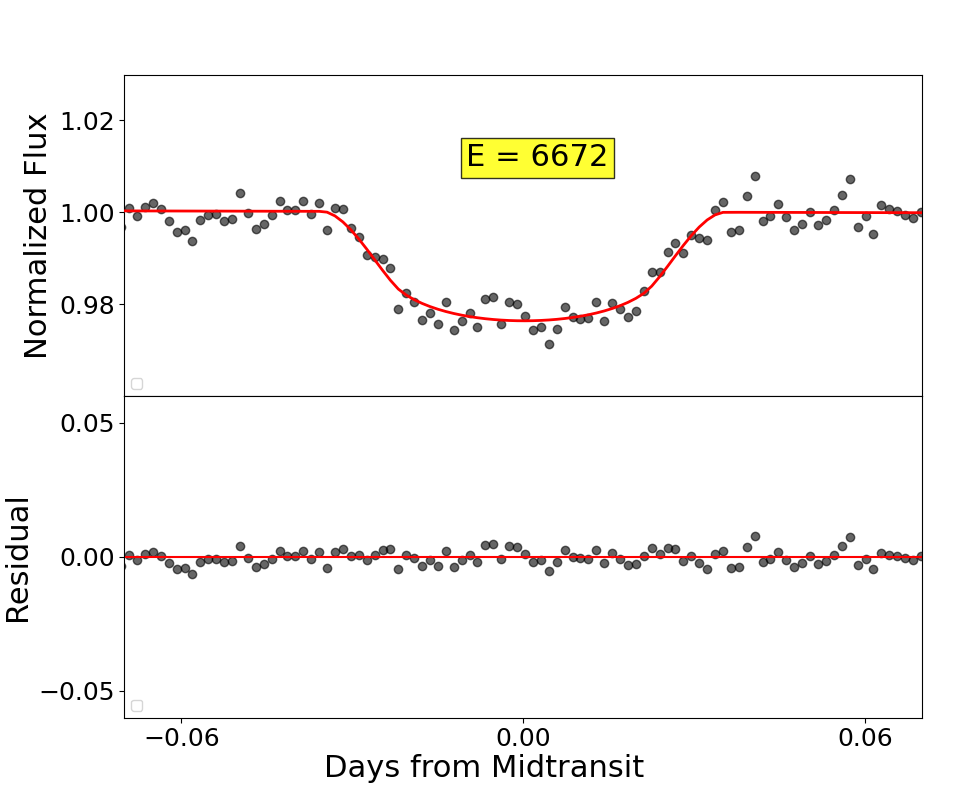}
    \hspace{0.0\textwidth}
    \includegraphics[width=0.23\textwidth, height=3.5cm]{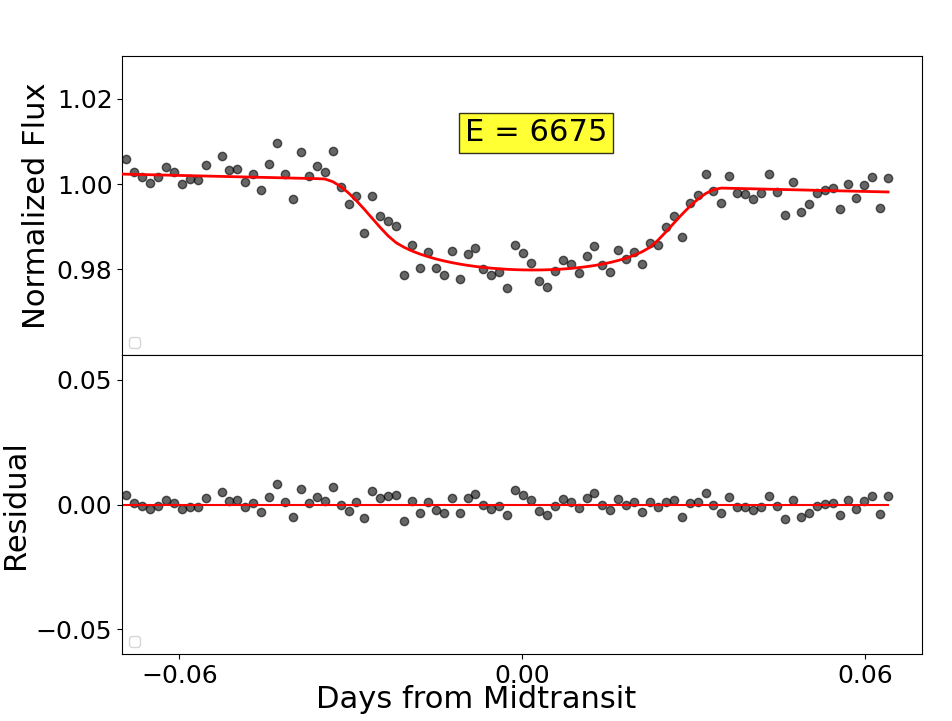}
    
    \caption{Stellar spot affected transit light curves of WASP-19b from different sectors of TESS. The solid lines represent the best-fit models, and the residuals are shown below each light curve.}
    \label{fig:TESS_spot_affected_LCs}
\end{figure}

\begin{figure}
    \centering
    \includegraphics[width=0.23\textwidth,
    height=3.5cm]{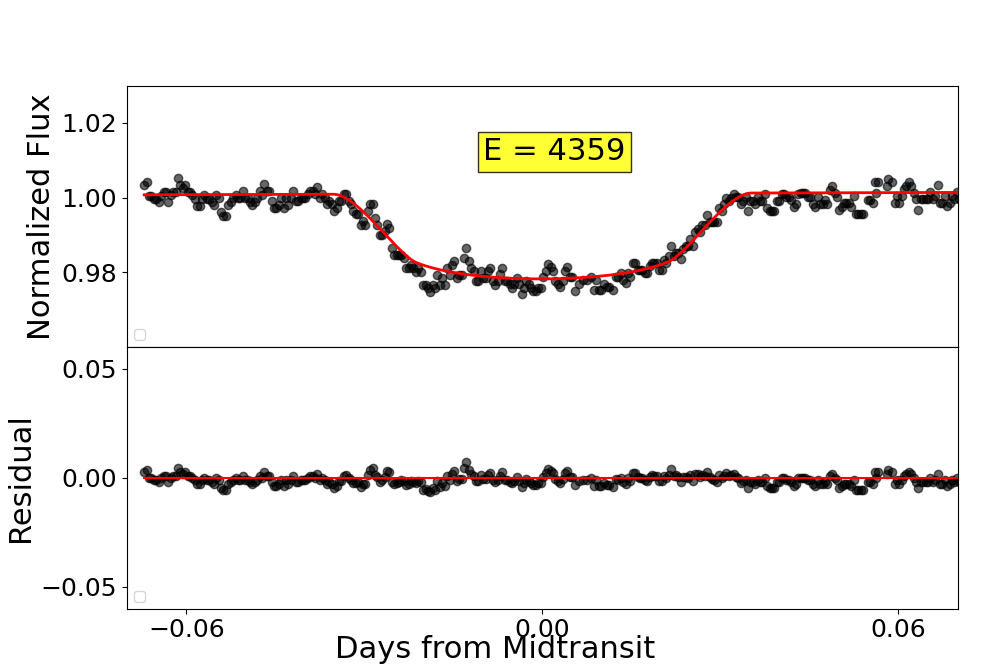}
    \hspace{0.0\textwidth} % Space between figures
    \includegraphics[width=0.23\textwidth,
    height=3.5cm]{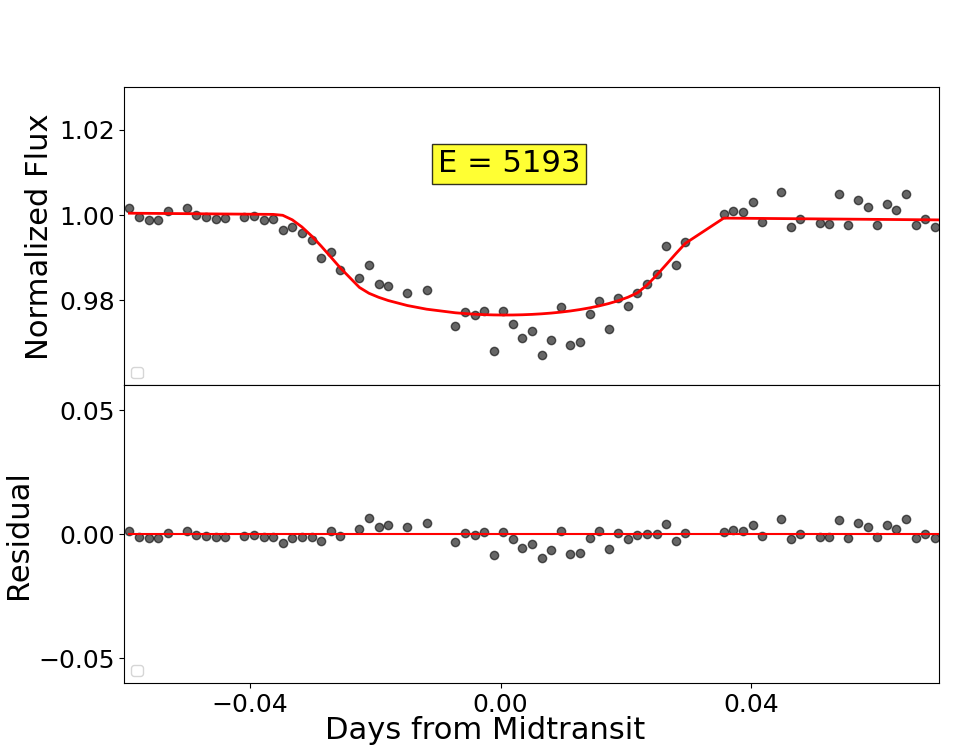}
    \hspace{0.0\textwidth}
\includegraphics[width=0.23\textwidth,
    height=3.5cm]{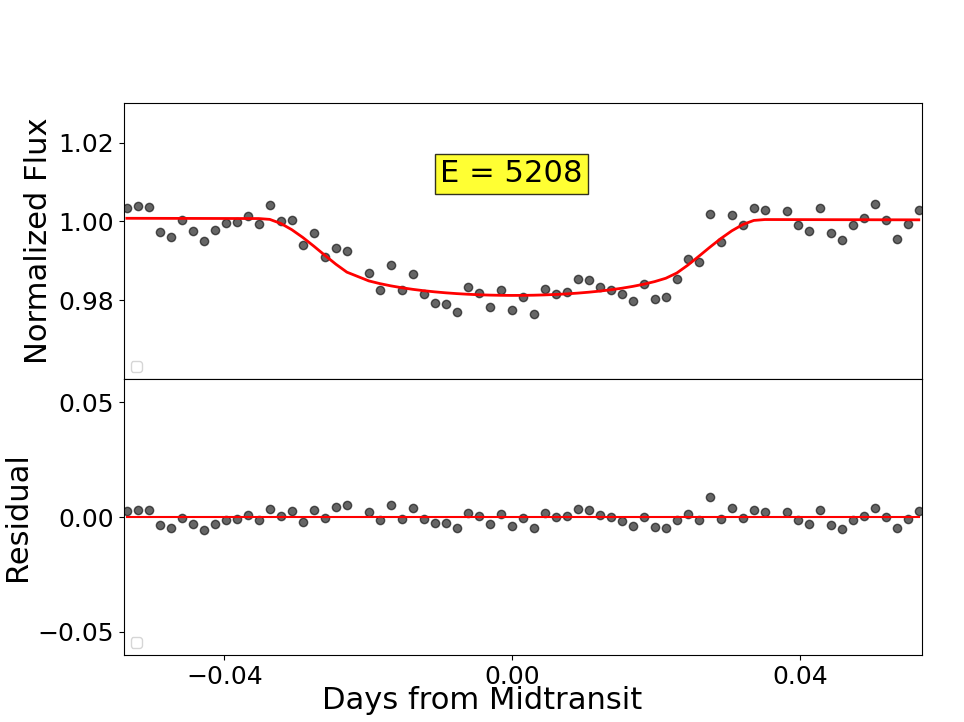}
    \hspace{0.0\textwidth}
    \includegraphics[width=0.23\textwidth,
    height=3.5cm]{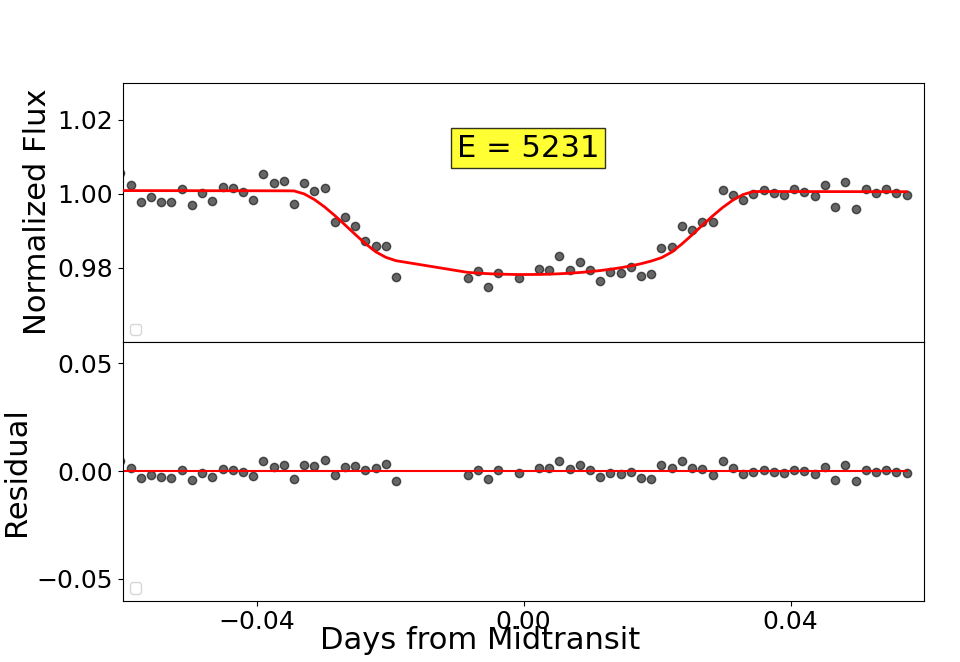}    

 \vspace{0.5cm} % Space between rows

    \includegraphics[width=0.23\textwidth, height=3.5cm]{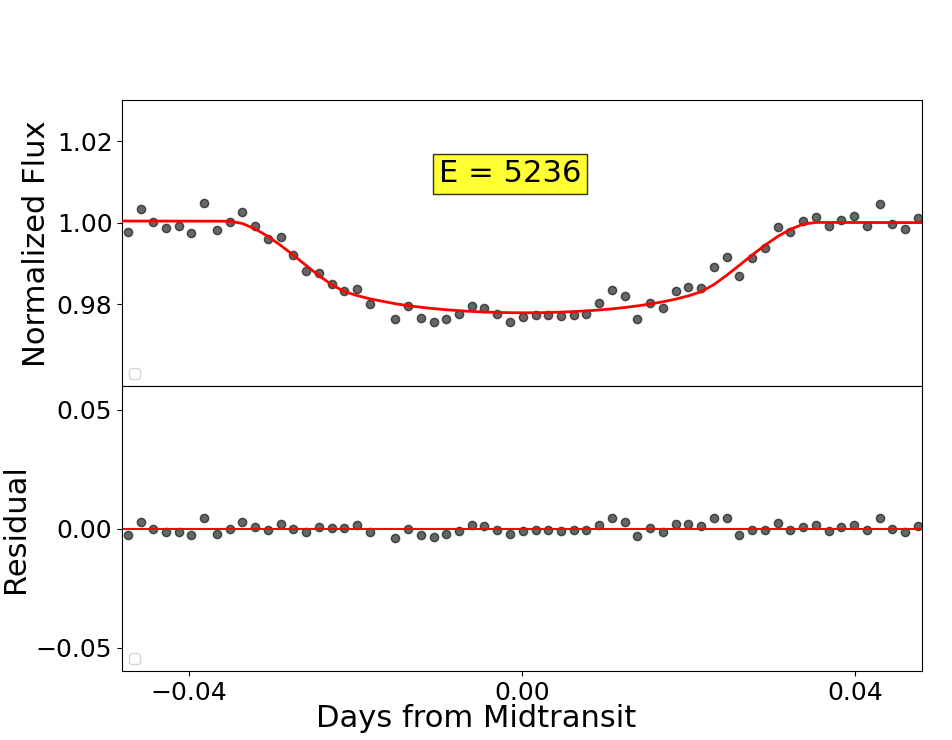}
    \hspace{0.0\textwidth}
    \includegraphics[width=0.23\textwidth, height=3.5cm]{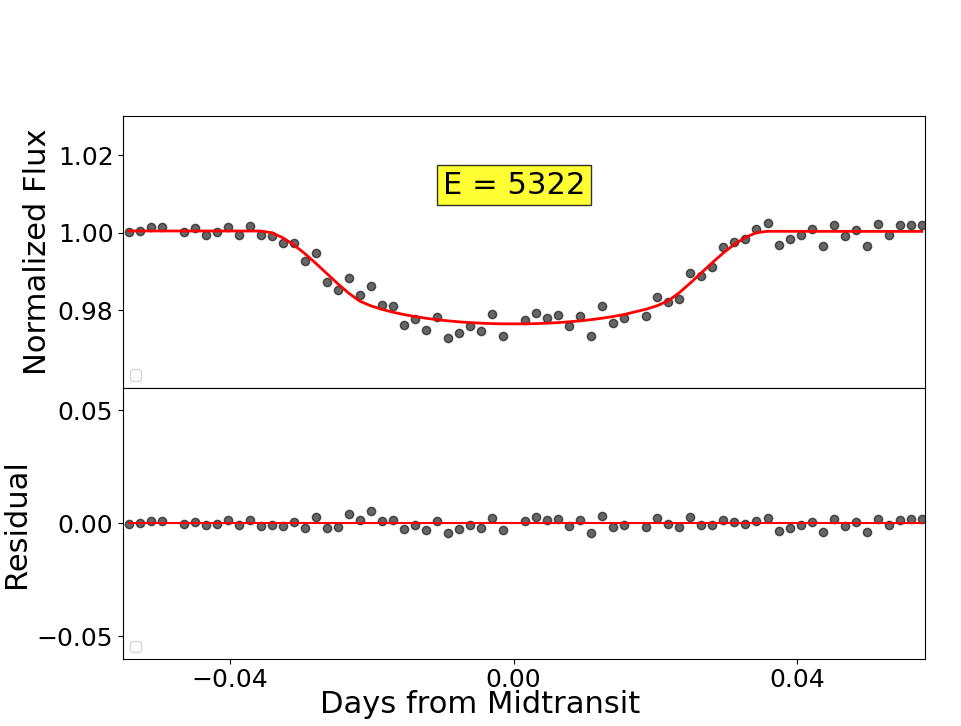}
    \hspace{0.0\textwidth}
    \includegraphics[width=0.23\textwidth, height=3.5cm]{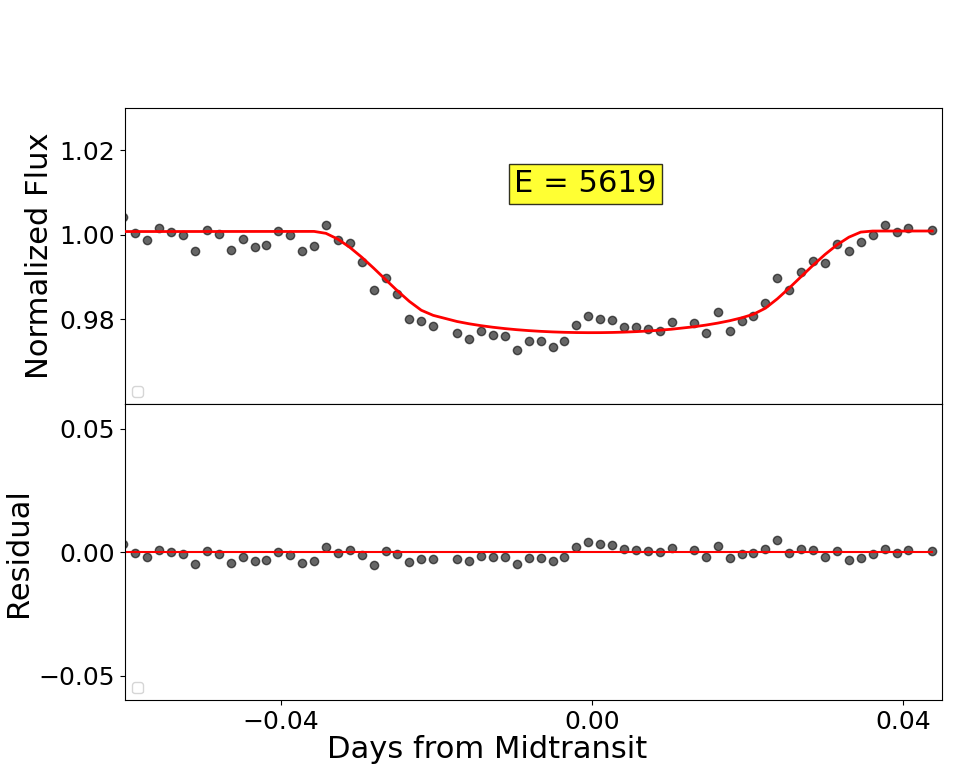}
    \hspace{0.0\textwidth}
    \includegraphics[width=0.23\textwidth, height=3.5cm]{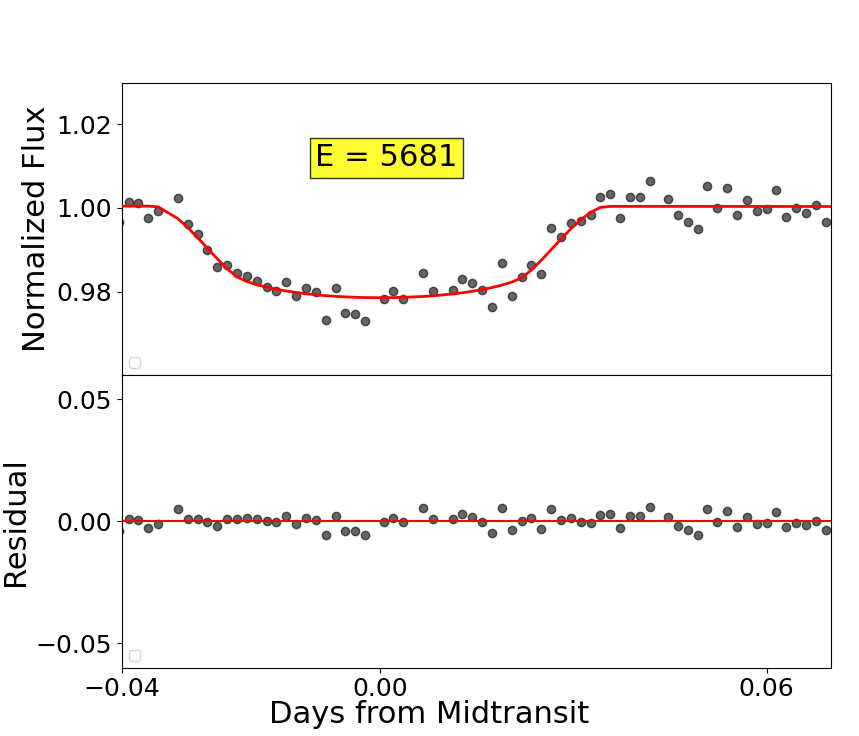}

     \vspace{0.5cm}
    
    \includegraphics[width=0.23\textwidth, height=3.5cm]{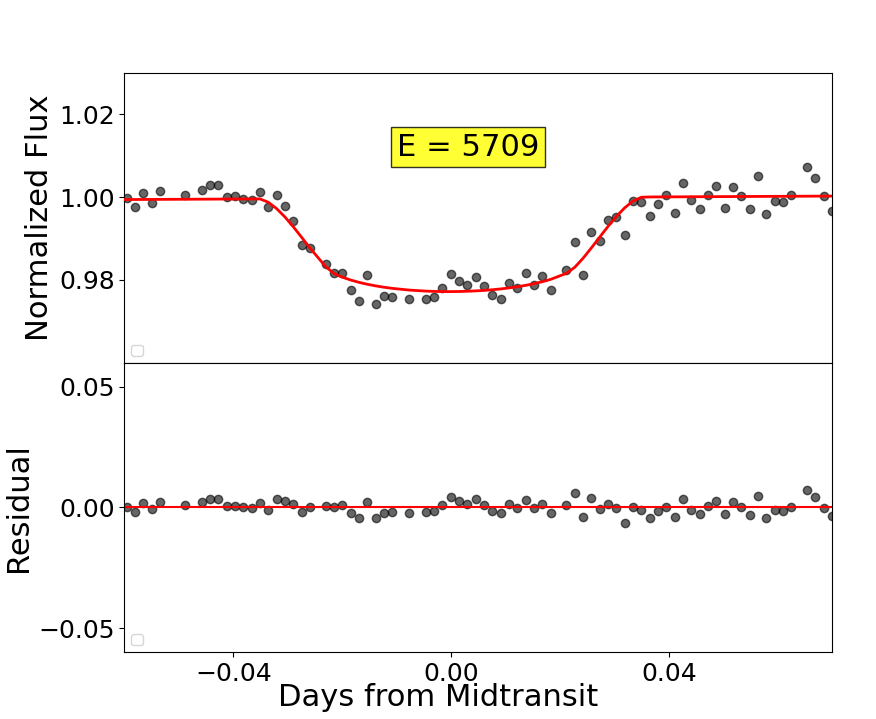}
     \hspace{0.0\textwidth}
    \includegraphics[width=0.23\textwidth, height=3.5cm]{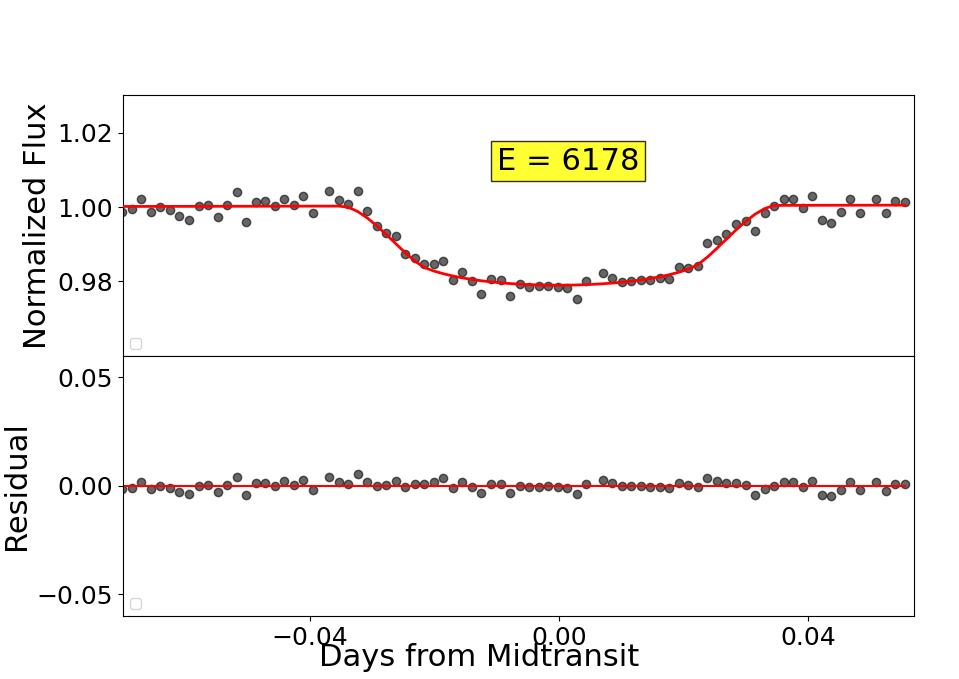}
    \hspace{0.0\textwidth}
    \includegraphics[width=0.23\textwidth, height=3.5cm]{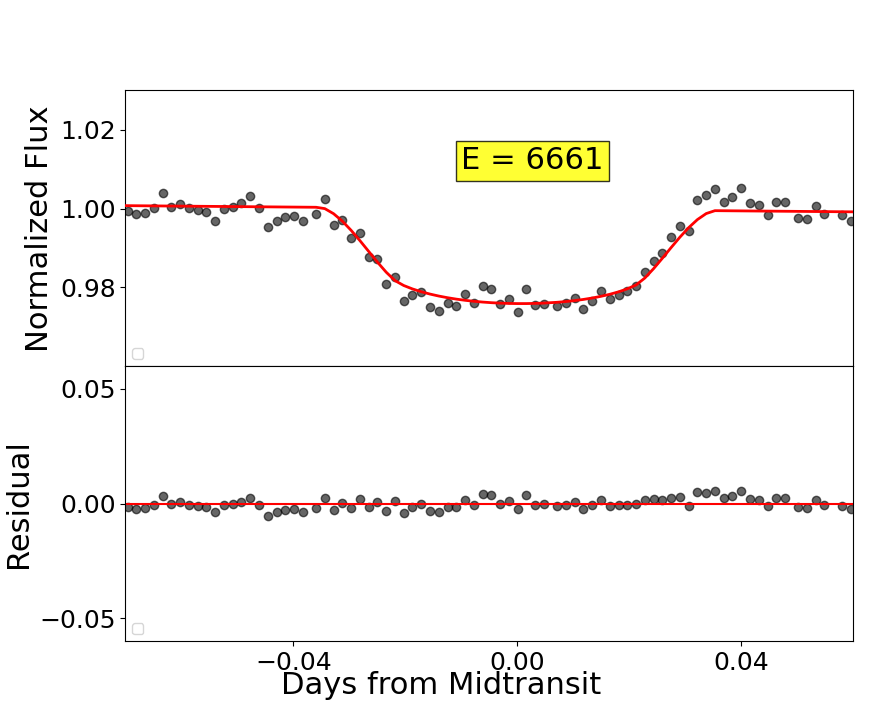}
     \hspace{0.0\textwidth}
    \includegraphics[width=0.23\textwidth, height=3.5cm]{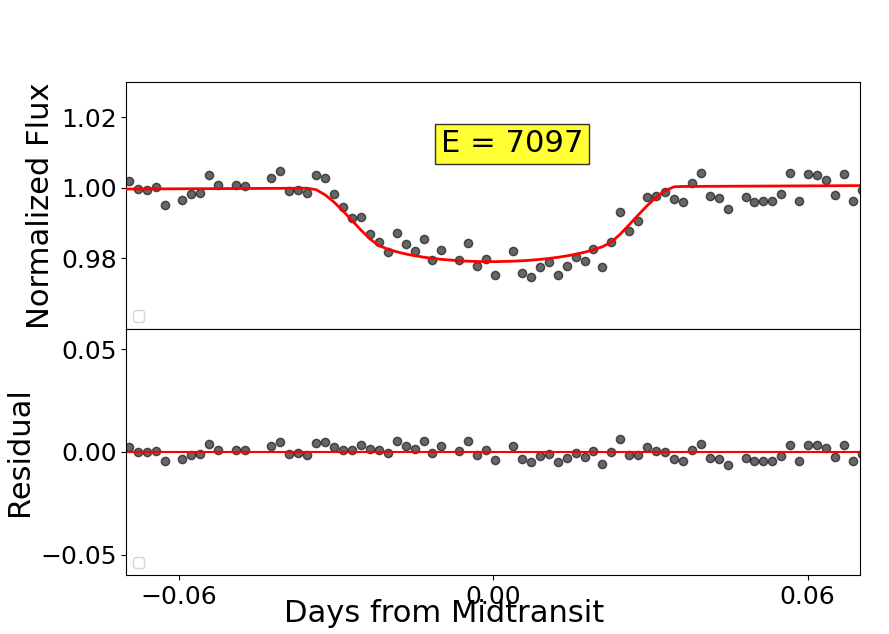}
    
    \vspace{0.5cm}
    
    \includegraphics[width=0.23\textwidth, height=3.5cm]{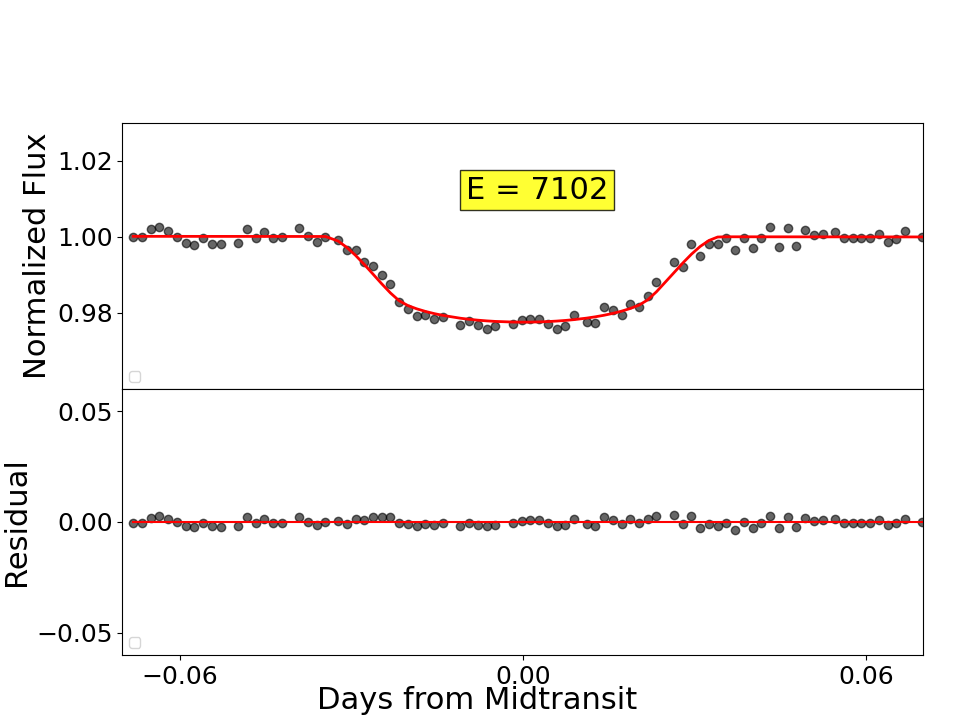}
    \hspace{0.0\textwidth} % Space between figures
    \includegraphics[width=0.23\textwidth, height=3.5cm]{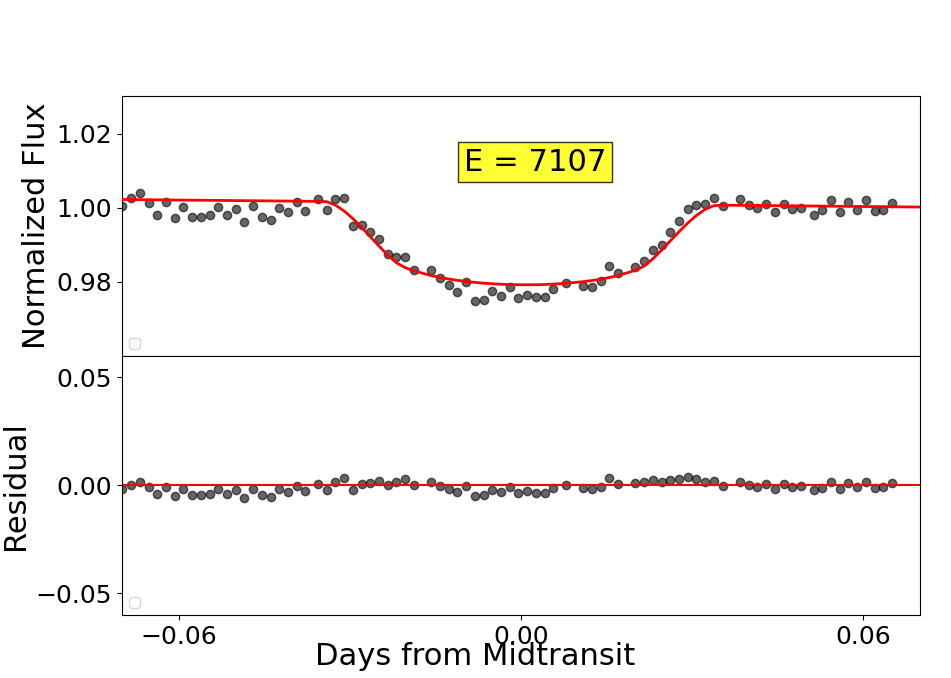}
    \hspace{0.0\textwidth}
    \includegraphics[width=0.23\textwidth, height=3.5cm]{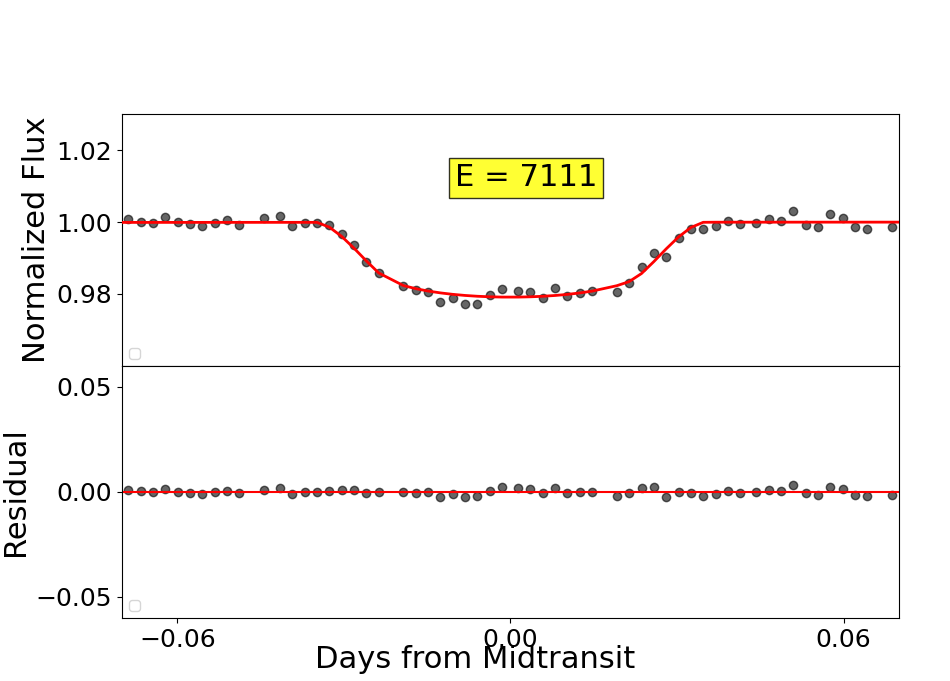}
    \hspace{0.0\textwidth}
    \includegraphics[width=0.23\textwidth, height=3.5cm]{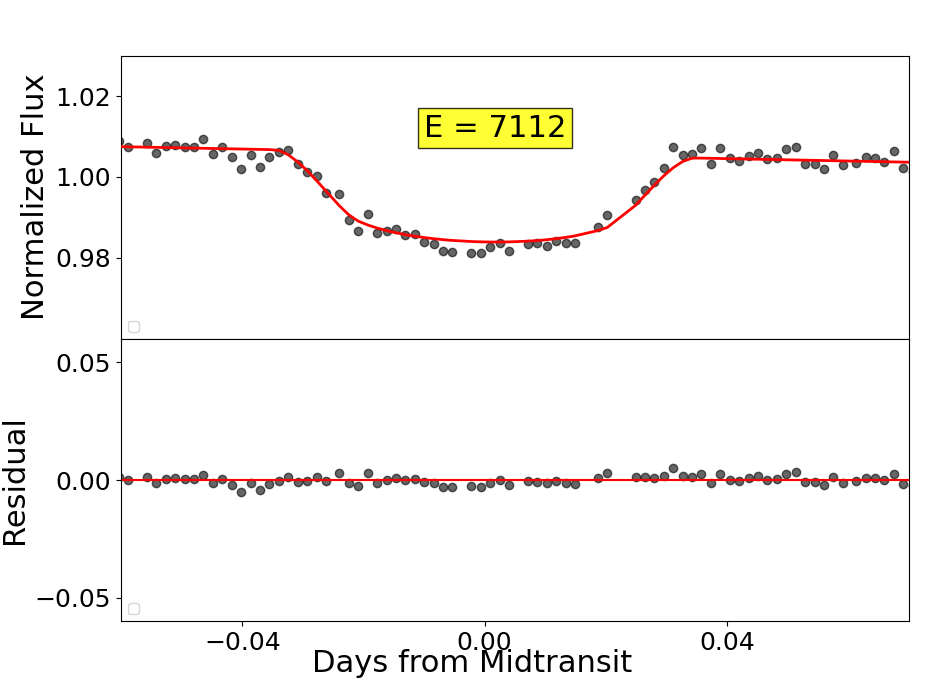}

    \caption{Stellar spot affected transit light curves of WASP-19b from ETD. The solid lines represent the best-fit models, and the residuals are shown below each light curve.}
    \label{fig:ETD_spot_affected_LCs}
\end{figure}

We have represented the \texttt{TAP} model fits for all  116 TESS light curves of WASP-19b in Figures \ref{fig:TESS1}--\ref{fig:TESS5}. Similarly, 65 light curves from ETD and 24 light curves from Exoclock have been represented in Figure \ref{fig:ETD1}--\ref{fig:ETD3} and Figure \ref{fig:Exoclock}. We have also represented all 24 stellar spot affected light curves from TESS and with their residuals in Figure \ref{fig:TESS_spot_affected_LCs} and 16 stellar spot affected light curves from ETD in Figure \ref{fig:ETD_spot_affected_LCs}.

%******************* Individual TESS Sector 2%*******************

\bibliography{wasp19b}{}
\bibliographystyle{aasjournal}

%% This command is needed to show the entire author+affiliation list when
%% the collaboration and author truncation commands are used.  It has to
%% go at the end of the manuscript.
%\allauthors

%% Include this line if you are using the \added, \replaced, \deleted
%% commands to see a summary list of all changes at the end of the article.
%\listofchanges

\end{document}